\documentclass{tlp}
\usepackage{aopmath}
 \usepackage{stmaryrd}
\usepackage{url}
\usepackage{amssymb}
\usepackage{natbib}

\begin{document}

\title[RedAlert]{RedAlert: Determinacy Inference for Prolog }

\author[J. Kriener and A. King]
{JAEL KRIENER and ANDY KING  \\
School of Computing, University of Kent, CT2~7NF, UK. 
}
\maketitle

\def\post {\mathsf{post}}
\def\mydownarrow {{\downarrow \!}}
\def\neck {\leftarrow}
\def \M { O}

\def \con {Con^{\downarrow}}
\def \conseqdef { \{ (\con-\{false\})^n \mid n >= 0\} \cup \{ \omega \}}
\def \conseq  { Con_{seq}^{\downarrow}}

\def \Tprog {\mathcal{F}_{P}}
\def \Tpred {\mathcal{F}_{H}}
\def \Tg {\mathcal{F}_{G}}

\def \Dp {\mathcal{D}_P}
\def \Dh {\mathcal{D}_H}
\def \Dg {\mathcal{D}_G}

\def \adelta {\abs{\delta}}

\def \se   {{success environment}}
\def \ses  {{success environments}}
\def \de   {{determinacy environment}}
\def \des  {{determinacy environments}}
\def \ade  {{abstract determinacy environment}}
\def \ades {{abstract determinacy environments}}

\newcommand{\closed} [1] {\mydownarrow \{ #1 \}}
\newcommand{\eval} [1] {\llbracket #1 \rrbracket}
\newcommand{\abs} [1] {#1^{\alpha}}
\newcommand{\dk} [1] {#1^{DK}}
\newcommand{\abstr} [2] {\alpha_{#1} (#2)}
\newcommand{\concr} [2] {\gamma_{#1} (#2)}

\newtheorem{definition}{Definition} % [section]
\newtheorem{example}{Example} % [section]

\pagerange{\pageref{firstpage}--\pageref{lastpage}}

\setcounter{page}{1}

\label{firstpage}

\begin{abstract}
This paper revisits the problem of determinacy inference addressing the problem of how to uniformly handle $cut$. To this end a new semantics is introduced for $cut$, which is abstracted to systematically derive a backward analysis that derives conditions sufficient for a goal to succeed at most once. The method is conceptionally simpler and easier to implement than existing techniques, whilst improving the latter's handling of $cut$. Formal arguments substantiate correctness and experimental work, and a tool called 'RedAlert'
demonstrates the method's generality and applicability.

\end{abstract}


\section{Introduction} 
The question of determinacy is constantly on the mind of a
good Prolog programmer.
It is almost as important to know that a goal
will not compute an answer multiply, as it is
to know that it will compute the right answer.
To this effect, Prolog programmers often use the $cut$ to literally cut off
all choice points that may lead to additional answers, once a goal has suceeded. A $cut$ that is
used to (brutely) enforce determinacy in this way is termed a ``red cut'' \citep{OKeefe}.
O'Keefe also distinguishes between further uses of $cut$, namely ``green cut'' and ``blue cut'',
which are used to avoid repeating tests in clause selection
and
exploring clauses which would ultimately fail.
Such classifications have been introduced to facilitate
reasoning about the determinising effects of $cut$ in different contexts. 
Since these issues are subtle, they
motivate developing semantically justified tools which
aid the programmer in reasoning about determinacy in the presence of $cut$.

In light of this close connection between
determinacy and $cut$, it is clear
that $cut$ ought to play a prominent role in determinacy analysis.   
This was recognised by \citet{Sahlin91}, twenty years ago, who proposed
an analysis which checks whether a goal can succeed more than once.
The analysis abstracts away from the instantiation of arguments within a call which
weakens its applicability. 
\citet{Mogensen96} recognised the need to ground the work of Sahlin on
a formal semantics, yet his work illustrates
the difficulty of 
constructing and then abstracting a semantics for $cut$.
Very recently \citet{Schneider-KampGSST10} have shown how a semantics, carefully crafted to
facilate abstraction, can be applied to check termination of logic
programs with $cut$ on classes of calls.  This begs the question whether a semantics can be distilled
which is ameniable to inferring determinacy conditions.
A good answer to this question will provide the basis for a tool that supports the software
development process by providing determinacy conditions in the presence of $cut$.

\subsection{Existing methods for determinacy inference}
The issue of inferring determinacy in logic programs has been considered before
\citep{esop/LuK05,iclp/KingLG06}, though neither of the works adequately addressed
the $cut$. \citet{iclp/KingLG06} for example present a method for infering determinacy conditions initially for $cut$-free Prolog programs by using suspension analysis in a constraint-based framework. Their motivation is to overcome a limitation of the method presented by \citet{esop/LuK05} that arises from the way in which the order of the literals in the clause influences the strength of the determinacy conditions inferred. To demonstrate this problem, consider the following example:

\begin{center}
\begin{verbatim}
diag([],[],_).
diag([(X,Y)|Xs],[(Y,X)|Ys],[_|Ds]) :- diag(Xs,Ys,Ds).

vert([],[],_).
vert([(X,Y)|Xs],[(X1,Y)|Ys],[_|Ds]) :- {X1 = -X}, vert(Xs,Ys,Ds).

rot(Xs,Ys) :- diag(Xs,Zs,Ys), vert(Zs,Ys,Xs).
\end{verbatim}
\end{center}

\noindent(The constraint notation in the second clause of \verb|vert| is needed to render the predicate multi-modal.)
The method presented by \citet{esop/LuK05} infers the groundness of \verb|Xs| as a sufficient condition for the determinacy of \verb|rot(Xs,Ys)|. It does not detect that the groundness of \verb|Ys|, too, is sufficient for determinacy. This is because the method only considers the left-to-right flow of information from one goal to the next. For instance, 
%if \verb|Xs| is ground in the call \verb|rot(Xs,Ys)|, then it follows that \verb|Xs| is ground in the call to \verb|diag(Xs,Zs,Ys)|, which is sufficient for this goal to be deterministic. Execution of this goal will in turn ground \verb|Zs|, which is sufficient for the call to \verb|vert(Zs,Ys,Xs)| to be deterministic. However, 
if \verb|rot(Xs,Ys)| is called with \verb|Ys| ground, then when the call \verb|diag(Xs,Zs,Ys)| is encountered, neither \verb|Xs| nor \verb|Zs| are ground, hence the call is possibly non-deterministic and therefore the method concludes that only groundness of \verb|Xs| is sufficient for determinacy of \verb|rot(Xs,Ys)|.

In response, \citet{iclp/KingLG06} propose a framework in which the order of the literals in a clause does not impose the implicit assumption that the determinacy of a goal is not affected by the bindings subsequently made by a later goal. To demonstrate, notice that if \verb|Ys| is ground then the execution of \verb|vert(Zs,Ys,Xs)| grounds \verb|Zs|, which is sufficient for the earlier goal \verb|diag(Xs,Zs,Ys)| to be deterministic as well. They achieve this by delaying execution of a goal until a mutual exclusion condition between its clauses is fulfilled and then using suspension inference \citep{tocl/GenaimKing08} to infer a determinacy condition for the goals that constitute the body of a clause. This allows them to infer the determinacy condition $\mathtt{Xs} \vee \mathtt{Ys}$ for the goal \verb|rot(Xs,Ys)|. Notice, however, the irony in solving a problem that arises from the failure to abstract away from the temporal order of execution by adding temporal complexity into the program. 
%Furthermore, suspension inference can only infer conditions that fall within the monotone fragment of Boolean logic and hence the conditions cannot express interdependencies that can, in fact, be captured with positive Boolean formulae.
%As well as limiting expressivity, deriving a monotonic formula from a more expressive condition is not without cost.

\subsection{Limitations of existing methods}

However, the limitations of \citep{iclp/KingLG06} become sharply apparent when considering the way that the framework is extended to $cut$: Their method is extended by strengthening the determinacy condition for a predicate to ensure that calls before a $cut$ are invoked with ground arguments only. While this treatment is sufficient to handle green and blue $cut$s, it means that a $cut$ will invariably strengthen the determinacy conditions derived. This is
unsatisfactory when considering red $cut$s, given that they are used to ensure determinacy. In that case, the presence of $cut$ ought to have a weakening effect on determinacy conditions. 
To demonstrate, consider the following pair of predicates:

\begin{center}
\begin{verbatim}
memberchk(X,L) :- member(X,L), !.
member(X,[X|_]). 
member(X,[_|L]) :- member(X,L). 
\end{verbatim}
\end{center}

\noindent In the framework of \citet{iclp/KingLG06}, \verb|memberchk| inherits its determinacy conditions from \verb|member| and (if necessary) strengthens them to ensure that the arguments in the call to \verb|member| are ground. In this situation, the determinacy condition derived for \verb|member| is $false$, which cannot be strengthened within the domain of boolean constraints. Therefore the determinacy condition derived for \verb|memberchk| is $false$ as well. However, it should be obvious that the effect of the red $cut$ in this situation is to make \verb|memberchk| deterministic \textit{independently of the determinacy of} \verb|member|. 
%The determinacy condition that ought to be derived for it is $true$.
This example demonstrates that in the presence of $cut$, determinacy conditions on predicates cannot be derived by a straightforward compositional method where parent predicates inherit their conditions from their sub-predicates. Rather, the method needs to allow for weakening and disregarding of determinacy information in the transition from parent to sub-predicates. Aiming to develop a uniform technique for handling $cut$ along these lines, this paper makes the following contributions:
\begin{itemize}

\item it presents a concise semantics for Prolog with $cut$, based on a $cut$-normal
form, that constitutes the basis for a correctness argument (and
as far as we are aware
the sequence ordering underpinning the semantics is itself novel);

\item it presents and proves correct a method for inferring determinacy conditions on Prolog predicates which abstracts over the order of their execution and is both conceptually simpler and easier to implement than previous techniques; 

\item it reports experimental work that demonstrates precision improvements
over existing methods; correctness
proofs are given in \citep{Appendix}.

\end{itemize}

\section{Preliminaries} 
%This section defines background for the semantic constructions below. (All proofs can be found in \cite{Appendix}.)
\subsection{Computational domains}
The basic domain underlying the semantics presented in the next section is the set of constraints, $Con$, containing diagonalization constraints of the form $\vec{x} = \vec{y}$, expressing constraints on and bindings to program variables. $Con$ is pre-ordered by the entailment relation, $\models$, and closed under disjunction and conjunction. We assume the existence of an extensive projection of $\theta$ onto $\vec{x}$, denoted by $\overline{\exists}_{\vec{x}} (\theta)$. 

\subsubsection{$\con$}
Our concrete domain is the set of closed non-empty sets of constraints ($\con$), which represent program states by capturing all possible bindings to the program variables consistent with a specific set of constraints on the same. The elements of $\con$ are constructed thus: For any set of constraints $\Theta$, $\closed{\Theta} = \{ \phi \mid \exists \theta \in \Theta _{.} \phi \models \theta \}$, i.e. the set of all constraints that entail some constraints in $\Theta$. (Observe that $\closed{false} = \{false\}$.) In this construction, unification is straightforwardly modeled by intersection: The result of unifying variable $A$ with constant $c$ at state $\closed{\Phi}$ is simply $\closed{A \!= \!c} \cap \closed{\Phi}$.
$\con$ is partially ordered by $\subseteq$ and $\langle \con,\ \subseteq,\ \{false\},\ \closed{true},\ \bigcup,\ \bigcap \rangle$ is a complete lattice. (Notice that $\emptyset \notin \con$.)
\\
Two projections, one an over-, the other an under-approximation, are defined on $\con$ as follows: $ \overline{\exists}_{\vec{x}} (\Theta) = \{\overline{\exists}_{\vec{x}} (\theta) \mid \theta \in \Theta  \}$, $ \overline{\forall}_{\vec{x}} (\Theta) = \{ \psi \in \Theta \mid \overline{\exists}_{\vec{x}}( \psi) = \psi \}$. Notice that both projections on $\con$ are defined in terms of an arbitrary existential projection on the elements of $Con$. %Maybe add that this is deliberate?
Each of these two is required later on to ensure soundness: The denotational and success set semantics (Sects. 3.1 and 3.2) need to be over-approximations to be correct. Intuitively, they need to capture \textit{all} possible solutions, even at the cost of letting a few impossible ones slip in. The determinacy semantics (Sect. 3.3) needs to be an under-approximation, which in that context has the effect of strengthening the determinacy condition. Weakening would lead to a loss of soundness there.
A renaming operator $\rho_{\vec{x},\vec{y}}$ is defined on $\con$ thus: $\rho_{\vec{x},\vec{y}} (\Theta) = \overline{\exists}_{\vec{y}}(\overline{\exists}_{\vec{x}}(\Theta) \cup \{\vec{x} = \vec{y}\})$. (Notice here that $\rho_{\vec{x},\vec{y}}(\Theta) = \rho_{\vec{x},\vec{y}}( \overline{\exists}_{\vec{x}} (\Theta)) $.)
For a single constraint $\theta$, $vars(\theta)$ is the set of all variables occurring in $\theta$. 
\\
Similar to the notion of definiteness defined by \cite{BakSon93}, a constraint $\theta$ \textit{fixes} those variables, in respect to which it cannot be strengthened:
\begin{center}
$ fix(\theta) = \{ y \mid  \forall \psi_{.}( (\psi \models \theta \wedge \psi \neq false) \rightarrow \overline{\exists}_{\vec{y}}(\theta) \models \overline{\exists}_{\vec{y}} (\psi) ) \} $
\end{center} 
Put simply, $fix(\theta)$ is the set of variables that are fixed or
grounded by $\theta$. 

In addition to these fairly standard constructions, we define two binary operators on $\con$ to express more complex relations between its elements:      
Given $\Theta_1$, $\Theta_2 \in \con$  their mutual exclusion ($mux$) is the union of all
those $\phi \in Con$, which fix a set of variables, on which $\Theta_1$ and $\Theta_2$ are inconsistent:
\begin{center}
$ mux( \Theta_1, \Theta_2) = \{ \phi \mid \exists Y \subseteq fix(\phi)_{.}(\overline{\exists}_{Y} (\Theta_1) \cap \overline{\exists}_{Y} (\Theta_2) = \{false\} )\}$
\end{center}
For example, given two sets $\Theta_1 = \closed{A\!=\!c,B\!=\!d}$, $\Theta_2 = \closed{A\!=\!e,B\!=\!d}$, their mutual exclusion will contain all constraints which fix the variable $A$ to any constant $f$: $mux(\Theta_1,\Theta_2) = \closed{A\!=\!f}$. Notice that, since $\Theta_1$ and $\Theta_2$ do not disagree on $B$, fixing $B$ will not distinguish between them and $B$ is therefore not constrained in $mux(\Theta_1,\Theta_2)$. 
Observe that for $\Theta_1$, $\Theta_2 \in \con$, $ mux( \Theta_1, \Theta_2) \in \con$, i.e. the $mux$ of two closed sets is closed and that $mux(\Theta_1,\Theta_2) = \closed{true}$ if $\Theta_1$ or $\Theta_2$ is $\{false\}$.
%,- $false$ is incompatible with everything.

Given $\Theta_1$, $\Theta_2 \in \con$, their implication is defined as the union of all those elements of $\con$ which, when combined with $\Theta_1$, form subsets of $\Theta_2$:
\begin{center}
$\Theta_1 \rightarrow \Theta_2 = \bigcup \{\Phi \mid \Phi \cap \Theta_1 \subseteq \Theta_2 \}$  
\end{center}
For example, given two sets $\Theta_1 = \closed{B\!=\!d}$ and $\Theta_2 = \closed{A\!=\!c,B\!=\!d}$, $\Theta_1 \rightarrow \Theta_2 = \closed{A\!=\!c}$. 
Notice that this construction mirrors material implication on boolean formulae in that the following statements are true for any $\Theta$:
$\closed{true} \rightarrow \Theta = \Theta$,
$\Theta \rightarrow \closed{true} = \closed{true}$,
$\closed{false} \rightarrow \Theta = \closed{true}$,
$\Theta \rightarrow \closed{false} = \closed{false}$.
Notice also that it is possible to recover $\Theta_2$ from $\Theta_1 \rightarrow \Theta_2$ by simply intersecting the latter with $\Theta_1$: $\Theta_1 \rightarrow \Theta_2$ is, in a sense, a systematic weakening of $\Theta_2$ by $\Theta_1$. 

\subsubsection{$\conseq$}
To model the indeterministic behaviour of Prolog semantically, we extend $\con$ to finite sequences of its elements which do not contain the set $\{false\}$, the elements of which are denoted by
$\vec{\Theta}$. Concatenation is denoted `$:$', e.g., $\Theta_1 : [\Theta_2, \Theta_3] = [\Theta_1, \Theta_2, \Theta_3]$. To obtain a top element we add a single
infinite sequence, $\omega = [\closed{true}, \closed{true}, \ldots ]$ and
define  $\conseq$ = $\conseqdef$. $Sub_{\ell} (\vec{\Theta})$ denotes the set of all subsequences of $\vec{\Theta}$ of length $\ell$. Eg: $Sub_2([\Theta_1, \Theta_2, \Theta_3]) = \{[\Theta_1, \Theta_2], [\Theta_2, \Theta_3], [\Theta_1, \Theta_3] \}$. Given a sequence of elements of $\con$, $\Theta^*$, $trim(\Theta^*)$ is the result of removing all instances of $\{ false \}$ from $\Theta^*$. 

$\conseq$ can be partially ordered by a prefix-ordering (as is done by \cite{DebrayM88}). However, under that ordering, the presence of $cut$ poses problems in defining suitable monotonic semantic operators. Therefore, we define a partial order on $\conseq$ ($\sqsubseteq$) thus: $\forall \vec{\Theta}_1, \vec{\Theta}_2 \in {\conseq}_{.}( \vec{\Theta}_1 \sqsubseteq \vec{\Theta}_2) \ iff \ \exists \vec{\Phi} \in {Sub_m}(\vec{\Theta}_2) . (\vec{\Theta}_1 \subseteq_{pw} \vec{\Phi})$ where $|\vec{\Theta}_1| = m$ and $\subseteq_{pw}$ is point-wise comparison on sequences of equal length. The lattice $\langle \conseq, \sqsubseteq, [], \omega, \bigsqcup, \bigsqcap \rangle$ is complete (see Appendix), with $\bigsqcap$ and $\bigsqcup$ defined as follows (note that $\bigsqcap$ is needed only to define the fixpoints):\\
$
\vec{\Theta}_1 \sqcap \vec{\Theta}_2  = \left\{ \begin{array}{l l}
\vec{\Theta}_2 & if\ \vec{\Theta}_1 = \omega \\
\vec{\Theta}_1 & if\ \vec{\Theta}_2 = \omega \\
\vec{\Theta}_2 \sqcap \vec{\Theta}_1 & if\ n < m\\
trim( \bigcup_{pw} \{\vec{\Theta}_1 \cap_{pw} \vec{\Phi} \mid \vec{\Phi} \in Sub_m ( \vec{\Theta}_2 )\} ) & otherwise \\
\end{array}	\right.
$
\\ where $|\vec{\Theta}_1| = m$, $|\vec{\Theta}_2| = n$ and $\cup_{pw}$ and $\cap_{pw}$ are point-wise union and intersection, which require their operands to be equal length. $\bigsqcap S $ is defined as the lifting of $\sqcap$ to sets in the natural way. From this we can define $\bigsqcup S = \bigsqcap \{\vec{\Theta} \mid \forall \vec{\Phi} \in S _{.} \vec{\Phi} \sqsubseteq \vec{\Theta} \}$ in the normal way.  
The operators $\downarrow$, $ \overline{\exists}_{\vec{x}} $, $ \overline{\forall}_{\vec{x}} $ and $\rho_{\vec{x},\vec{y}} $ are all lifted straightforwardly to the elements of $\conseq$ as the results of applying the same operations to each member of a given $\vec{\Theta}$. Eg: $\mydownarrow \overline{\exists}_{\vec{x}} ([\Theta_1, \Theta_2]) = [\mydownarrow \overline{\exists}_{\vec{x}}(\Theta_1), \mydownarrow \overline{\exists}_{\vec{x}}(\Theta_2)]$. 
$\bigcup \vec{\Theta}$ denotes the union of all the elements of $\vec{\Theta}$, which itself is an element of $\con$. 
Finally, to save some space in the presentation of the definition of $\Tg$ in Section 3.1, a mixed $\cap$ is defined thus: $(\Phi : \vec{\Phi}) \cap \Theta = (\Phi \cap \Theta) : (\vec{\Phi} \cap \Theta)$. 
%\newpage
\subsection{Cut normal form}

To simplify the presentation of the semantics, we require each predicate in the analysed program to be defined in a single definition of the form $p(\vec{x}) \neck G_1 ; G_2, !, G_3; G_4$. For example, the \verb|memberchk| and \verb|member| predicates can be transformed to:
\small
\begin{verbatim}
memberchk(X, L) :- false; (member(X, L), !, true); false. 
member(X, L) :- L = [X| _]; (false, !, true); (L = [_| L_1], member(X, L_1)).
\end{verbatim}
\normalsize
where {\tt true} and {\tt false} abbreviate $\post(true)$ and $\post(false)$ respectively. This does not introduce a loss of generality. (For details on this transformation see Appendix.)

\subsection{Syntax and stratification}

Given this normal form, the syntax of our programs is defined as follows: 
\[
\begin{array}{@{}l@{\;}c@{\;}l@{}}
Head & ::= & p(\vec{x}) \indent \indent (\text{where } \vec{x} \text{ is a vector of distinct variables})\\
Goal & ::= & \post(\theta) \mid Head \mid  Goal,Goal \\
Predicate &::= & Head \neck Goal\ ;\ Goal\ ,\ !\ ,\ Goal\ ;\ Goal\\
Program &::= &\epsilon \mid Predicate.Program\\
\end{array}
\]
where $\post(\phi)$ indicates that $\phi$ is added to the current constraint store. Again, $vars(G)$ is the set of variables in a goal $G$. Further, $heads(P)$ contains the heads of the predicates defined in $P$.

One would expect that an off-the-shelf denotational semantics
could be taken and abstracted to distill a form of determinacy inference.
However, the non-monotonic nature of $cut$ poses a problem for the definition of such a semantics. In particular, $cut$ can be used to define inconsistent predicates, eg: $p \neck false\ ;\ p,\ !,\ false\ ;\ true$. To construct a denotational semantics, we have to address the problem posed by predicates like $p$, which cannot be assigned a consistent semantics.

\cite{AptBW88} address a parallel problem in the context of negation by banning the use of such viciously circular definitions. To this end, they introduce the notion of stratification with respect to negation. In their view, negation is used `safely', if all predicates falling under the scope of a negation are defined independently of the predicate in which that negation occurs. Given the similarity between $cut$ and $not$, it is natural to adopt a similar approach towards our analogous problem. We define stratification with respect to $cut$, assuming that $cut$ is used safely, if only predicates that are defined independently of the context of a $cut$, can decide whether it is reached or not: A program $P$ is $cut$-stratified, if there exists a partition $P = P_1 \cup \ldots \cup P_n$ such that the following two conditions are met for all $1 <= i <= n$:	1. For all $p(\vec{x}) \neck G_1 ; G_2, !, G_3; G4$ in $P_i$, all calls in $G_2$ are to predicates in $\bigcup_{j<i} P_j$.
	2. For all $p(\vec{x}) \neck G_1 ; G_2, !, G_3; G4$ in $P_i$, all calls in $G_1$, $G_3$ and $G_4$ are to predicates in $\bigcup_{j<=i} P_j$. 
Henceforth, we shall simply write `stratified' to mean `$cut$-stratified'. Notice that this restriction is almost purely theoretical. In the worst case, a $cut$ after a recursive call produces a situation like or similar to that of the predicate $p$ above, which has no stable semantics and in practice introduces an infinite loop. In the best case, such a $cut$ is simply redundant. Either way, we have not been able to find such a $cut$ in an actual Prolog program, nor have we been able to come up with an example in which such a $cut$ is put to good use. 
%Notice that this restriction is purely theoretically, since it only rules out predicate definitions like that of $p$ above. 

\section{Semantics} 
Given these preliminaries, we can now define a denotational semantics for Prolog with $cut$ (section 3.1), over $\conseq$, which is expressive enough to capture multiple answers, and a determinacy semantics (section 3.3), over $\con$, suitable for abstraction to boolean conditions.
The success set semantics presented in between these two (section 3.2) provides a link between them.  

\subsection{Denotational semantics}

To establish a basis for arguing the determinacy semantics presented in the following sections correct, we define a denotational semantics for Prolog with $cut$. The driving intuition here is, that the semantics of a program $P$ is a mapping from goals called in the context of $P$ to sequences of possible answer substitutions. 
%When the sequence returned is empty, the goal has no answers under $P$. 
The context is provided by an environment ($\mu$), henceforth called a \se \hspace{1ex}to distinguish it from other types of environments, which is a mapping from predicate heads and $\conseq$ to $\conseq$: $ Env ::= Head \rightarrow \conseq \rightarrow \conseq$. The notation $\mu[p(\vec{y}) \mapsto \vec{\Theta}]$ denotes the result of updating $\mu$ with a new assignment from $p(\vec{y})$ to $\vec{\Theta}$. For a given program $P$, the set $E_{P}$ of \ses \hspace{1ex}is point-wise partially ordered by: $\mu_1 \sqsubseteq \mu_2\ iff\ \forall p(\vec{y}), \vec{\Theta}_{.} (\mu_1 (p(\vec{y})) (\vec{\Theta}) \sqsubseteq  \mu_2 (p(\vec{y})) (\vec{\Theta}))$. For any program $P$ the lattice $\langle E_P, \sqsubseteq, \mu_{\bot}, \mu_{\top}, \bigsqcup, \bigsqcap \rangle$ is complete, where:
\[
\begin{array}{@{}l@{\;}l@{\;}l@{}@{}l@{\;}l@{\;}l@{}l@{}}
\mu_{\bot} &=& \lambda p(\vec{y}) \vec{\Theta}_{.} [] &
\mu_{\top} &=& \lambda p(\vec{y}) \vec{\Theta}_{.} \omega \\
\mu_1 \sqcup \mu_2 & = & \mu_3\ \! s_{.}t_{.}\ \! \forall \vec{\Theta}, p(\vec{y}) \! \in \! heads(P)_{.}( \mu_3 (p(\vec{y})&) \vec{\Theta}\ &=& \mu_1 (p(\vec{y})) \vec{\Theta} \sqcup \mu_2 (p(\vec{y})) \vec{\Theta} )  \\
\mu_1 \sqcap \mu_2 & = & \mu_3\ \! s_{.}t_{.}\ \! \forall \vec{\Theta}, p(\vec{y}) \! \in \! heads(P)_{.}( \mu_3 (p(\vec{y})&) \vec{\Theta}\ &=& \mu_1 (p(\vec{y})) \vec{\Theta} \sqcap \mu_2 (p(\vec{y})) \vec{\Theta} )  \\
\end{array}
\]
And $\bigsqcup$ and $\bigsqcap$ are lifted to sets of environments in the normal way. 

\begin{definition}
For a given stratified program $P$, its semantics - $\mu_P$ - is defined as a fixpoint of $\Tprog$:
\[
\begin{array}{@{}l@{\;}c@{\;}l@{}}
\Tprog &::& Program \rightarrow Env \rightarrow Env\\
\Tprog \eval{\epsilon} \mu & = & \mu \\
\Tprog \eval{P.Ps} \mu & = & \Tprog \eval{Ps} (\mu [p(\vec{y}) \mapsto (\Tpred \eval{P} \mu) (p(\vec{y})) ])
\\ & where & P = p(\vec{y}) \neck B 
\\[1.5ex]
\Tpred &::& Predicate \rightarrow Env \rightarrow Env\\
\Tpred \eval{p(\vec{y}) \neck B} \mu & = &  \mu [p(\vec{y}) \mapsto \lambda \vec{\Theta}. \downarrow \overline{\exists}_{\vec{y}} (\Tg \eval{G_1} \mu \vec{\Theta} : \vec{\Psi}) ]
\\ & where  & \vec{\Psi} = \left\{ \begin{array}{l l}
\Tg \eval{G_3 } \mu [\Phi] 
& if \ \Tg \eval{ G_2} \mu \vec{\Theta} = \Phi : \vec{\Phi} \\
\Tg \eval{G_4} \mu \vec{\Theta} 
& otherwise
 \end{array} \right .
\\ & and & B = G_1 ; G_2,!,G_3 ; G_4
\\[1.5ex]
\Tg &::& Goal \rightarrow Env \rightarrow \conseq \rightarrow \conseq\\
\Tg \eval{G} \mu [] & = & [] \\
\Tg \eval{\post(\phi)} \mu (\Theta : \vec{\Theta}) & = & trim(\closed{\phi} \cap \Theta : \Tg \eval{\post(\phi)} \mu \vec{\Theta})\\
\Tg \eval{ p(\vec{x})} \mu (\Theta : \vec{\Theta}) & = & (\downarrow \rho_{\vec{y}, \vec{x}}(\ \mu\ p(\vec{y})\ (\downarrow \rho_{\vec{x}, \vec{y}}([\Theta])) ))\! \cap \! \Theta : \Tg \eval{ p(\vec{x})} \mu \vec{\Theta}
\\ &  where & p(\vec{y}) \in dom(\mu)
\\ & and & vars(\vec{x}) \cap vars(\vec{y}) = \emptyset\\
\Tg \eval{G_1, G_2} \mu (\Theta : \vec{\Theta}) & = &
\Tg \eval{G_2} \mu (\Tg \eval{ G_1} \mu (\Theta : \vec{\Theta}))
\\
\end{array}
\]
\end{definition}
%\newpage
Observe that given a stratified program $P = P_1 \cup \ldots \cup P_n$, $\Tprog$ is monotonic, under our sub-sequence order, within each stratum $P_i$. By Tarski's theorem, $\Tprog \eval{P_i}$ has a least fixed point. $\mu_P$ can therefore be defined as the result of evaluating all strata in order from lowest to highest, starting with $\mu_{\bot}$ and then taking the least fixed point of the previous stratum as input to the evaluation of the next stratum.    
\\
\indent The crucial part is in $\Tpred$, which updates the assignments in the \se \hspace{1ex}and reflects the possible indeterminacy in a predicate by splitting the resulting sequence up into the possibility resulting from executing $G_1$ and that resulting from either executing $G_3$ or $G_4$, depending on the success of $G_2$. Given a call to a predicate, $\Tg$ imposes onto each open possibility (i.e. each member of $\vec{\Theta}$) the constraints associated with that predicate in the given $\mu$.
The constraints are determined by the application of $\mu$ to that predicate,
after first applying projection and renaming operations required to match formal
and actual parameters. Information about other variables, which is lost in that process, is recovered by intersecting the result of the predicate call with the previous state of computation. The effect of this is, that constraints on the variables that the predicate is called on are strengthened in accordance with its definition, while those on all other variables are preserved. 
Given a goal of the form `$\post(\phi)$' or `$G_1, G_2$', $\Tg$ does what you would expect: In the former case, it imposes $\phi$ onto each open possibility in the current state of computation, filtering out those possibilities which fail as a result. In the latter case, it successively evaluates $G_1$ and $G_2$. Notice further that given an empty sequence (i.e. a failed state of computation), $\Tg$ simply returns an empty sequence, regardless of its other parameters. 

\begin{example}
To illustrate, suppose
\verb|member(A,S)| and \verb|memberchk(A,S)| are called at a point in a program
where there is only one possible set of bindings
$\Theta = \closed{A=3 \wedge S=[3,2,3]}$. \\
$\Tg \eval{member(A,S)}\ \mu\ [\Theta]
= [\Theta \cap \closed{S\!=\![A|\_]},\Theta]$ \\
$\Tg \eval{memberchk(A,S)}\ \mu\ [\Theta] = [\Theta \cap \closed{S\!=\![A|\_]}]$
\end{example}

\subsection{Success set semantics}

For the purposes of the determinacy inference, a coarser representation of the constraints under which a goal can succeed is given by the following pair of functions.
%required. This is achieved by the following pair of functions, with $S_G \eval{G}$ soundly over-approximating $\Tg \eval{G}$.
\begin{definition}
For a given program $P$, $S_G: Goal \to Con^{\mydownarrow}$ and $S_H: Head \to Con^{\mydownarrow} $ are defined as the least maps, such that:  
\[
\begin{array}{@{}l@{\;}c@{\;}l@{}}
S_G \eval{ \post(\phi)} & = & \closed{\phi} 
\\
S_G \eval{p(\vec{x})} & = & \downarrow \rho_{\vec{y},\vec{x}} (S_H \eval{p(\vec{y})})  
\\ & where & p(\vec{y}) \neck  B \in P  
\\ & and & vars(\vec{x}) \cap vars(\vec{y}) = \emptyset
\\
S_G \eval{G_1, G_2}  & = & S_G \eval{G_1} \cap S_G \eval{G_2}
\\[2ex]
S_H \eval{p(\vec{y})} & = & \downarrow \overline{ \exists}_{\vec{y}} ( S_G \eval{G_1} \cup S_G \eval{G_2, G_3} \cup S_G \eval{G_4} )
\\ & where & p(\vec{y}) \neck B \in P \ and \ B = G_1\ ;\ G_2\ ,\ !\ ,\ G_3\ ;\ G_4 
\end{array}
\]
\end{definition}

\begin{example}
To illustrate consider again \verb|member| and \verb|memberchk|:
$S_G \eval{memberchk(A,S)}$ = \\
\mbox{$S_G \eval{member(A,S)}$} = 
$\closed{S\!=\![A|\_]} \cup 
\closed{S\!=\![\_,A|\_]} \cup 
\closed{S\!=\![\_,\_,A|\_]} \cup \ldots$.
%Notice that the last component of the union is $\closed{true}$, because it is the existential (therefore extenisve) projection onto $\langle A,S \rangle$ of a set that constraints the variables in the recursive call.\\
\end{example}

%\normalsize
Theorem 1 states that $S$ is a sound over-approximation of $\mathcal{F}$:
\begin{theorem}
$ \bigcup \Tg \eval{G} \mu_P \vec{\Theta} \subseteq ( \bigcup \vec{\Theta} ) \cap S_G \eval{G}$ \indent Proof: See Appendix.
\end{theorem}

\subsection{Determinacy semantics}
With these in place, we can construct and prove correct a group of functions to derive a set of constraints which guarantee the determinacy of a goal in the context of a program $P$, its determinacy condition, henceforth abbreviated to `dc'. As before, the context is provided as an environment: A \de \hspace{1ex}($\delta$) is a mapping from predicate heads to $\con$: $DEnv ::= Head \rightarrow \con$. Again, $\delta[p(\vec{y}) \mapsto \Theta]$ is an update operation. As above, the set $E^d_{P}$ of \des \hspace{1ex}for a program $P$ is partially ordered point-wise by: $\delta_1 \sqsubseteq \delta_2\ iff\ \forall p(\vec{y})_{.} (\delta_1 (p(\vec{y})) \subseteq \delta_2 (p(\vec{y})))$.
The lattice $\langle E^d_P, \sqsubseteq, \delta_{\bot}, \delta_{\top}, \bigsqcup, \bigsqcap \rangle$ is complete, with: 
\[
\begin{array}{@{}l@{\;}l@{\;}l@{}@{}l@{\;}l@{\;}l@{}l@{}}
\delta_{\bot} &= &\lambda p(\vec{y}) _{.} \{ false \} &
\delta_{\top} &=& \lambda p(\vec{y}) _{.} \closed{true} \\
\delta_1 \sqcup \delta_2 &=& \delta_3\ such\ that\ \forall p(\vec{y}) \in heads(P) . (\delta_3 (p(\vec{y}&)) &=& \delta_1 (p(\vec{y})) \cup \delta_2 (p(\vec{y}))) \\
\delta_1 \sqcap \delta_2 &=& \delta_3\ such\ that\ \forall p(\vec{y}) \in heads(P) . (\delta_3 (p(\vec{y}&)) &=& \delta_1 (p(\vec{y})) \cap \delta_2 (p(\vec{y}))) \\
\end{array}
\]
And again, $\bigsqcup$ and $\bigsqcap$ are lifted to sets in the normal way.

\begin{definition} 
The determinacy semantics - $\delta_P$ - of a program $P$ is the greatest fixpoint of $\Dp \eval{P}$: 
\[
\begin{array}{@{}l@{\;}c@{\;}l@{}}
\Dp & :: & Program \rightarrow DEnv \rightarrow DEnv \\
\Dp \eval{\epsilon} \delta & = & \delta \\
\Dp \eval{P.Ps} \delta & = & \Dp \eval{Ps} (\delta[p(\vec{y}) \mapsto (\Dh \eval{P} \delta) (p(\vec{y}))])\\
 & where & P = p(\vec{y}) \neck B
\\[1.5ex]
\Dh & :: & Predicate \rightarrow DEnv \rightarrow DEnv \\
\Dh \eval{p(\vec{y}) \neck B} \delta & = & \delta[p(\vec{y}) \mapsto  \downarrow \overline{\forall}_{\vec{y}}( \Dg \eval{ G_1} \delta \\
& & \indent \indent \indent \indent \indent \cap  (S_G \eval{ G_2} \rightarrow \Dg \eval{G_3} \delta) \\
& & \indent \indent \indent \indent \indent \cap  \Dg \eval{G_4} \delta \cap \Theta_1 \cap \Theta_2)] 
\\ & where & \Theta_1 = mux( S_G \eval{G_1} , S_G \eval{G_4})  
\\ & and & \Theta_2 = mux( S_G \eval{G_1}, S_G \eval{ G_2, G_3})
\\ & and & p(\vec{y}) \neck G_1\ ;\ G_2\ ,\ !\ ,\ G_3\ ;\ G_4  \in P 
%\end{array}
%\]
%\newpage
%\[
%\begin{array}{@{}l@{\;}c@{\;}l@{}}
\\[1.5ex]
\Dg & :: & Goal \rightarrow DEnv \rightarrow \con \\
\Dg \eval{ \post(\phi)} \delta & = & \closed{ true} \\
\Dg \eval{ p(\vec{x})} \delta  & = & \downarrow \rho_{\vec{y},\vec{x} }  \overline{\forall}_{\vec{y}}( \delta (p(\vec{y})) )
\\ & where & p(\vec{y}) \in dom(\delta)\\
\Dg \eval{ G_1,G_2 } \delta & = & (S_G \eval{ G_2} \rightarrow \Dg \eval{ G_1} \delta) \cap (S_G \eval{ G_1} \rightarrow \Dg \eval{ G_2} \delta) \\
\end{array}
\]
\end{definition}

Given a goal of the form `$\post(\phi)$', $\Dg$ returns $\closed{true}$ since the goal cannot introduce indeterminacy in the computation. As before, given a predicate call, $\Dg$ applies the projection and renaming necessary to match 
parameters before calling $\Dh$. Notice that the projection used here is $\overline{\forall}$, since an under-approximation is required to derive a sufficient condition. $\Dh$ maps predicates defined in $cut$ normal form to a condition that entails: (a) the dc for $G_1$, (b) the dc for $G_3$ weakened by the success set of $G_2$ - the intuition here being that the dc for $G_3$ will only be relevant if $G_2$ can succeed and therefore its dc can be weakened by the success set of $G_2$ - (c) the dc for $G_4$, and finally mutual exclusion conditions for the two possibilities arising from the structure of the predicate definition. (The case that needs to be excluded is that of $G_1$ succeeding and subsequently $G_2$ and $G_3$ succeeding or subsequently $G_2$ failing and $G_4$ succeeding.) Finally, when given a compound goal `$G_1, G_2$', $\Dg$ returns a condition that entails both the dc for $G_2$ weakened by the success set of $G_1$ and the dc for $G_1$ weakened by the success set of $G_2$. The intuition here is, that the temporal order of execution is irrelevant. Weakening the dc for $G_2$ by the success set of $G_1$ is intuitive, since one can safely assume that $G_1$ will have succeeded at the point when determinacy of $G_2$ needs to be enforced. But similarly, when enforcing determinacy on $G_1$, one can safely assume that $G_2$ \textit{will} succeed, since both $G_1$ and $G_2$ need to succeed for the compound goal to succeed.
  
\begin{example}
Consider again \verb|member| and \verb|memberchk|.
Observe that
$\Dg \eval{member(A,S)}\ \delta = \{false\} $ 
since $mux(S_G \eval{G_1}, S_G \eval{G_4}) = \{false\}$ 
is a component of $\Dh \eval{member(X,L)} \delta$, 
where
$G_1 = (L=[X|\_])$ and
$G_4 = (L=[\_|L_1],member(X,L_1))$.
\verb|member| is therefore inferred to be non-deterministic for exactly the right reason: There is no groundedness condition on its parameters such that only one of its clauses can succeed. \\
$\Dg \eval{memberchk(A,S)}\ \delta 
=  \downarrow \rho_{\vec{y},\vec{x} }  \overline{\forall}_{\vec{y}}(\closed{true} \cap (S_G \eval{member(A,S)} \rightarrow \closed{true}) \cap \closed{true} \cap mux(\{false\},\{false\}) \cap mux(\{false\}, S_G \eval{member(A,S),true} ) )\\
 = \closed{true}$\\
The crucial observation here is, that $\Dg \eval{member(A,S)}\ \delta$ is not required in this construction at all;
%Only $S_G \eval{member(A,S)}$ is required.
\verb|memberchk| does not simply inherit its condition from \verb|member|.
\end{example}

\noindent Theorem 2 states that, in the context of a stratified program $P$, the condition given by $\Dg \eval{G} \delta_P$ is indeed sufficient to guarantee the determinacy of a call to $G$:

\begin{theorem}
If $\Theta \subseteq \Dg \eval{G} \delta_P$ then
$|\Tg \eval{G} \mu_P [ \Theta ] | \leq 1$ for stratified $P$ (i.e. $P = P_0 \cup \ldots \cup P_n$).

Proof: See Appendix
\end{theorem}

\section{Abstraction}
In order to synthesize a determinacy inference from the above determinacy semantics, we systematically under-approximate sets of constraints with boolean formulae that express groundness conditions. 
$Pos$, however, is augmented with a constant for falsity, so as to express unsatisfiable requirements. The abstract domain $\langle Pos_{\bot}, \models, true, false, \wedge, \vee \rangle$ is a complete lattice \citep{armstrong98two} and to
define the abstraction of a single atomic constraint we introduce: 
\[
\begin{array}{@{}l@{\;}c@{\;}l@{}}
\abstr{\vec{x}}{\theta} & = &  \big( \bigwedge (vars(\vec{x}) \cap fix(\theta))  \wedge \neg \bigvee (vars(\vec{x}) \setminus fix(\theta)) \big) \vee \bigwedge vars(\vec{x}) 
\end{array}
\]
\noindent For example, if $\theta = A\!=\!c$
then
$\abstr{\langle A \rangle}{\theta} = A$, while $\abstr{\langle A,B,C \rangle}{\theta} =
(A \wedge \neg B \wedge \neg C) \vee (A \wedge B \wedge C)$. Notice that finiteness is achieved by limiting the scope to a finite vector of variables $\vec{x}$. A Galois connection can then be established thus:
\[
\begin{array}{@{}l@{\;}l@{\quad }l@{}@{\;}l}
\alpha_{\vec{x}} & :: \con \rightarrow Pos_{\bot} & \gamma_{\vec{x}} & :: Pos_{\bot} \rightarrow \con \\
\abstr{\vec{x}}{\Theta} & =  \bigvee \{ \abstr{\vec{x}}{\theta} \mid \theta \in \Theta \wedge \theta \neq false \} & \concr{\vec{x}}{f} & = \bigcup \{ \Theta \in \con \mid \abstr{\vec{x}}{\Theta} \models f  \}
\end{array}
\]
For instance, if $\Theta = \closed{A\!=\!c,B\!=\!d}$ then $\abstr{\langle A,B \rangle}{\Theta} = A \wedge B$. \\
\\
The following two propositions and two axioms establish relations between the concrete notions of implication, mutual exclusion and the projections and their abstract counterparts. (Notice that abstract implication is simply boolean implication.) 
\begin{paragraph}{Abstract Implication}
Proposition 1 establishes the link between concrete ($\rightarrow$) and abstract ($\Rightarrow$) implication as follows:
\begin{proposition} \rm
If
$\Theta_1 \subseteq \concr{\vec{x}}{f_1}$
and
$\concr{\vec{x}}{f_2} \subseteq \Theta_2$
then
$\concr{\vec{x}}{f_1 \Rightarrow f_2} \subseteq \Theta_1 \rightarrow \Theta_2$
Proof:
See Appendix. 
\end{proposition}
\end{paragraph}
\begin{paragraph}{Abstract Mutual Exclusion}
In order to construct an abstract mutual exclusion operator
we need to approximate elements of $\con$.
We do so with depth-$k$ abstractions which are finite
sets $\Theta^{DK} \subseteq Con$ such that each atomic constraint
$\theta$ 
of the form $x\!=\!t$
occurring in $\Theta^{DK}$
has a term $t$ whose depth does not exceed $k$.
From these we synthesize boolean requirements sufficient for mutual exclusion thus:
\[
 \abs{mux_{\vec{x}}} (\Theta_1^{DK}\!, \Theta_2^{DK}) = \vee \left\{ \wedge\! Y \left| 
 \begin{array}{@{}l@{}}
  Y \subseteq vars(\vec{x}) \quad \wedge \\ 
\forall \theta_1\! \in\! {\Theta_1^{DK}}\!, \theta_2\! \in\! {\Theta_2^{DK}}\!_{.} (\overline{\exists}_Y (\theta_1) \wedge \overline{\exists}_Y (\theta_2) = \bot ) 
\end{array} \right. \right\} 
\]
Notice, again, that $\abs{mux_{\vec{x}}} (\Theta_1^{DK}\!, \Theta_2^{DK}) = true$ if either of $\Theta_1^{DK}$ or $\Theta_2^{DK}$ is $\{false\}$.

\begin{example}
Consider
$\abs{mux_{\langle X,L \rangle}}(\{L\!=\![]\}, S_G \eval{G_4}^{DK})$ 
where
$G_4 = (L=[\_|L_1],member(X,L_1))$.
If depth $k\!=\!3$, then $S_G\eval{G_4}^{DK} = \{ \theta_1, \theta_2 \}$
where
$\theta_1 = (L_1= [X|\_] \wedge L = [\_|L_1])$
and
$\theta_2 = (L_1= [\_,X|\_] \wedge L = [\_|L_1])$. 
In this situation $\abs{mux_{\langle X,L \rangle}}(\{L \!=\! []\}, S_G\eval{G_4}^{DK})$ is $L \vee (L \wedge X) \!=\! L$.
\end{example}

\noindent Proposition 2 states how this abstract construction and the concrete one are related:
\begin{proposition} 
$ \concr{\vec{x}}{\abs{mux_{\vec{x}}} (\Theta_1^{DK}, \Theta_2^{DK})} \subseteq mux(\Theta_1, \Theta_2)$ \indent Proof:  See Appendix.
\end{proposition}
\end{paragraph}
\begin{paragraph}{Abstract Projections}
Had we defined a specific concrete projection on single constraints, we could synthesis abstract ones in the standard way \citep{CousotC79}. However, since both concrete projection operators on $\con$ are defined in terms of an arbitrary projection on single constraints, we follow \citet[Sect.7.1.1]{Gi92} in simply requiring the following to hold for any such projection:
\begin{center}
$\overline{\exists}_{\vec{x}} ( \concr{} {f}) \subseteq \concr{}{\abs{\overline{\exists}_{\vec{x}}}(f)}$ \qquad
$\concr{}{ \abs{\overline{\forall}_{\vec{x}}}(f)} \subseteq  \overline{\forall}_{\vec{x}} ( \concr{} {f})$ 
\end{center}

\noindent In addition to the above two axioms, a requirement on the relation between concrete and abstract renaming functions in the context of universal projection is stipulated: 
\begin{center}
$\concr{vars(\vec{x})}{ \abs{\rho_{\vec{y},\vec{x} } } \abs{\overline{\forall}_{\vec{y}}}(f)} \subseteq  \rho_{\vec{y},\vec{x}} \overline{\forall}_{\vec{x}} ( \concr{vars(\vec{y})} {f})  $
\end{center}
\end{paragraph}
\subsection{ Abstract success semantics}
The last construction that needs to be abstracted in order to mechanise the determinacy semantics presented above is the success set construction $S$. 
\begin{definition} \rm
The abstract success semantics is defined as the least maps $\abs{S_G}$, $\abs{S_H}$ such that: 
\[
\begin{array}{@{}l@{\;}c@{\;}l@{}}
\abs{S_G} \eval{ \post(\phi)} & = & \alpha_{vars (\phi)} (\phi) 
\\
\abs{S_G} \eval{p(\vec{x})} & = & \downarrow \abs{\rho_{\vec{y},\vec{x}}} (\abs{ \overline{\exists}_{\vec{y}}}(\abs{S_H} \eval{p(\vec{y})}))  
\\ & where & p(\vec{y}) \neck  B \in P  
\\
\abs{S_G} \eval{G_1, G_2}  & = & \abs{ S_G} \eval{G_1} \wedge \abs{S_G} \eval{G_2}
\\[3ex]
\abs{S_H} \eval{p(\vec{y})} & = & \downarrow \abs{\overline{ \exists}_{\vec{y}}} ( \abs{S_G} \eval{G_1} \vee \abs{S_G} \eval{G_2, G_3} \vee \abs{S_G} \eval{G_4} )
\\ & where & p(\vec{y}) \neck B \in P \ and \ B = G_1\ ;\ G_2\ ,\ !\ ,\ G_3\ ;\ G_4 
\end{array}
\]
\end{definition}

\noindent Proposition 3 formalises the connection between $\abs{S}$ and its concrete counterpart:
\begin{proposition} $S_G \eval{G} \subseteq \concr{vars(G)}{\abs{S_G} \eval{G}}$ \indent Proof: standard.
\end{proposition}

\noindent Depth-$k$ abstractions can be derived analogously to groundness
dependencies and therefore we omit these details.

\subsection{Determinacy inference}

Finally, an \ade \hspace{1ex}($\adelta$) is a mapping from predicate heads to Boolean formulae representing groundness conditions on the arguments of the predicate sufficient to guarantee determinacy of a call to that predicate: $ADEnv ::= Head \rightarrow Pos_{\bot}$.
As in the case of \des, the set of \ades \hspace{1ex}for a given program ($\abs{E_P}$) is partially ordered point-wise by $\adelta_2 \sqsubseteq \adelta_1\ iff\ \forall p(\vec{y})_{.}(\adelta_1(p(\vec{y})) \models \adelta_2(p(\vec{y})) )$. The lattice $\langle \abs{E_P}, \sqsubseteq, \adelta_{\bot}, \adelta_{\top}, \bigsqcup, \bigsqcap \rangle $ is complete, where $\adelta_{\top} = \lambda p(\vec{y})_{.} true$, $\adelta_{\bot} = \lambda p(\vec{y})_{.} false$ and $\bigsqcup$ and $\bigsqcap$ are constructed analogously to the case of concrete environments.  For a given program $P$, its abstract determinacy semantics -- $\adelta_P$ -- is defined as the greatest fixed point of $\abs{\Dp} \eval{P} \adelta_{\top}$, where $\abs{\Dp}$ is given by the following construction which, unsurprisingly, is very similar in structure to the definition of $\Dp$: (We write $(S_G \eval{G})^{DK}$ as $\dk{S_G} \eval{G}$.)
\begin{definition} \rm
\[
\begin{array}{@{}l@{\;}c@{\;}l@{}}
\abs{\Dp} & :: & Program \rightarrow ADEnv \rightarrow ADEnv \\
\abs{\Dp} \eval{\epsilon} \adelta & = & \adelta \\
\abs{\Dp} \eval{P . Ps} \adelta & = & \Dp \eval{Ps} (\adelta [p(\vec{y}) \mapsto (\abs{\Dh} \eval{P} \adelta) (p(\vec{y})  ) ] ) \\
 & where & P = p(\vec{y}) \neck B 
\\[1.5ex]
\abs{\Dh} & :: & Predicate \rightarrow ADEnv \rightarrow ADEnv \\
\abs{\Dh} \eval{p(\vec{y}) \neck B} \adelta & = & \adelta[p(\vec{y}) \mapsto  \abs{\overline{\forall}_{\vec{y}}}( \abs{\Dg} \eval{ G_1} \adelta \\
 & & \indent \indent \indent \indent \indent \wedge (\abs{S_G} \eval{ G_2} \Rightarrow \abs{\Dg} \eval{G_3} \adelta)\\
 & & \indent \indent \indent \indent \indent \wedge \abs{\Dg}  \eval{G_4} \adelta \wedge f_1 \wedge f_2 ]\\
 & where & f_1 = \abs{mux_{vars(\vec{y})}} ( \dk{S_G} \eval{G_1} , \dk{S_G} \eval{G_4})  
\\ & and & f_2 = \abs{mux_{vars(\vec{y})} }  ( \dk{S_G} \eval{G_1}, \dk{S_G} \eval{ G_2, G_3})
\\ & and & B =  G_1 ; G_2,!,G_3 ; G_4
%\\[1.5ex]
\end{array}
\]
\newpage
\[
\begin{array}{@{}l@{\;}c@{\;}l@{}}
\abs{\Dg} & :: & Goal \rightarrow ADEnv \rightarrow Pos_{\bot}\\
\abs{\Dg} \eval{ \post(\phi)} \adelta & = & true
\\
\abs{\Dg} \eval{p(\vec{x})} \adelta & = & \abs{\rho_{\vec{y},\vec{x} }}  \abs{\overline{\forall}_{\vec{y}}} ( \adelta (p(\vec{y})) )
\\ & where & p(\vec{y}) \in  dom(\adelta)
\\
\abs{\Dg} \eval{ G_1,G_2 } \adelta & = & (\abs{S_G} \eval{ G_2} \Rightarrow \abs{\Dg} \eval{ G_1} \adelta) \wedge (\abs{S_G} \eval{ G_1} \Rightarrow \abs{\Dg} \eval{ G_2} \adelta) 
\end{array}
\]
\end{definition}

%\begin{example}
%Consider one more time \verb|member(A,S)| and \verb|memberchk(A,S)|: \\
%$\abs{\Dg} \eval{member(A,S)} \adelta \!=\!
%% \abs{\rho_{\langle A,S \rangle},{\langle X,L \rangle} }  \abs{\overline{\forall}_{\langle S,L \rangle}} true \wedge (false \Rightarrow true) \wedge (() \Rightarrow ()) \wedge  \abs{mux_{\langle X,S \rangle}} ( \{false\} , \dk{S_G} \eval{G_4}) \wedge  \abs{mux_{\langle X,S \rangle}} ( \{false\} , \dk{S_G} \eval{G_4}) )$ \\
%$
%$\abs{\Dg} \eval{memberchk(A,S)} \adelta \!=\! 
%%\abs{\rho_{\vec{\langle A,S \rangle},\vec{\langle X,L \rangle} }}  \abs{\overline{\forall}_{\langle S,L \rangle}}(true \wedge (\abs{S_G} \eval{member(X,L)} \Rightarrow true) \wedge true \wedge  \abs{mux_{\langle X,L \rangle}} (\{false\}\!, \{false\}) \wedge \abs{mux_{\langle X,L \rangle}} (\{false\}\!, \dk{S_G} \eval{member(X,L),true}) )\\
%= \abs{\rho_{\langle A,S \rangle,\langle X,L \rangle }}  \abs{\overline{\forall}_{\langle S,L \rangle}}( \abs{S_G} \eval{member(X,L)}) = $
%\end{example}

\noindent Theorem 3 states that each parallel application of $\Dp$ and $\abs{\Dp}$ preserves the correspondence between the dc and its abstract counterpart and
Corollary 1 states a direct consequence of this, namely that the same correspondence holds between the greatest fixpoints of these constructions. 
\begin{theorem}
 $\forall i \in \mathbb{N}: \concr{vars(G)}{\abs{\Dg} \eval{G} \adelta_i} \subseteq \Dg \eval{G} \delta_i$, where $\adelta_i$ (resp. $\delta_i$) are the results of $i$ applications of $\abs{\Dp} \eval{P}$ (resp. $\Dp \eval{P}$) to $ \adelta_{\top}$ (resp. $\delta_{\top}$). \indent Proof: See Appendix.
\end{theorem}

\begin{corollary}
$\concr{vars(G)}{\abs{\Dg} \eval{G} \adelta_P} \subseteq \Dg \eval{G} \delta_P$ \indent Proof: Straightforward.
\end{corollary}

\noindent These two statements establish, in effect, that $\adelta_P$ is correct with respect to (i.e. is a sound under-approximation of) $\delta_P$. The significance of this is, that the correctness of $\Dg \eval{G} \delta_P$ as a determinacy condition for $G$, which was proved in the last section, is carried over to $\abs{\Dg} \eval{G} \adelta_P$. Since the latter is finite and can be mechanised, an implementation is therefore proven to give a correct (if possibly overly strong) determinacy condition for a goal $G$ in the context of a stratified program $P$.

\section{Implementation}
The determinacy inference specified in the previous section is realised as a tool called `RedAlert', using a simple bottom-up fixpoint engine in the style of those discussed by \citet{Cod-Son:JonesFest02}. Boolean formulae are represented in CNF as lists of lists of non-ground variables.
In this way, renaming is straightforward and
conjunction is reduced to list-concatenation \citep{HoweKing01}. However, disjunction, implication and existential quantifier elimination are performed by enumerating prime implicants \citep{3094}, which reduces these operations to incremental SAT.
The solver is called through a foreign language interface following \cite{Cod08}.
It is interesting to note, that we have not found any of the benchmarks to be non-stratified, though even if this were the case, a problematic $cut$ could be discarded albeit at the cost of precision. 

In the case of the \verb|memberchk| predicate mentioned in the introduction, the implementation does indeed infer $true$ as its determinacy condition, as desired. To discuss a more interesting case, consider the partition predicate of quicksort. 
\small
%qs(A, C, B) :- 
%      iff(A, []), herb(A, []), iff(B, [C]), herb(B, C)
%   ;  false, !, false
%   ;  iff(A, [D, E]), herb(A, [D|E]), pt(E, D, H, I), 
%      iff(F, [D, G]), herb(F, [D|G]), qs(H, C, F), qs(I, G, B).   
%pt(A, G, B, C) :- 
%      iff(A, []), herb(A, []), iff(B, []), herb(B, []), 
%      iff(C, []), herb(C, [])
%   ;  iff(A, [D, E]), herb(A, [D|E]), iff(B, [D, F]), 
%      herb(B, [D|F]), req(D, []), req(G, []), !, pt(E, G, F, C)
%   ;  pt_0(A, G, B, C).
%pt_0(A, F, G, D) :-
%      iff(A, [B, C]), herb(A, [B|C]), iff(D, [B, E]), 
%      herb(D, [B|E]), pt(C, F, G, E)
%   ;  false, !, false
%   ;  false.
\begin{verbatim}pt([], _, [], []).
pt([X | Xs], M, [X | L], G) :- X =< M, !, pt(Xs, M, L, G).
pt([X | Xs], M, L, [X | G]) :- pt(Xs, M, L, G).\end{verbatim} \normalsize
The method presented in \cite{iclp/KingLG06} handles this $cut$ by enforcing monotonicity on the predicate. To this end, the negation of the constraint before the $cut$ ($X > M$) is conceptually added to the last clause and the $cut$ then disregarded. The groundness requirement inferred in this way for $pt(w,x,y,z)$ is 
%$x \wedge (w \vee y) \wedge (w \vee z)$.
$(w \wedge x) \vee (x \wedge y \wedge z)$.
The determinacy condition inferred for the same predicate by the method presented in this paper is: 
%$(w \vee y) \wedge (w \vee z)
$w \wedge (y \vee z)$, which is clearly an improvement, though still sufficient. 
Improvements similar to this can be observed when analysing a number of benchmark programs. 
Table 1 summarises the results of this comparison on 22 benchmarks (which are available at 
\small \url{http://www.cs.kent.ac.uk/people/staff/amk/cut-normal-form-benchmarks.zip}\normalsize). Under `\textit{org}' is the number of predicate definitions in the original program. To give a measure of the impact of the $cut$ normal form transformation, under `\textit{new}' is the number of new predicates introduced by it. Under `\textit{impr}' is the number of predicates in the original benchmark (excluding any newly introduced ones) on which the determinacy inference is improved by our method over \cite{iclp/KingLG06}. Under `\textit{mean}' is the mean size of improvement (i.e. the mean number of variables which occur in the previous determinacy condition but not in the new one). The results show
a uniform improvement. Note that randc, dialog, neural and boyer give
precision improvements but no determinancy conditions are inferred which involve
strictly fewer variables.
The runtime for the groundness analysis, the depth-$k$ analysis and the backwards analysis, that propagates determinacy requirements against the control flow, are all under a second for all benchmarks (and
not even SCCs are considered in the bottom-up fixpoint calculations). However, the overall runtime is up to an order of magnitude greater, due to the time required to calculate the mutual exclusion conditions. This is because the definition of the abstract mutual exclusion in section 4 is inherently exponential in the arity of a predicate.  This is currently the bottleneck. 

\begin{table}
\begin{tabular}{|@{}r|@{}r|@{}r|@{}r|@{}r||@{}r|@{}r|@{}r|@{}r|@{}r|}
\cline{1-10}
{\it benchmark} & {\it org} & {\it new} &  {\it impr} & {\it mean} &
{\it benchmark} & {\it org} & {\it new} &  {\it impr} & {\it mean} \\
\cline{1-10}
asm		&	44&	157&		5&	0.6&
peval		&	108&	14&		2&	1\\
\, crypt\_wamcc	&	11&	12&		2&	2&
nandc		&	12&	5&		2&	0\\
semi		&	22&	19&		0&	0&
life		&	10&	11&		7&	1.85\\
qsort		&	3&	1&		1&	1&
ronp		&	16&	5&		4&	1\\
browse		&	15&	7&		1&	2&
tsp		&	23&	2&		10&	1.4\\
ga		&	58&	102&		2&	1.5&
flatten		&	27&	25&		6&	1.5\\
dialog		&	30&	11&		3&	0 &
neural		&	34&	23&		3&	0\\
unify		&	26&	33&		3&	1.33 &
nbody		&	48&	34&		11&	2\\
peep		&	20&	189&		0&	0&
boyer		&	26&	95&		4&	0\\
read		&	42&	89&		0&	0&
qplan		&	65&	41&		7&	2.57\\
reducer		&	31&	57&		9&	2&
simple\_analyzer&	60&	50&		9&	2.22\\ \cline{1-10}
\end{tabular}
\caption{Comparison }\end{table}

\section{Related Work} 
\begin{paragraph}{Determinacy inference and analysis}
As mentioned above, \citet{esop/LuK05} and \citet{iclp/KingLG06} address the problem of inferring determinacy conditions on a predicate. Since their limitations have been discussed above, we will not repeat them here.
\citet{DawsonRRS93} present a method for inferring determinacy information from a program by adding constraints to the clauses of a predicate which allow the inference of mutual exclusion conditions between these clauses rather than determinacy conditions for a whole predicate. 
\citet{Sahlin91} presents a method for determinacy analysis, based on a partial evaluation technique for full Prolog which detects whether there are none, one or more than one ways a goal can succeed. This approach has been developed by \citet{Mogensen96} (see below). 
\citet{CharlierRH94} present a top-down framework for abstract interpretation of Prolog which is based on sequences of substitutions and can be instantiated to derive an analysis equivalent to that of \citet{Sahlin91}. 
\end{paragraph}
\begin{paragraph}{Denotational semantics for Prolog with $cut$}
\citet{Mogensen96} constructs a denotational semantics for Prolog with $cut$ based on streams of substitutions as the basis for a formal correctness argument for the determinacy analysis. The problem of constructing a denotational semantics for Prolog with $cut$ has been addressed before by \citet{billaud90simple}, \citet{DebrayM88} and \citet{Vink89} a good 20 years ago, around the same time that \citet{AptBW88} first published their theory of non-monotonic reasoning, introducing the idea of stratification.
\citet{billaud90simple} constructs an elegant denotational semantics based on streams of states of computation and proves it correct with respect to an operational semantics. \citet{DebrayM88} construct a more complex semantics over a domain of sequences of substitutions, comparable to our $\conseq$, which is partially ordered, in contrast to $\conseq$, by a prefix-ordering, rather than a sub-sequence-ordering. Both proceed by first defining a semantics for $cut$-free Prolog and then extending it to $cut$. In both cases, they argue monotonicity for the former of these constructions and appear to assume that it carries over to the latter.
Finally \citet{Vink89}, too, presents a denotational semantics of Prolog with $cut$. His approach is probably closest to ours, using environments to represent the context provided by a program in a similar fashion. However, as in the case of \cite{DebrayM88}, no argument is provided for the monotonicity of their semantic operators, which casts some doubt over the question whether the semantics is well-defined.
Common to all these approaches is the view of $cut$ as essentially an independent piece of syntax. This view requires $cut$ to be treated on a par with success and failure, having an evaluation by itself, which creates the need for complex constructions involving the introduction and later elimination of $cut$-flags into the streams or sequences, to semantically simulate the effect that $cut$ has on a computation. In contrast, we view $cut$ as essentially relational. In our view, a $cut$ has no semantics of its own, but only affects the evaluation of the goals in the context where it occurs. This reliefs us of the need for systematically introducing and eliminating $cut$-flags. 

\end{paragraph}

\section{Conclusions}
This paper has presented a determinacy inference for Prolog with $cut$, which treats $cut$ in a uniform way, while being more elegant and powerful than previously existing methods. The inference has been proved correct with respect to a novel denotational semantics for Prolog with $cut$. We have demonstrated the viability of the method by reporting on the performance of an implementation thereof and evaluating it against a comparable existing method.   

\paragraph{Acknowledgements}
This work was inspired by the cuts that are ravaging the UK, but funded by a ACM-W scholarship and a DTA bursary. We thank Lunjin Lu and Samir Genaim for discussions that provided the backdrop for this work. We thank Michel Billaud for sending us copies of his early work and 
for his comments on the wider literature. We also thank an anonymous reviewer for invaluable help with the proofs in the appendix.

\bibliographystyle{acmtrans}
\bibliography{cutpaper}

\newpage
\section{Appendix - Proofs}
\subsection{$\conseq$ is a complete lattice}

\subsubsection{Relation on $\conseq$ is a partial order}

\paragraph{The relation is reflexive: $\vec{\Theta} \sqsubseteq \vec{\Theta}$}
$
\\
Observe\ that:\ \forall \vec{\Theta} \in \conseq (\vec{\Theta} \subseteq_{pw} \vec{\Theta} \wedge \vec{\Theta} \in Sub_{|\vec{\Theta}|} ) \\ 
hence\ \forall \vec{\Theta} \in \conseq (\vec{\Theta} \sqsubseteq \vec{\Theta}) \\
by\ selecting\ \Phi = \Theta
$

\paragraph{The relation is transitive: $\vec{\Theta_1} \sqsubseteq \vec{\Theta_2} \wedge  \vec{\Theta_2} \sqsubseteq \vec{\Theta_3} \rightarrow  \vec{\Theta_1} \sqsubseteq \vec{\Theta_3}$}
$
\\
\forall \vec{\Theta}_1,\vec{\Theta}_2,\vec{\Theta}_3 \in \conseq ((\vec{\Theta}_1 \sqsubseteq \vec{\Theta}_2 \wedge \vec{\Theta}_2 \sqsubseteq \vec{\Theta}_3 ) \rightarrow (\vec{\Theta}_1 \sqsubseteq \vec{\Theta}_3))
\\
let\ |\vec{\Theta}_1| = l,\ |\vec{\Theta}_2| = m,\ |\vec{\Theta}_3| = n,\\
l <= m <= n \\
(\vec{\Theta}_1 \sqsubseteq \vec{\Theta}_2) \rightarrow \exists \vec{\Phi}_1 \in Sub_l(\vec{\Theta}_2) _{.} (\vec{\Theta}_1 \subseteq_{pw} \vec{\Phi}_1)  \\
(\vec{\Theta}_2 \sqsubseteq \vec{\Theta}_3) \rightarrow \exists \vec{\Phi}_2 \in Sub_m(\vec{\Theta}_3) _{.} (\vec{\Theta}_2 \subseteq_{pw} \vec{\Phi}_2)  \\
since\ \vec{\Theta}_2 \subseteq_{pw} \vec{\Phi}_2\ and\ \exists \vec{\Phi}_1 \in Sub_l(\vec{\Theta}_2) _{.} (\vec{\Theta}_1 \subseteq_{pw} \vec{\Phi}_1):\  \exists \vec{\Phi}_3 \in Sub_l(\vec{\Phi}_2) _{.}(\vec{\Theta}_1 \subseteq_{pw} \vec{\Phi}_3) \\
Sub_l(\vec{\Phi}_2) \subseteq Sub_l( \vec{\Theta}_3)\\
hence\  \exists \vec{\Phi}_3 \in Sub_l(\vec{\Theta}_3) _{.}(\vec{\Theta}_1 \subseteq_{pw} \vec{\Phi}_3) \\  
therefore\ \vec{\Theta}_1 \sqsubseteq \vec{\Theta}_3
$

\paragraph{The relation is anti-symmetric:}
$
\forall \vec{\Theta}_1, \vec{\Theta}_2 \in \conseq (\vec{\Theta}_1 \sqsubseteq \vec{\Theta}_2 \wedge \vec{\Theta}_2 \sqsubseteq \vec{\Theta}_1 \rightarrow \vec{\Theta}_1 = \vec{\Theta}_2)\\
let\ |\vec{\Theta}_1| = m,\ |\vec{\Theta}_2| = n \\
(\vec{\Theta}_1 \sqsubseteq \vec{\Theta}_2) \rightarrow \exists \vec{\Phi}_1 \in Sub_m(\vec{\Theta}_2)\ such\ that\ \vec{\Theta}_1 \subseteq_{pw} \vec{\Phi}_1\\
(\vec{\Theta}_2 \sqsubseteq \vec{\Theta}_1) \rightarrow \exists \vec{\Phi}_2 \in Sub_n(\vec{\Theta}_1)\ such\ that\ \vec{\Theta}_2 \subseteq_{pw} \vec{\Phi}_2\\
|\vec{\Phi}_1| = m \ and\ |\vec{\Phi}_1| <= n \ hence\ m <= n\\
|\vec{\Phi}_2| = n \ and\ |\vec{\Phi}_2| <= m \ hence\ n <= m\\
hence\ m = n \ (by\ anti-symmetry\ of\ <=)\\
hence\ \vec{\Phi}_1 = \vec{\Theta}_2 \ and\ \vec{\Phi_2} = \vec{\Theta}_1\\
hence\ \vec{\Theta}_1 \subseteq_{pw} \vec{\Theta}_2 \ and\ \vec{\Theta}_2 \subseteq_{pw} \vec{\Theta}_1\\
therefore:\\
 \vec{\Theta}_1 = \vec{\Theta}_2 \ (by\ anti-symmetry\ of\ \subseteq_{pw})
$

\subsubsection{The meet of two sequences is unique and therefore well defined:}

First note that by the definition of $ \sqcap $,  $\vec{\Theta} \sqcap \vec{\Psi} \sqsubseteq \vec{\Theta}$ and  $\vec{\Theta} \sqcap \vec{\Psi} \sqsubseteq \vec{\Psi}$.\\
Then show: 
$
\forall \vec{\Theta}, \vec{\Psi}, \vec{\Gamma} \in \conseq: \vec{\Gamma} \sqsubseteq \vec{\Theta} \wedge \vec{\Gamma} \sqsubseteq \vec{\Psi} \rightarrow \vec{\Gamma} \sqsubseteq (\vec{\Theta} \sqcap \vec{\Psi})\\
\\
|\vec{\Theta}| = n,\ |\vec{\Psi}| = m,\ |\vec{\Gamma}| = k \\
\vec{\Gamma} \sqsubseteq \vec{\Theta} \rightarrow \exists \vec{\Theta}_1\ \in Sub_k(\vec{\Theta}) _{.}(\vec{\Gamma} \subseteq_{pw} \vec{\Theta}_1)\\
\vec{\Gamma} \sqsubseteq \vec{\Psi} \rightarrow \exists \vec{\Psi}_1\ \in Sub_k(\vec{\Psi}) _{.} (\vec{\Gamma} \subseteq_{pw} \vec{\Psi}_1)\\
|\vec{\Theta}_1| = k,\ |\vec{\Psi}_1| = k \\
$
assume (without loss of generality): $n>=m$, then: $|\vec{\Theta} \sqcap \vec{\Psi}| = l,\ l<=m$\\
since $\vec{\Gamma} \sqsubseteq \vec{\Theta}$ and $\vec{\Gamma} \sqsubseteq \vec{\Psi}$, $k <= m\ (and\ k<=n)\\$
since $\vec{\Gamma} \subseteq_{pw} \vec{\Theta}_1\ and\ \vec{\Gamma} \subseteq_{pw} \vec{\Psi}_1$, $\vec{\Gamma} \subseteq_{pw} (\vec{\Theta}_1 \cap_{pw} \vec{\Psi}_1)$\\
hence $\vec{\Gamma} \sqsubseteq (\vec{\Theta}_1 \cap_{pw} \vec{\Psi}_1)\\
(\vec{\Psi}_1 \in Sub_k(\vec{\Psi})) \rightarrow (\vec{\Psi}_1 \sqsubseteq \vec{\Psi}) \\
(\vec{\Theta}_1 \in Sub_k(\vec{\Theta})) \rightarrow (\vec{\Theta}_1 \sqsubseteq \vec{\Theta}) \\
(\vec{\Theta}_1 \cap_{pw} \vec{\Psi}_1) \in  \{\vec{X} \cap_{pw} \vec{\Psi}_1 \mid \vec{X} \in Sub_k(\vec{\Theta}) \}\ (since\ \vec{\Theta}_1 \in Sub_k(\vec{\Theta}))\\
(\vec{\Theta}_1 \cap_{pw} \vec{\Psi}_1) \subseteq_{pw} \bigcup_{pw} \{\vec{X} \cap_{pw} \vec{\Psi}_1 \mid \vec{X} \in Sub_k(\vec{\Theta}) \}\\
(note\ that\ since\ \vec{\Gamma} \in \conseq,\ \vec{\Gamma}\ does\ not\ contain\ \{false\}\\
and\ since\ \vec{\Gamma} \subseteq_{pw} (\vec{\Theta}_1 \cap_{pw} \vec{\Psi}_1),\ \vec{\Theta}_1 \cap_{pw} \vec{\Psi}_1\ does\ not\ contain\ \{false\} \\
hence\ (\vec{\Theta}_1 \cap_{pw} \vec{\Psi}_1) = trim(\vec{\Theta}_1 \cap_{pw} \vec{\Psi}_1))\\
(\vec{\Theta}_1 \cap_{pw} \vec{\Psi}_1) \sqsubseteq (\vec{\Theta} \sqcap \vec{\Psi}_1)\\
(\vec{\Theta} \sqcap \vec{\Psi}_1) \sqsubseteq (\vec{\Theta} \sqcap \vec{\Psi})\ (since\ \vec{\Psi}_1 \sqsubseteq \vec{\Psi}\ and\ \sqcap\ is\ monotonic)\\
therefore\ \vec{\Gamma} \sqsubseteq (\vec{\Theta} \sqcap \vec{\Psi})
$
\subsection{Cut-normal form}
We transform Prolog predicates that are defined by any number of clauses, none of which contains a disjunction, into this form by constructing $G_1, G_2, G_3$  and $ G_4$ as follows:\\

\begin{tabular}{l | p{11cm}}
$G_1$:& If no clause precedes the clause containing the first $cut$, set $G_1$ to $\post(false)$.\\
 & Else, if a single clause precedes the clause containing the first $cut$, set $G_1$ to the body of this clause.\\
 & Otherwise, define an auxiliary predicate to wrap up all clauses preceding the clause containing the first $cut$ and set $G_1$ to a call to that predicate. \\ \hline 
$G_2$:& If there is no $cut$ in the predicate, set $G_2$ to $\post(false)$.\\
& Else, if no atom precedes the first $cut$, set $G_2$ to $\post(true)$.\\
& Otherwise, set $G_2$ to the compound goal before the first $cut$.\\
\end{tabular}\\

\begin{tabular}{l | p{11cm}}
$G_3$:& If there is no $cut$ in the predicate, set $G_3$ to any goal, e.g. $\post(true)$.\\
 & Else, if no goal follows the first $cut$, set $G_3$ to $\post(true)$.\\
 & Else, if the compound goal following the first $cut$ does not contain another $cut$, set $G_3$ to that goal.\\
 & Otherwise, define an auxiliary predicate to wrap up the compound goal following the first $cut$ and set $G_3$ to a call to that predicate. \\ \hline
$G_4$:& If no clause follows the clause containing the first $cut$, set $G_4$ to $\post(false)$.\\
 & Else, if a single, $cut$-free clause follows the clause containing the first $cut$, set $G_4$ to the body of this clause.\\
 & Otherwise, define an auxiliary predicate to wrap up all clauses following the clause containing the first $cut$ and set $G_4$ to a call to that predicate.
\end{tabular}\\
\normalsize

\subsection{Theorem 1: $ \bigcup (\Tg \eval{G} \mu_P \vec{\Theta}) \subseteq \bigcup( \vec{\Theta}) \cap S_G \eval{G}$}

Notice first that the following things hold:\\
$\bigcup(\vec{\Psi}) \subseteq \bigcup(trim(\vec{\Psi}))$\\
$\mydownarrow (\Theta \cup \Phi) = \mydownarrow \Theta \cup \mydownarrow \Phi$ \\
$\mydownarrow (\Theta \cap \Phi) = \mydownarrow \Theta \cap \mydownarrow \Phi$ \\
$\overline{\exists}_{\vec{y}} (\Theta \cup \Phi) = \overline{\exists}_{\vec{y}}(\Theta) \cup \overline{\exists}_{\vec{y}} (\Phi)  $\\
$\overline{\exists}_{\vec{y}} (\Theta \cap \Phi) \subseteq \overline{\exists}_{\vec{y}}(\Theta) \cap \overline{\exists}_{\vec{y}} (\Phi)  $\\
$\rho_{\vec{x},\vec{y}} (\Theta \cup \Phi) = \rho_{\vec{x},\vec{y}} \Theta \cup \rho_{\vec{x},\vec{y}} \Phi $\\
$\rho_{\vec{x},\vec{y}} (\Theta \cap \Phi) \subseteq \rho_{\vec{x},\vec{y}} \Theta \cap \rho_{\vec{x},\vec{y}} \Phi $\\
$\overline{\exists}_{\vec{y}} (\mydownarrow \overline{\exists}_{\vec{y}} (\Theta)) = \overline{\exists}_{\vec{y}}(\Theta)$\\
\\
Proof by induction on length of $\vec{\Theta}$:
\\
Base Case: $\vec{\Theta} = []$\\
$
\bigcup (\Tg \eval{G} \mu_P [])  = \bigcup([]) = \emptyset\\
\bigcup ([]) \cap S_G \eval{G} = \emptyset \cap S_G \eval{G} = \emptyset\\
\emptyset \subseteq \emptyset \\
$
therefore: $\bigcup (\Tg \eval{G} \mu_P []) \subseteq \bigcup ([]) \cap S_G \eval{G}$\\ 
\\
Induction Step:\\
Assume: $\bigcup( \Tg \eval{G} \mu_P \vec{\Theta}) \subseteq \bigcup (\vec{\Theta}) \cap S_G \eval{G}$\\
Show: $\bigcup (\Tg \eval{G} \mu_P (\Theta:\vec{\Theta})) \subseteq \bigcup (\Theta:\vec{\Theta}) \cap S_G \eval{G}$\\
\\
Induction on structure of G:\\
Two base cases: (1) $G = \post(\phi)$, (2) $G = p(\vec{x})$\\
\\
(1) $G = \post(\phi)$\\
Assume: $\bigcup( \Tg \eval{\post(\phi)} \mu_P \vec{\Theta}) \subseteq \bigcup (\vec{\Theta}) \cap S_G \eval{\post(\phi)}$\\
Show: $\bigcup( \Tg \eval{\post(\phi)} \mu_P (\Theta:\vec{\Theta})) \subseteq \bigcup (\Theta:\vec{\Theta}) \cap S_G \eval{\post(\phi)}$\\
$
\bigcup (\Theta:\vec{\Theta}) \cap S_G \eval{\post)(\phi)}\\
\ \ = (\Theta \cap S_G \eval{\post(\phi)}) \cup (\bigcup (\vec{\Theta}) \cap S_G \eval{\post(\phi)})\\
\ \ = (\Theta \cap \closed{\phi}) \cup (\bigcup (\vec{\Theta}) \cap S_G \eval{\post(\phi)})\\
$

$
\bigcup( \Tg \eval{\post(\phi)} \mu_P (\Theta:\vec{\Theta}))\\
\ \ = \bigcup(trim(\closed{\phi} \cap \Theta : \Tg \eval{\post(\phi)} \mu_P \vec{\Theta} ))\\
\ \ \subseteq \bigcup(\closed{\phi} \cap \Theta : \Tg \eval{\post(\phi)} \mu_P \vec{\Theta}) \\
\ \ = (\closed{\phi} \cap \Theta) \cup \bigcup(\Tg \eval{\post(\phi)} \mu_P \vec{\Theta})\\ 
\ \ \subseteq (\closed{\phi} \cap \Theta) \cup (\bigcup (\vec{\Theta}) \cap S_G \eval{\post(\phi)}) \\
$
therefore: $\bigcup( \Tg \eval{\post(\phi)} \mu_P (\Theta:\vec{\Theta})) \subseteq \bigcup (\Theta:\vec{\Theta}) \cap S_G \eval{\post(\phi)}\\
$
\\
(2) $G = p(\vec{x})$\\
Assume (without loss of generality): $p(\vec{y}) \neck G_1 ; G_2,!,G_3 ; G_4 \in P$\\
Assume: $\bigcup( \Tg \eval{p(\vec{x})} \mu_P \vec{\Theta}) \subseteq \bigcup (\vec{\Theta}) \cap S_G \eval{p(\vec{x})}$\\
Show: $\bigcup (\Tg \eval{p(\vec{x})} \mu_P (\Theta:\vec{\Theta})) \subseteq \bigcup (\Theta:\vec{\Theta}) \cap S_G \eval{p(\vec{x})}$\\

$
\bigcup (\Theta:\vec{\Theta}) \cap S_G \eval{p(\vec{x})} \\
\ \ = (\Theta \cap S_G \eval{p(\vec{x})}) \cup (\bigcup (\vec{\Theta}) \cap S_G \eval{p(\vec{x})})\\
\ \ = (\Theta \cap \mydownarrow \rho_{\vec{y},\vec{x}} \overline{\exists}_{\vec{y}}(S_H \eval{p(\vec{y})})) \cup (\bigcup (\vec{\Theta}) \cap S_G \eval{p(\vec{x})})\\
\ \ = (\Theta \cap \mydownarrow \rho_{\vec{y},\vec{x}} \overline{\exists}_{\vec{y}} (\mydownarrow \overline{\exists}_{\vec{y}} (S_G \eval{G_1} \cup S_G \eval{G_2,G_3} \cup S_G \eval{G_4}))) \cup (\bigcup (\vec{\Theta}) \cap S_G \eval{p(\vec{x})})\\
\ \ = (\Theta \cap \mydownarrow \rho_{\vec{y},\vec{x}} \overline{\exists}_{\vec{y}} (S_G \eval{G_1} \cup S_G \eval{G_2,G_3} \cup S_G \eval{G_4})) \cup (\bigcup (\vec{\Theta}) \cap S_G \eval{p(\vec{x})})\\
\ \ = (\Theta \cap (\mydownarrow \rho_{\vec{y},\vec{x}} \overline{\exists}_{\vec{y}} (S_G \eval{G_1})  \cup \mydownarrow \rho_{\vec{y},\vec{x}} \overline{\exists}_{\vec{y}}(S_G \eval{G_2,G_3}) \cup \mydownarrow \rho_{\vec{y},\vec{x}} \overline{\exists}_{\vec{y}}(S_G \eval{G_4})) \\
\indent \cup (\bigcup (\vec{\Theta}) \cap S_G \eval{p(\vec{x})})\\
$

$
\bigcup (\Tg \eval{p(\vec{x})} \mu_P (\Theta:\vec{\Theta})) \\
\ \ = \bigcup (\mydownarrow \rho_{\vec{y},\vec{x}} \overline{\exists}_{\vec{y}}(\mu\ (p(\vec{y}))\ \mydownarrow \rho_{\vec{x},\vec{y}} \overline{\exists}_{\vec{x}} ([\Theta])) \cap \Theta : \Tg \eval{ p(\vec{x})} \mu \vec{\Theta})\\
\ \ = (\bigcup(\mydownarrow \rho_{\vec{y},\vec{x}} \overline{\exists}_{\vec{y}}(\mu\ (p(\vec{y}))\ \mydownarrow \rho_{\vec{x},\vec{y}} \overline{\exists}_{\vec{x}} ([\Theta]))) \cap \Theta) \cup \bigcup (\Tg \eval{ p(\vec{x})} \mu \vec{\Theta})\\
\ \ = (\bigcup(\mydownarrow \rho_{\vec{y},\vec{x}} \overline{\exists}_{\vec{y}} (\mydownarrow \overline{\exists}_{y}(\Tg \eval{G_1} \mu [{\Theta'}] : \vec{\Psi}) )) \cap \Theta) \cup \bigcup( \Tg \eval{ p(\vec{x})} \mu \vec{\Theta})\\
\indent where\ \vec{\Psi} = \left\{ \begin{array}{l l}
\Tg \eval{G_3 } \mu [\Phi] 
& if \ \Tg \eval{ G_2} \mu [{\Theta'} ]= \Phi : \vec{\Phi} \\
\Tg \eval{G_4} \mu [{\Theta'}] 
& if \  \Tg \eval{G_2} \mu [{\Theta'}] = [] \\
\end{array}	\right.\\
\indent and\ {\Theta'} = \mydownarrow \rho_{\vec{x},\vec{y}} \overline{\exists}_{\vec{x}} (\Theta)\\$
To be on the safe side, consider the sequence resulting from appending \textit{both} possibilities for $\vec{\Psi}$, the union of which is certainly a superset of the above:\\
$\subseteq (\bigcup(\mydownarrow \rho_{\vec{y},\vec{x}} \overline{\exists}_{\vec{y}}(\mydownarrow \overline{\exists}_{y}(\Tg \eval{G_1} \mu [{\Theta'}] : \Tg \eval{G_3 } \mu [\Phi] : \Tg \eval{G_4} \mu [{\Theta'}]) )) \cap \Theta) \cup \bigcup(\Tg \eval{ p(\vec{x})} \mu \vec{\Theta})\\
\indent where\ \Phi : \vec{\Phi} = \Tg \eval{ G_2} \mu [{\Theta'}]\\$
Again, changing this to include all, rather than only the first, possibilities for $\Tg \eval{G_2} \mu [\Theta']$ will result in a safe over-approximation, i.e. a superset of the above:
$\subseteq (\bigcup(\mydownarrow \rho_{\vec{y},\vec{x}} \overline{\exists}_{\vec{y}}(\mydownarrow \overline{\exists}_{y}(\Tg \eval{G_1} \mu [{\Theta'}] : \Tg \eval{G_3 } \mu (\Tg \eval{ G_2} \mu [{\Theta'}]) : \Tg \eval{G_4} \mu [{\Theta'}]) )) \cap \Theta)
\\ \indent \cup \bigcup(\Tg \eval{ p(\vec{x})} \mu \vec{\Theta})\\
\ \ = (\bigcup(\mydownarrow \rho_{\vec{y},\vec{x}} \overline{\exists}_{\vec{y}}(\Tg \eval{G_1} \mu [{\Theta'}] : \Tg \eval{G_3 } \mu (\Tg \eval{ G_2} \mu [{\Theta'}]) : \Tg \eval{G_4} \mu [{\Theta'}]) ) \cap \Theta) \cup \bigcup(\Tg \eval{ p(\vec{x})} \mu \vec{\Theta})\\
\ \ = (\bigcup(\mydownarrow \rho_{\vec{y},\vec{x}} \overline{\exists}_{\vec{y}}(\Tg \eval{G_1} \mu [{\Theta'}]) : \mydownarrow \rho_{\vec{y},\vec{x}} \overline{\exists}_{\vec{y}} ( \Tg \eval{G_3 } \mu (\Tg \eval{ G_2} \mu [{\Theta'}]))  : \mydownarrow \rho_{\vec{y},\vec{x}} \overline{\exists}_{\vec{y}}( \Tg \eval{G_4} \mu [{\Theta'}]) ) 
\\ \indent \cap \Theta) \cup \bigcup(\Tg \eval{ p(\vec{x})} \mu \vec{\Theta})\\
\ \ = ((\bigcup(\mydownarrow \rho_{\vec{y},\vec{x}} \overline{\exists}_{\vec{y}}(\Tg \eval{G_1} \mu [{\Theta'}])) \cup \bigcup ( \mydownarrow \rho_{\vec{y},\vec{x}} \overline{\exists}_{\vec{y}} ( \Tg \eval{G_3 } \mu (\Tg \eval{ G_2} \mu [{\Theta'}]))) \cup \bigcup( \mydownarrow \rho_{\vec{y},\vec{x}} \overline{\exists}_{\vec{y}}( \Tg \eval{G_4} \mu [{\Theta'}]) )) \\
\indent \cap \Theta)  \cup \bigcup(\Tg \eval{ p(\vec{x})} \mu \vec{\Theta})\\
\ \ = (\mydownarrow \rho_{\vec{y},\vec{x}} \overline{\exists}_{\vec{y}}(\bigcup(\Tg \eval{G_1} \mu [{\Theta'}])) \cup  \mydownarrow \rho_{\vec{y},\vec{x}} \overline{\exists}_{\vec{y}} (\bigcup ( \Tg \eval{G_3 } \mu (\Tg \eval{ G_2} \mu [{\Theta'}]))) \cup  \mydownarrow \rho_{\vec{y},\vec{x}} \overline{\exists}_{\vec{y}}( \bigcup(\Tg \eval{G_4} \mu [{\Theta'}]) ) \\
\indent \cap \Theta) \cup \bigcup(\Tg \eval{ p(\vec{x})} \mu \vec{\Theta})\\
$
since:  $\bigcup(\Tg \eval{G_1} \mu [{\Theta'}]) \subseteq S_G \eval{G_1} \cap \Theta'$\\
and: $\bigcup ( \Tg \eval{ G_2} \mu [{\Theta'}]) \subseteq S_G \eval{G_2} \cap \Theta'$\\
hence: $\bigcup ( \Tg \eval{G_3} \mu (\Tg \eval{ G_2} \mu [{\Theta'}])) \subseteq S_G \eval{G_3} \cap (S_G \eval{G_2} \cap \Theta')$\\
hence: $\bigcup ( \Tg \eval{G_3} \mu (\Tg \eval{ G_2} \mu [{\Theta'}])) \subseteq (S_G \eval{G_3} \cap S_G \eval{G_2}) \cap \Theta'$\\
hence: $\bigcup ( \Tg \eval{G_3 } \mu (\Tg \eval{ G_2} \mu [{\Theta'}])) \subseteq S_G \eval{G_2,G_3} \cap \Theta'$\\
and: $\bigcup(\Tg \eval{G_4} \mu [{\Theta'}]) \subseteq S_G \eval{G_4} \cap \Theta'$\\
using these, therefore, the above superset of $\bigcup (\Tg \eval{p(\vec{x})} \mu_P (\Theta:\vec{\Theta}))$ is a subset of:\\
$\subseteq ((\mydownarrow \rho_{\vec{y},\vec{x}} \overline{\exists}_{\vec{y}}(S_G \eval{G_1} \cap \Theta') \cup  \mydownarrow \rho_{\vec{y},\vec{x}} \overline{\exists}_{\vec{y}} (S_G \eval{G_2,G_3} \cap \Theta') \cup \mydownarrow \rho_{\vec{y},\vec{x}} \overline{\exists}_{\vec{y}}( S_G \eval{G_4} \cap \Theta' ))\cap \Theta) \\
\indent \cup \bigcup(\Tg \eval{ p(\vec{x})} \mu \vec{\Theta})\\$
since: ${\Theta'} = \mydownarrow \rho_{\vec{x},\vec{y}} \overline{\exists}_{\vec{x}} (\Theta)$, the following holds: $\mydownarrow \rho_{\vec{y},\vec{x}} \overline{\exists}_{\vec{y}}({\Theta'}) = \mydownarrow \rho_{\vec{y},\vec{x}} \overline{\exists}_{\vec{y}}(\mydownarrow \rho_{\vec{x},\vec{y}} \overline{\exists}_{\vec{x}} (\Theta)) 
\\ \indent = \mydownarrow \overline{\exists}_{\vec{x}}(\Theta) \supseteq \Theta $\\
intersecting this with $\Theta$ therefore gives $\Theta$ itself: $\mydownarrow \rho_{\vec{y},\vec{x}} \overline{\exists}_{\vec{y}}({\Theta'}) \cap \Theta = \Theta $
distributing the projections and collecting and intersection the occurrences of $\mydownarrow \rho_{\vec{y},\vec{x}} \overline{\exists}_{\vec{y}}({\Theta'})$ and $\Theta$ above therefore gives: \\
$\subseteq ((\mydownarrow \rho_{\vec{y},\vec{x}} \overline{\exists}_{\vec{y}}(S_G \eval{G_1}) \cup  \mydownarrow \rho_{\vec{y},\vec{x}} \overline{\exists}_{\vec{y}} (S_G \eval{G_2,G_3} ) \cup  \mydownarrow \rho_{\vec{y},\vec{x}} \overline{\exists}_{\vec{y}}( S_G \eval{G_4})) \cap \Theta) \cup \bigcup(\Tg \eval{ p(\vec{x})} \mu \vec{\Theta})\\
\subseteq ((\mydownarrow \rho_{\vec{y},\vec{x}} \overline{\exists}_{\vec{y}}(S_G \eval{G_1}) \cup  \mydownarrow \rho_{\vec{y},\vec{x}} \overline{\exists}_{\vec{y}} (S_G \eval{G_2,G_3} ) \cup  \mydownarrow \rho_{\vec{y},\vec{x}} \overline{\exists}_{\vec{y}}( S_G \eval{G_4})) \cap \Theta) \\
\indent \cup (\bigcup (\vec{\Theta}) \cap S_G \eval{p(\vec{x})}) \\
 = \bigcup (\Theta:\vec{\Theta}) \cap S_G \eval{p(\vec{x})}\\
$

\noindent Induction Step: $G = G_1, G_2$\\
Assume: $\bigcup( \Tg \eval{G_1,G_2} \mu_P \vec{\Theta}) \subseteq \bigcup (\vec{\Theta}) \cap S_G \eval{G_1,G_2}$\\
And:  $\bigcup (\Tg \eval{G_1} \mu_P \vec{\Phi}) \subseteq \bigcup (\vec{\Phi}) \cap S_G \eval{G_1}$\\
And:  $\bigcup (\Tg \eval{G_2} \mu_P \vec{\Phi}) \subseteq \bigcup (\vec{\Phi}) \cap S_G \eval{G_2}$\\
Show: $\bigcup (\Tg \eval{G_1,G_2} \mu_P (\Theta:\vec{\Theta})) \subseteq \bigcup (\Theta:\vec{\Theta}) \cap S_G \eval{G_1,G_2}$\\

$
\bigcup (\Theta:\vec{\Theta}) \cap S_G \eval{G_1,G_2}\\
\ \ =  (\Theta \cap S_G \eval{G_1,G_2}) \cup (\bigcup (\vec{\Theta}) \cap S_G \eval{G_1,G_2}) \\
\ \ =  (\Theta \cap S_G \eval{G_1} \cap S_G \eval{G_2}) \cup (\bigcup (\vec{\Theta}) \cap S_G \eval{G_1,G_2})\\
$

$
\bigcup (\Tg \eval{G_1,G_2} \mu_P (\Theta:\vec{\Theta}))\\
\ \ = \bigcup (\Tg \eval{G_2} \mu (\Tg \eval{ G_1} \mu (\Theta : \vec{\Theta})))\\
\ \ \subseteq \bigcup (\Tg \eval{ G_1} \mu (\Theta : \vec{\Theta}) \cap S_G \eval{G_2}\\
\ \ \subseteq \bigcup (\Theta:\vec{\Theta}) \cap S_G \eval{G_1} \cap S_G \eval{G_2}\\
\ \ = \bigcup (\Theta:\vec{\Theta}) \cap S_G \eval{G_1,G_2}\\
$
QED

\subsection{Theorem 2: For $\Theta \in \con$ and stratified $P = P_0 \cup \ldots \cup P_n$: $\Theta \subseteq \Dg \eval{G} \delta_P \Rightarrow |\Tg \eval{G} \mu_P [ \Theta ] | \leq 1$.}

\subsubsection{Lemma 1: $(\Tg \eval{G} \mu \vec{\Theta}) \cap \Psi = \Tg \eval{G} \mu (\vec{\Theta} \cap \Psi)$}

Proof by nested induction on: \\
1. $\mu$,\\
2. $|\vec{\Theta}|$,\\
3. structure of $G$\\
\\
1 Base Case: $\mu = \mu_{\bot}$ \\
Show: $(\Tg \eval{G} \mu_{\bot} \vec{\Theta}) \cap \Psi = \Tg \eval{G} \mu_{\bot} (\vec{\Theta} \cap \Psi)$\\
\\
1.1 Base Case: $\vec{\Theta} = []$ \\ 
Show: $(\Tg \eval{G} \mu_{\bot} []) \cap \Psi = \Tg \eval{G} \mu_{\bot} ([] \cap \Psi)$\\
$
([]) \cap \Psi = \Tg \eval{G} \mu_{\bot} ([])\\
\ [] = []
$\\
\\
1.2 Induction Step: $(\Theta : \vec{\Theta})$ \\
Assume: $(\Tg \eval{H} \mu_{\bot} \vec{\Theta}) \cap \Psi = \Tg \eval{H} \mu_{\bot} (\vec{\Theta} \cap \Psi)$\\
Show: $(\Tg \eval{G} \mu_{\bot} (\Theta : \vec{\Theta})) \cap \Psi = \Tg \eval{G} \mu_{\bot} ((\Theta : \vec{\Theta}) \cap \Psi)$\\
\\
1.2.1 Two Base Cases: (1) $G = \post(\phi)$, (2) $G = p(\vec{x})$\\
\\
(1) $G = \post(\phi)$\\
Show: $(\Tg \eval{\post(\phi)} \mu_{\bot} (\Theta : \vec{\Theta})) \cap \Psi = \Tg \eval{\post(\phi)} \mu_{\bot} ((\Theta : \vec{\Theta}) \cap \Psi)$\\
\\
$
(\Tg \eval{\post(\phi)} \mu_{\bot} (\Theta : \vec{\Theta})) \cap \Psi = \Tg \eval{\post(\phi)} \mu_{\bot} ((\Theta : \vec{\Theta}) \cap \Psi)\\
trim(\closed{\phi} \cap \Theta : \Tg \eval{\post(\phi)} \mu_{\bot} \vec{\Theta}) \cap \Psi = \Tg \eval{\post(\phi)} \mu_{\bot} ((\Theta \cap \Psi) : (\vec{\Theta} \cap \Psi))\\
(trim[\closed{\phi} \cap \Theta ] : trim (\Tg \eval{\post(\phi)} \mu_{\bot} \vec{\Theta})) \cap \Psi \\
\indent = trim(\closed{\phi} \cap \Theta \cap \Psi : \Tg \eval{\post(\phi)} \mu_{\bot} (\vec{\Theta} \cap \Psi))\\
(trim[\closed{\phi} \cap \Theta ] \cap \Psi) : (trim (\Tg \eval{\post(\phi)} \mu_{\bot} \vec{\Theta}) \cap \Psi)\\
\indent = trim[\closed{\phi} \cap \Theta \cap \Psi] : trim (\Tg \eval{\post(\phi)} \mu_{\bot} (\vec{\Theta} \cap \Psi))\\$
by assumption:$
(\Tg \eval{\post(\phi)} \mu_{\bot} \vec{\Theta}) \cap \Psi = \Tg \eval{\post(\phi)} \mu_{\bot} (\vec{\Theta} \cap \Psi)\\
trim (\Tg \eval{\post(\phi)} \mu_{\bot} \vec{\Theta}) \cap \Psi =  trim (\Tg \eval{\post(\phi)} \mu_{\bot} (\vec{\Theta} \cap \Psi))\\
trim(\closed{\phi} \cap \Theta ) \cap \Psi = trim(\closed{\phi} \cap \Theta \cap \Psi)
$ \\
\\
(2) $G = p(\vec{x})$\\
Show: $(\Tg \eval{p(\vec{x})} \mu_{\bot} (\Theta : \vec{\Theta})) \cap \Psi = \Tg \eval{p(\vec{x})} \mu_{\bot} ((\Theta : \vec{\Theta}) \cap \Psi)\\
(\Tg \eval{p(\vec{x})} \mu_{\bot} (\Theta : \vec{\Theta})) \cap \Psi \\
= (\mydownarrow \rho_{\vec{y}, \vec{x}}\overline{\exists}_{\vec{y}}(\mu_{\bot} (p(\vec{y}))\ \mydownarrow \rho_{\vec{x}, \vec{y}}\overline{\exists}_{\vec{x}}([\Theta])) \cap \Theta : \Tg \eval{ p(\vec{x})} \mu_{\bot}  \vec{\Theta}) \cap \Psi  \\
= (\mydownarrow \rho_{\vec{y}, \vec{x}}\overline{\exists}_{\vec{y}}([]) \cap \Theta : \Tg \eval{ p(\vec{x})} \mu_{\bot}  \vec{\Theta}) \cap \Psi  \\
= ([]) : ( \Tg \eval{ p(\vec{x})} \mu_{\bot}  \vec{\Theta}) \cap \Psi$\\
by assumption: $ (\Tg \eval{ p(\vec{x})} \mu_{\bot}  \vec{\Theta}) \cap \Psi = \Tg \eval{ p(\vec{x})} \mu_{\bot}  (\vec{\Theta} \cap \Psi)\\
$
\\
$
\Tg \eval{p(\vec{x})} \mu_{\bot} ((\Theta : \vec{\Theta}) \cap \Psi) \\
= \Tg \eval{p(\vec{x})} \mu_{\bot} ((\Theta \cap \Psi) : (\vec{\Theta} \cap \Psi) ) \\
= \mydownarrow \rho_{\vec{y}, \vec{x}}\overline{\exists}_{\vec{y}}(\mu_{\bot} (p(\vec{y}))\ \mydownarrow \rho_{\vec{x}, \vec{y}}\overline{\exists}_{\vec{x}}([\Theta \cap \Psi])) \cap \Theta \cap \Psi : \Tg \eval{ p(\vec{x})} \mu_{\bot} (\vec{\Theta} \cap \Psi) \\
= \mydownarrow \rho_{\vec{y}, \vec{x}}\overline{\exists}_{\vec{y}}([]) \cap \Theta \cap \Psi : \Tg \eval{ p(\vec{x})} \mu_{\bot} (\vec{\Theta} \cap \Psi)  \\
= ([]) :  \Tg \eval{ p(\vec{x})} \mu_{\bot} (\vec{\Theta} \cap \Psi)$\\
hence: $(\Tg \eval{p(\vec{x})} \mu_{\bot} (\Theta : \vec{\Theta})) \cap \Psi = \Tg \eval{p(\vec{x})} \mu_{\bot} ((\Theta : \vec{\Theta}) \cap \Psi)
$\\
\\
1.2.2 Induction Step: $G = G_1, G_2$\\
Assume: $(\Tg \eval{G_1} \mu_{\bot} (\Theta : \vec{\Theta})) \cap \Psi = \Tg \eval{G_1} \mu_{\bot} ((\Theta : \vec{\Theta}) \cap \Psi)$\\
And: $(\Tg \eval{G_2} \mu_{\bot} (\Theta : \vec{\Theta})) \cap \Psi = \Tg \eval{G_2} \mu_{\bot} ((\Theta : \vec{\Theta}) \cap \Psi)$\\
Show: $(\Tg \eval{G_1,G_2} \mu_{\bot} (\Theta : \vec{\Theta})) \cap \Psi = \Tg \eval{G_1,G_2} \mu_{\bot} ((\Theta : \vec{\Theta}) \cap \Psi)$\\ 
$(\Tg \eval{G_1,G_2} \mu_{\bot} (\Theta : \vec{\Theta})) \cap \Psi\\
= (\Tg \eval{G_2} \mu_{\bot} (\Tg \eval{G_1} \mu_{\bot} (\Theta : \vec{\Theta}))) \cap \Psi\\
= (\Tg \eval{G_2} \mu_{\bot} (\Tg \eval{G_1} \mu_{\bot} (\Theta : \vec{\Theta})) \cap \Psi)\\
= \Tg \eval{G_2} \mu_{\bot} (\Tg \eval{G_1} \mu_{\bot} ((\Theta : \vec{\Theta}) \cap \Psi))\\
= \Tg \eval{G_1,G_2} \mu_{\bot} ((\Theta : \vec{\Theta}) \cap \Psi)\\
$
\\
2 Induction Step: $\mu = \mu_{k+1}$\\
Assume: $(\Tg \eval{H} \mu_{k} \vec{\Delta}) \cap \Lambda = \Tg \eval{H} \mu_{k} (\vec{\Delta} \cap \Lambda)$\\
Show: $(\Tg \eval{G} \mu_{k+1} \vec{\Theta}) \cap \Psi = \Tg \eval{G} \mu_{k+1} (\vec{\Theta} \cap \Psi)$\\
where $\mu_{k+1} = \Tprog \eval{P} \mu_k$\\
\\
2.1 Base Case: $\vec{\Theta} = []$\\
Show: $(\Tg \eval{G} \mu_{k+1} []) \cap \Psi = \Tg \eval{G} \mu_{k+1} ([] \cap \Psi)\\
 ([]) \cap \Psi = \Tg \eval{G} \mu_{k+1} ([])\\
 ([]) = ([])\\
$
\\
2.2 Induction Step: $\vec{\Theta} = (\Theta : \vec{\Theta}) $\\
Assume: $(\Tg \eval{G} \mu_{k+1} \vec{\Theta}) \cap \Psi = \Tg \eval{G} \mu_{k+1} (\vec{\Theta} \cap \Psi)$\\
Show: $(\Tg \eval{G} \mu_{k+1} (\Theta :\vec{\Theta})) \cap \Psi = \Tg \eval{G} \mu_{k+1} ((\Theta : \vec{\Theta}) \cap \Psi)$\\
\\
2.2.1 Two Base Cases: (1) $G = \post(\phi)$, (2) $G = p(\vec{x})$\\
(1) $G = \post(\phi)$\\
Show: $(\Tg \eval{\post(\phi)} \mu_{k+1} (\Theta :\vec{\Theta})) \cap \Psi = \Tg \eval{\post(\phi)} \mu_{k+1} ((\Theta : \vec{\Theta}) \cap \Psi)$\\
\\
$
(\Tg \eval{\post(\phi)} \mu_{k+1} (\Theta : \vec{\Theta})) \cap \Psi = \Tg \eval{\post(\phi)} \mu_{k+1} ((\Theta : \vec{\Theta}) \cap \Psi)\\
trim(\closed{\phi} \cap \Theta : \Tg \eval{\post(\phi)} \mu_{k+1} \vec{\Theta}) \cap \Psi = \Tg \eval{\post(\phi)} \mu_{k+1} ((\Theta \cap \Psi) : (\vec{\Theta} \cap \Psi))\\
(trim[\closed{\phi} \cap \Theta ] : trim (\Tg \eval{\post(\phi)} \mu_{k+1} \vec{\Theta})) \cap \Psi \\
= trim(\closed{\phi} \cap \Theta \cap \Psi : \Tg \eval{\post(\phi)} \mu_{k+1} (\vec{\Theta} \cap \Psi))\\
(trim(\closed{\phi} \cap \Theta ) \cap \Psi) : (trim (\Tg \eval{\post(\phi)} \mu_{k+1} \vec{\Theta}) \cap \Psi) \\
= trim(\closed{\phi} \cap \Theta \cap \Psi) : trim (\Tg \eval{\post(\phi)} \mu_{k+1} (\vec{\Theta} \cap \Psi))$\\
by assumption: $\Tg \eval{\post(\phi)} \mu_{k+1} \vec{\Theta}) \cap \Psi = \Tg \eval{\post(\phi)} \mu_{k+1} (\vec{\Theta} \cap \Psi)$\\
hence: $trim (\Tg \eval{\post(\phi)} \mu_{k+1} \vec{\Theta}) \cap \Psi =  trim (\Tg \eval{\post(\phi)} \mu_{k+1} (\vec{\Theta} \cap \Psi))\\
trim(\closed{\phi} \cap \Theta ) \cap \Psi = trim(\closed{\phi} \cap \Theta \cap \Psi)$ \\
\\
(2) $G = p(\vec{x})$\\
Assume (without loss of generality): $p(\vec{y}) \neck G_1 ; G_2,!,G_3 ; G_4 \in P$\\
Show: $(\Tg \eval{p(\vec{x})} \mu_{k+1} (\Theta :\vec{\Theta})) \cap \Psi = \Tg \eval{p(\vec{x})} \mu_{k+1} ((\Theta : \vec{\Theta}) \cap \Psi)$\\
\\
$
(\Tg \eval{p(\vec{x})} \mu_{k+1} (\Theta :\vec{\Theta})) \cap \Psi \\
= (\mydownarrow \rho_{\vec{y}, \vec{x}}\overline{\exists}_{\vec{y}}(\mu_{k+1} (p(\vec{y}))\ \mydownarrow \rho_{\vec{x}, \vec{y}}\overline{\exists}_{\vec{x}}([\Theta])) \cap \Theta : \Tg \eval{ p(\vec{x})} \mu_{k+1}  \vec{\Theta}) \cap \Psi  \\
= (\mydownarrow \rho_{\vec{y}, \vec{x}}\overline{\exists}_{\vec{y}}(\mu_{k+1} (p(\vec{y}))\ \mydownarrow \rho_{\vec{x}, \vec{y}}\overline{\exists}_{\vec{x}}([\Theta])) \cap \Theta) \cap \Psi : (\Tg \eval{ p(\vec{x})} \mu_{k+1}  \vec{\Theta}) \cap \Psi \\
$
\\
$
\Tg \eval{p(\vec{x})} \mu_{k+1} ((\Theta : \vec{\Theta}) \cap \Psi) \\
=\Tg \eval{p(\vec{x})} \mu_{k+1} ((\Theta \cap \Psi):( \vec{\Theta}\cap \Psi)) \\
=(\mydownarrow \rho_{\vec{y}, \vec{x}}\overline{\exists}_{\vec{y}}(\mu_{k+1} (p(\vec{y}))\ \mydownarrow \rho_{\vec{x}, \vec{y}}\overline{\exists}_{\vec{x}}([\Theta \cap \Psi])) \cap \Theta \cap \Psi): (\Tg \eval{ p(\vec{x})} \mu_{k+1} (\vec{\Theta} \cap \Psi))$\\
by assumption: $(\Tg \eval{ p(\vec{x})} \mu_{k+1}  \vec{\Theta}) \cap \Psi = (\Tg \eval{ p(\vec{x})} \mu_{k+1} (\vec{\Theta} \cap \Psi))$\\
hence the question is whether the following holds:\\
$(\mydownarrow \rho_{\vec{y}, \vec{x}}\overline{\exists}_{\vec{y}}(\mu_{k+1} (p(\vec{y}))\ \mydownarrow \rho_{\vec{x}, \vec{y}}\overline{\exists}_{\vec{x}}([\Theta])) \cap \Theta) \cap \Psi \\
\indent = (\mydownarrow \rho_{\vec{y}, \vec{x}}\overline{\exists}_{\vec{y}}(\mu_{k+1} (p(\vec{y}))\ \mydownarrow \rho_{\vec{x}, \vec{y}}\overline{\exists}_{\vec{x}}([\Theta \cap \Psi])) \cap \Theta \cap \Psi)$\\
\\
$(\mydownarrow \rho_{\vec{y}, \vec{x}}\overline{\exists}_{\vec{y}}(\mu_{k+1} (p(\vec{y}))\ \mydownarrow \rho_{\vec{x}, \vec{y}}\overline{\exists}_{\vec{x}}([\Theta \cap \Psi])) \cap \Theta) \cap \Psi  \\
= \mydownarrow \rho_{\vec{y}, \vec{x}}\overline{\exists}_{\vec{y}}(\mydownarrow \overline{\exists}_{\vec{y}} (\Tg \eval{G_1} \mu_k\ \mydownarrow \rho_{\vec{x}, \vec{y}}\overline{\exists}_{\vec{x}}([\Theta \cap \Psi])) : \vec{\Delta})) \cap \Theta \cap \Psi
\\ where\ \vec{\Delta} = \left\{ \begin{array}{l l}
\Tg \eval{G_3 } \mu_k [\Lambda] \
& if\ \Tg \eval{ G_2} \mu_k\ \mydownarrow \rho_{\vec{x}, \vec{y}}\overline{\exists}_{\vec{x}}([\Theta \cap \Psi]) = \Lambda : \vec{\Lambda} \\
\Tg \eval{G_4} \mu_k\ \mydownarrow \rho_{\vec{x}, \vec{y}}\overline{\exists}_{\vec{x}}([\Theta \cap \Psi])  \
 &if\  \Tg \eval{G_2} \mu_k\ \mydownarrow \rho_{\vec{x}, \vec{y}}\overline{\exists}_{\vec{x}}([\Theta \cap \Psi]) = [] \\
\end{array}	\right.\\
= (\mydownarrow \rho_{\vec{y}, \vec{x}}\overline{\exists}_{\vec{y}}(\Tg \eval{G_1} \mu_k\  \mydownarrow \rho_{\vec{x}, \vec{y}}\overline{\exists}_{\vec{x}}([\Theta \cap \Psi]))  ) \cap \Theta \cap \Psi) : (\mydownarrow \rho_{\vec{y}, \vec{x}}\overline{\exists}_{\vec{y}}(\vec{\Delta})  \cap \Theta \cap \Psi)$\\
\\
Observe that for any $F$: $\mydownarrow \rho_{\vec{y}, \vec{x}}\overline{\exists}_{\vec{y}}(\Tg \eval{F} \mu_k\  \mydownarrow \rho_{\vec{x}, \vec{y}}\overline{\exists}_{\vec{x}}([\Theta])) = \mydownarrow \overline{\exists}_{\vec{x}}(\Tg \eval{F} \mu_k\  \mydownarrow \overline{\exists}_{\vec{x}}([\Theta]))$\\
hence: $(\mydownarrow \rho_{\vec{y}, \vec{x}}\overline{\exists}_{\vec{y}}(\Tg \eval{G_1} \mu_k\  \mydownarrow \rho_{\vec{x}, \vec{y}}\overline{\exists}_{\vec{x}}([\Theta \cap \Psi]))  ) \cap \Theta \cap \Psi \\
 =  (\mydownarrow \overline{\exists}_{\vec{x}}(\Tg \eval{G_1} \mu_k\  \mydownarrow \overline{\exists}_{\vec{x}}([\Theta \cap \Psi]))  ) \cap \Theta \cap \Psi\\
= (\mydownarrow \overline{\exists}_{\vec{x}}(\Tg \eval{G_1} \mu_k\ [ \mydownarrow \overline{\exists}_{\vec{x}}(\Theta \cap \Psi) \cap \mydownarrow \overline{\exists}_{\vec{x}}(\Theta) ])  ) \cap \Theta \cap \Psi\ \\
\indent \indent (\mbox{since } \mydownarrow \overline{\exists}_{\vec{x}}(\Theta \cap \Psi) \subseteq \mydownarrow \overline{\exists}_{\vec{x}}(\Theta))\\$
which by assumption is equal to:\\
$(\mydownarrow \overline{\exists}_{\vec{x}}(\Tg \eval{G_1} \mu_k\ [ \mydownarrow \overline{\exists}_{\vec{x}}(\Theta)]) \cap \mydownarrow \overline{\exists}_{\vec{x}}(\Theta \cap \Psi) ) \cap \Theta \cap \Psi\ \\
= (\mydownarrow \overline{\exists}_{\vec{x}}( \mydownarrow \overline{\exists}_{\vec{x}} (\Tg \eval{G_1} \mu_k\ [ \mydownarrow \overline{\exists}_{\vec{x}}(\Theta)]) \cap \mydownarrow \overline{\exists}_{\vec{x}}(\Theta \cap \Psi) ) \cap \Theta \cap \Psi\ \\
\indent \indent (\text{since } \mydownarrow \overline{\exists}_{\vec{x}}( A \cap B) = \mydownarrow \overline{\exists}_{\vec{x}} ( \mydownarrow \overline{\exists}_{\vec{x}}(A) \cap B) )\\
= ( \mydownarrow \overline{\exists}_{\vec{x}} (\Tg \eval{G_1} \mu_k\ [ \mydownarrow \overline{\exists}_{\vec{x}}(\Theta)]) \cap \mydownarrow \overline{\exists}_{\vec{x}}(\Theta \cap \Psi)  \cap \Theta \cap \Psi\ \\
\indent \indent (\text{since } \mydownarrow \overline{\exists}_{\vec{x}}( \mydownarrow \overline{\exists}_{\vec{x}}(A) \cap \mydownarrow \overline{\exists}_{\vec{x}}(B)) = \mydownarrow \overline{\exists}_{\vec{x}}(A) \cap \mydownarrow \overline{\exists}_{\vec{x}}(B))\\
= ( \mydownarrow \overline{\exists}_{\vec{x}} (\Tg \eval{G_1} \mu_k\ [ \mydownarrow \overline{\exists}_{\vec{x}}(\Theta)]) \cap \Theta \cap \Psi\ \\
\indent \indent (\text{since } \Theta \cap \Psi \subseteq \mydownarrow \overline{\exists}_{\vec{x}}(\Theta \cap \Psi) )\\
= (\mydownarrow \rho_{\vec{y}, \vec{x}}\overline{\exists}_{\vec{y}}(\Tg \eval{G_1} \mu_k\  \mydownarrow \rho_{\vec{x}, \vec{y}}\overline{\exists}_{\vec{x}}([\Theta]))  ) \cap \Theta \cap \Psi$ \\
by parallel reasoning:\\
 $ (\mydownarrow \rho_{\vec{y}, \vec{x}}\overline{\exists}_{\vec{y}}(\Tg \eval{G_4} \mu_k\  \mydownarrow \rho_{\vec{x}, \vec{y}}\overline{\exists}_{\vec{x}}([\Theta \cap \Psi]))  ) \cap \Theta \cap \Psi \\
\indent = (\mydownarrow \rho_{\vec{y}, \vec{x}}\overline{\exists}_{\vec{y}}(\Tg \eval{G_4} \mu_k\  \mydownarrow \rho_{\vec{x}, \vec{y}}\overline{\exists}_{\vec{x}}([\Theta]))  ) \cap \Theta \cap \Psi$\\
also by parallel reasoning:\\
 $(\mydownarrow \rho_{\vec{y}, \vec{x}}\overline{\exists}_{\vec{y}}(\Tg \eval{G_2} \mu_k\  \mydownarrow \rho_{\vec{x}, \vec{y}}\overline{\exists}_{\vec{x}}([\Theta \cap \Psi]))  ) \cap \Theta \cap \Psi\\
\indent = (\mydownarrow \rho_{\vec{y}, \vec{x}}\overline{\exists}_{\vec{y}}(\Tg \eval{G_2} \mu_k\  \mydownarrow \rho_{\vec{x}, \vec{y}}\overline{\exists}_{\vec{x}}([\Theta]))  ) \cap \Theta \cap \Psi$\\
\\  
hence if $(\mydownarrow \rho_{\vec{y}, \vec{x}}\overline{\exists}_{\vec{y}}(\Tg \eval{G_2} \mu_k\  \mydownarrow \rho_{\vec{x}, \vec{y}}\overline{\exists}_{\vec{x}}([\Theta \cap \Psi]))  ) \cap \Theta \cap \Psi = \Lambda : \vec{\Lambda}$\\
and $(\mydownarrow \rho_{\vec{y}, \vec{x}}\overline{\exists}_{\vec{y}}(\Tg \eval{G_2} \mu_k\  \mydownarrow \rho_{\vec{x}, \vec{y}}\overline{\exists}_{\vec{x}}([\Theta]))  ) \cap \Theta \cap \Psi = \Phi : \vec{\Phi}$\\
then $\Lambda = \Phi$\\
hence $(\mydownarrow \rho_{\vec{y}, \vec{x}}\overline{\exists}_{\vec{y}}(\Tg \eval{G_3} \mu_k\  [\Lambda]))  ) \cap \Theta \cap \Psi  = (\mydownarrow \rho_{\vec{y}, \vec{x}}\overline{\exists}_{\vec{y}}(\Tg \eval{G_3} \mu_k\  [\Phi]))  ) \cap \Theta \cap \Psi$\\
\\    
now say $\vec{\Gamma} = \left\{ \begin{array}{l l}
\Tg \eval{G_3 } \mu_k [\Phi] \
& if\ \Tg \eval{ G_2} \mu_k\ \mydownarrow \rho_{\vec{x}, \vec{y}}\overline{\exists}_{\vec{x}}([\Theta]) = \Phi : \vec{\Phi} \\
\Tg \eval{G_4} \mu_k\ \mydownarrow \rho_{\vec{x}, \vec{y}}\overline{\exists}_{\vec{x}}([\Theta])  \
 &if\  \Tg \eval{G_2} \mu_k\ \mydownarrow \rho_{\vec{x}, \vec{y}}\overline{\exists}_{\vec{x}}([\Theta]) = [] \\
\end{array}	\right.$\\
then $\vec{\Gamma} = \vec{\Delta}$\\
hence: $\mydownarrow \rho_{\vec{y}, \vec{x}}\overline{\exists}_{\vec{y}}(\mydownarrow \overline{\exists}_{\vec{y}} (\Tg \eval{G_1} \mu_k\ \mydownarrow \rho_{\vec{x}, \vec{y}}\overline{\exists}_{\vec{x}}([\Theta \cap \Psi])) : \vec{\Delta})) \cap \Theta \cap \Psi \\
= \mydownarrow \rho_{\vec{y}, \vec{x}}\overline{\exists}_{\vec{y}}(\mydownarrow \overline{\exists}_{\vec{y}} (\Tg \eval{G_1} \mu_k\ \mydownarrow \rho_{\vec{x}, \vec{y}}\overline{\exists}_{\vec{x}}([\Theta])) : \vec{\Gamma})) \cap \Theta \cap \Psi$\\
hence: $(\mydownarrow \rho_{\vec{y}, \vec{x}}\overline{\exists}_{\vec{y}}(\mu_{k+1} (p(\vec{y}))\ \mydownarrow \rho_{\vec{x}, \vec{y}}\overline{\exists}_{\vec{x}}([\Theta])) \cap \Theta) \cap \Psi\\
 = (\mydownarrow \rho_{\vec{y}, \vec{x}}\overline{\exists}_{\vec{y}}(\mu_{k+1} (p(\vec{y}))\ \mydownarrow \rho_{\vec{x}, \vec{y}}\overline{\exists}_{\vec{x}}([\Theta \cap \Psi])) \cap \Theta \cap \Psi)$\\
therefore: $(\Tg \eval{p(\vec{x})} \mu_{k+1} (\Theta :\vec{\Theta})) \cap \Psi = \Tg \eval{p(\vec{x})} \mu_{k+1} ((\Theta : \vec{\Theta}) \cap \Psi)$\\
\\
2.2.2 Induction Step $G = G_1, G_2$\\
Assume: $(\Tg \eval{G_1} \mu_{k+1} (\Theta : \vec{\Theta})) \cap \Psi = \Tg \eval{G_1} \mu_{k+1} ((\Theta : \vec{\Theta}) \cap \Psi)$\\
And: $(\Tg \eval{G_2} \mu_{k+1} (\Theta : \vec{\Theta})) \cap \Psi = \Tg \eval{G_2} \mu_{k+1} ((\Theta : \vec{\Theta}) \cap \Psi)$\\
Show: $(\Tg \eval{G_1,G_2} \mu_{k+1} (\Theta : \vec{\Theta})) \cap \Psi = \Tg \eval{G_1,G_2} \mu_{k+1} ((\Theta : \vec{\Theta}) \cap \Psi)$\\
\\
$ 
(\Tg \eval{G_1,G_2} \mu_{k+1} (\Theta : \vec{\Theta})) \cap \Psi \\
= (\Tg \eval{G_2} \mu_{k+1} (\Tg \eval{G_1} \mu_{k+1} (\Theta : \vec{\Theta}))) \cap \Psi \\
= (\Tg \eval{G_2} \mu_{k+1} (\Tg \eval{G_1} \mu_{k+1} (\Theta : \vec{\Theta})) \cap \Psi) \\
= \Tg \eval{G_2} \mu_{k+1} (\Tg \eval{G_1} \mu_{k+1} ((\Theta : \vec{\Theta}) \cap \Psi))\\
= \Tg \eval{G_1,G_2} \mu_{k+1} ((\Theta : \vec{\Theta}) \cap \Psi)\\
$
QED

\subsubsection{Lemma 2: $\Tg \eval{G} \mu [\Theta] = \Tg \eval{G} \mu [\Theta \cap S_G \eval{G}]$}

Proof in two stages: \\
(a) $\Tg \eval{G} \mu [\Theta \cap S_G \eval{G}] \sqsubseteq \Tg \eval{G} \mu [\Theta]$\\
(b) $\Tg \eval{G} \mu [\Theta] \sqsubseteq \Tg \eval{G} \mu [\Theta \cap S_G \eval{G}]$\\
\\
(a) by monotonicity of $\Tg$: \\
$ [\Theta \cap S_G \eval{G}] \sqsubseteq [\Theta] \Rightarrow \Tg \eval{G} \mu [\Theta \cap S_G \eval{G}] \sqsubseteq \Tg \eval{G} \mu [\Theta]$\\
\\
(b) $\Tg \eval{G} \mu [\Theta] \sqsubseteq \Tg \eval{G} \mu [\Theta \cap S_G \eval{G}]$\\
Proof by nested induction on:\\
1. $\mu$,\\
2. structure of $G$:\\
\\
1 Base Case: $\Tg \eval{G} \mu_{\bot} [\Theta] \sqsubseteq \Tg \eval{G} \mu_{\bot} [\Theta \cap S_G \eval{G}]$\\
\\
induction on structure of $G$:\\
1.1 Two Base Cases: (1) $G = \post(\phi)$, (2) $G = p(\vec{x})$\\
(1) $G = \post(\phi)$\\
Show: $\Tg \eval{\post(\phi)} \mu_{\bot} [\Theta] \sqsubseteq \Tg \eval{\post(\phi)} \mu_{\bot} [\Theta \cap S_G \eval{\post(\phi)}]$\\
\\
$\Tg \eval{\post(\phi)} \mu_{\bot} [\Theta] = trim([\Theta \cap \closed{\phi}])\\
\Tg \eval{\post(\phi)} \mu_{\bot} [\Theta \cap S_G \eval{\post(\phi)}] \\
\indent = trim([\Theta \cap S_G \eval{\post(\phi)} \cap \closed{\phi}])\\
\indent = trim([\Theta \cap \closed{\phi} \cap \closed{\phi}]) \\
\indent = trim([\Theta \cap \closed{\phi}])$\\
\\
(2) $G =  p(\vec{x})$\\
Show: $\Tg \eval{p(\vec{x})} \mu_{\bot} [\Theta] \sqsubseteq \Tg \eval{p(\vec{x})} \mu_{\bot} [\Theta \cap S_G \eval{p(\vec{x})}]$\\
\\
$\Tg \eval{p(\vec{x})} \mu_{\bot} [\Theta] = []\\
\Tg \eval{p(\vec{x})} \mu_{\bot} [\Theta \cap S_G \eval{p(\vec{x})}] = []$\\
\\
1.2 Induction Step: $G = G_1, G_2$\\
Assume: $\Tg \eval{G_1} \mu_{\bot} [\Theta_1] \sqsubseteq \Tg \eval{G_1} \mu_{\bot} [\Theta_1 \cap S_G \eval{G_1}]$\\
And: $\Tg \eval{G_2} \mu_{\bot} [\Theta_2] \sqsubseteq \Tg \eval{G_2} \mu_{\bot} [\Theta_2 \cap S_G \eval{G_2}]$\\
Show: $\Tg \eval{G_1, G_2} \mu_{\bot} [\Theta] \sqsubseteq \Tg \eval{G_1, G_2} \mu_{\bot} [\Theta \cap S_G \eval{G_1, G_2}]$\\
\\
$\Tg \eval{G_1, G_2} \mu_{\bot} [\Theta] = \Tg \eval{G_2} \mu_{\bot} (\Tg \eval{G_1} \mu_{\bot} [\Theta])$\\
by assumption: $\Tg \eval{G_1} \mu_{\bot} [\Theta] \sqsubseteq \Tg \eval{G_1} \mu_{\bot} [\Theta \cap S_G \eval{G_1}] $\\
hence: $\Tg \eval{G_2} \mu_{\bot} (\Tg \eval{G_1} \mu_{\bot} [\Theta]) \sqsubseteq \Tg \eval{G_2} \mu_{\bot} (\Tg \eval{G_1} \mu_{\bot} [\Theta \cap S_G \eval{G_1}])$\\
by assumption: \\
$\Tg \eval{G_2} \mu_{\bot} (\Tg \eval{G_1} \mu_{\bot} [\Theta \cap S_G \eval{G_1}]) \\
\indent \sqsubseteq \Tg \eval{G_2} \mu_{\bot} ((\Tg \eval{G_1} \mu_{\bot} [\Theta \cap S_G \eval{G_1}]) \cap S_G \eval{G_2}$)\\
by Lemma 1: $(\Tg \eval{G_1} \mu_{\bot} [\Theta \cap S_G \eval{G_1}]) \cap S_G \eval{G_2} \\
\indent = \Tg \eval{G_1} \mu_{\bot} [\Theta \cap S_G \eval{G_1} \cap S_G \eval{G_2}]$\\
hence: $\Tg \eval{G_2} \mu_{\bot} ((\Tg \eval{G_1} \mu_{\bot} [\Theta \cap S_G \eval{G_1}]) \cap S_G \eval{G_2})\\
\indent = \Tg \eval{G_2} \mu_{\bot} (\Tg \eval{G_1} \mu_{\bot} [\Theta \cap S_G \eval{G_1} \cap S_G \eval{G_2}])\\
\indent = \Tg \eval{G_1, G_2} \mu_{\bot} [\Theta \cap S_G \eval{G_1,G_2}]$\\
therefore: $\Tg \eval{G_1, G_2} \mu_{\bot} [\Theta ] \sqsubseteq \Tg \eval{G_1, G_2} \mu_{\bot} [\Theta \cap S_G \eval{G_1,G_2}]$\\
\\
2 Induction Step:\\
Assume: $\Tg \eval{H} \mu_k [\Theta'] \sqsubseteq \Tg \eval{H} \mu_k [\Theta' \cap S_G \eval{H}]$\\
Show: $\Tg \eval{G} \mu_{k+1} [\Theta] \sqsubseteq \Tg \eval{G} \mu_{k+1} [\Theta \cap S_G \eval{G}]$\\
where $\mu_{k+1} = \Tprog \eval{P} \mu_k$\\
\\
induction on structure of G:\\
2.1 Two Base Cases: (1) $G = \post(\phi)$, (2) $G = p(\vec{x})$\\
(1) $G = \post(\phi)$\\
Show: $\Tg \eval{\post(\phi)} \mu_{k+1} [\Theta] \sqsubseteq \Tg \eval{\post(\phi)} \mu_{k+1} [\Theta \cap S_G \eval{\post(\phi)}]$\\
where $\mu_{k+1} = \Tprog \eval{P} \mu_k$\\
\\
$\Tg \eval{\post(\phi)} \mu_{k+1} [\Theta] = trim([\Theta \cap \closed{\phi}])\\
\Tg \eval{\post(\phi)} \mu_{k+1} [\Theta \cap S_G \eval{\post(\phi)}]\\
\indent = trim([\Theta \cap S_G \eval{\post(\phi)} \cap \closed{\phi}])\\
\indent = trim([\Theta \cap \closed{\phi} \cap \closed{\phi}]) \\
\indent = trim([\Theta \cap \closed{\phi}])$\\
\\
(2) $G = p(\vec{x})$\\
Assume (without loss\ of generality): $p(\vec{y}) \neck G_1 ; G_2,!,G_3 ; G_4 \in P$\\
Show: $\Tg \eval{p(\vec{x})} \mu_{k+1} [\Theta] \sqsubseteq \Tg \eval{p(\vec{x})} \mu_{k+1} [\Theta \cap S_G \eval{p(\vec{x})}]$\\
where: $\mu_{k+1} = \Tprog \eval{P} \mu_k$\\
$\Tg \eval{p(\vec{x})} \mu_{k+1} [\Theta \cap S_G \eval{p(\vec{x})}] \\
\indent = \mydownarrow \rho_{\vec{y}, \vec{x}}\overline{\exists}_{\vec{y}}(\mu_{k+1} (p(\vec{y}))\ \mydownarrow \rho_{\vec{x}, \vec{y}}\overline{\exists}_{\vec{x}}([\Theta \cap S_G \eval{p(\vec{x})}])) \cap \Theta \cap S_G \eval{p(\vec{x})}\\
\mu_{k+1} (p(\vec{y}))\ \mydownarrow \rho_{\vec{x}, \vec{y}}\overline{\exists}_{\vec{x}}([\Theta \cap S_G \eval{p(\vec{x})}]) = \mydownarrow \overline{\exists}_{\vec{y}} (\Tg \eval{G_1} \mu_k \ \mydownarrow \rho_{\vec{x}, \vec{y}}\overline{\exists}_{\vec{x}}([\Theta \cap S_G \eval{p(\vec{x})}]) : \vec{\Psi})$\\ 
where\\
$ \vec{\Psi} = \left\{ \begin{array}{l l}
\Tg \eval{G_3 } \mu_k [\Phi] 
& if \ \Tg \eval{ G_2} \mu_k\ \mydownarrow \rho_{\vec{x}, \vec{y}}\overline{\exists}_{\vec{x}}([\Theta \cap S_G \eval{p(\vec{x})}]) = \Phi : \vec{\Phi} \\
\Tg \eval{G_4} \mu_k\ \mydownarrow \rho_{\vec{x}, \vec{y}}\overline{\exists}_{\vec{x}}([\Theta \cap S_G \eval{p(\vec{x})}]) 
& if \  \Tg \eval{G_2} \mu_k\ \mydownarrow \rho_{\vec{x}, \vec{y}}\overline{\exists}_{\vec{x}}([\Theta \cap S_G \eval{p(\vec{x})}]) = [] \\
\end{array}	\right. $\\
now: $S_G \eval{p(\vec{x})} = \mydownarrow \rho_{\vec{y},\vec{x}}  \overline{\exists}_{\vec{y}}(S_H \eval{p(\vec{y})}) \\
\indent = \mydownarrow \rho_{\vec{y},\vec{x}}  \overline{\exists}_{\vec{y}}( \mydownarrow \overline{ \exists}_{\vec{y}} ( S_G \eval{G_1} \cup S_G \eval{G_2, G_3} \cup S_G \eval{G_4} ) ) \\
\indent = \mydownarrow \rho_{\vec{y},\vec{x}}  \overline{\exists}_{\vec{y}}( S_G \eval{G_1} \cup S_G \eval{G_2, G_3} \cup S_G \eval{G_4} )  \\
\indent = \mydownarrow \rho_{\vec{y},\vec{x}} \overline{\exists}_{\vec{y}}( S_G \eval{G_1})  \cup \mydownarrow \rho_{\vec{y},\vec{x}} \overline{\exists}_{\vec{y}}(S_G \eval{G_2, G_3}) \cup \mydownarrow \rho_{\vec{y},\vec{x}} \overline{\exists}_{\vec{y}}(S_G \eval{G_4} )\\
\indent \indent (\text{because } \overline{\exists}\ \text{distributes over } \cup)$ \\
Since $S_G \eval{p(\vec{x})}$ is the union of these three components, it is a superset of each of them, hence: $ S_G \eval{p(\vec{x})} \supseteq \mydownarrow \rho_{\vec{y},\vec{x}} \overline{\exists}_{\vec{y}}( S_G \eval{G_1})$\\
and: $S_G \eval{p(\vec{x})} \supseteq \mydownarrow \rho_{\vec{y},\vec{x}} \overline{\exists}_{\vec{y}}(S_G \eval{G_2, G_3})$\\
and: $ S_G \eval{p(\vec{x})} \supseteq  \mydownarrow \rho_{\vec{y},\vec{x}} \overline{\exists}_{\vec{y}}(S_G \eval{G_4} )$\\
Intersecting each side with $\Theta$ preserves the order,
hence: $ \Theta \cap S_G \eval{p(\vec{x})} \supseteq  \Theta \cap \mydownarrow \rho_{\vec{y},\vec{x}} \overline{\exists}_{\vec{y}}( S_G \eval{G_1})$\\
and: $ \Theta \cap S_G \eval{p(\vec{x})} \supseteq  \Theta \cap \mydownarrow \rho_{\vec{y},\vec{x}} \overline{\exists}_{\vec{y}}(S_G \eval{G_2, G_3})$\\
and: $ \Theta \cap S_G \eval{p(\vec{x})} \supseteq  \Theta \cap \mydownarrow \rho_{\vec{y},\vec{x}} \overline{\exists}_{\vec{y}}(S_G \eval{G_4} )$\\
Again, projecting and renaming both sides in the same way preserves the order,
hence: $ \mydownarrow \rho_{\vec{x}, \vec{y}}\overline{\exists}_{\vec{x}}(\Theta \cap S_G \eval{p(\vec{x})}) \supseteq  \mydownarrow \rho_{\vec{x}, \vec{y}}\overline{\exists}_{\vec{x}}(\Theta \cap \mydownarrow \rho_{\vec{y},\vec{x}} \overline{\exists}_{\vec{y}}( S_G \eval{G_1}))$\\
and: $\mydownarrow \rho_{\vec{x}, \vec{y}}\overline{\exists}_{\vec{x}}( \Theta \cap S_G \eval{p(\vec{x})}) \supseteq  \mydownarrow \rho_{\vec{x}, \vec{y}}\overline{\exists}_{\vec{x}}(\Theta \cap \mydownarrow \rho_{\vec{y},\vec{x}} \overline{\exists}_{\vec{y}}(S_G \eval{G_2, G_3}))$\\
and: $\mydownarrow \rho_{\vec{x}, \vec{y}}\overline{\exists}_{\vec{x}}( \Theta \cap S_G \eval{p(\vec{x})}) \supseteq   \mydownarrow \rho_{\vec{x}, \vec{y}}\overline{\exists}_{\vec{x}}(\Theta \cap \mydownarrow \rho_{\vec{y},\vec{x}} \overline{\exists}_{\vec{y}}(S_G \eval{G_4} ))$\\
Now, since the following holds in general:\\
\indent \indent $\mydownarrow \rho_{\vec{x}, \vec{y}}\overline{\exists}_{\vec{x}}(\Gamma_1 \cap \mydownarrow \rho_{\vec{y},\vec{x}} \overline{\exists}_{\vec{y}}(\Gamma_2 )) \supseteq \mydownarrow \rho_{\vec{x}, \vec{y}}\overline{\exists}_{\vec{x}}(\Gamma_1) \cap \Gamma_2$,\\
performing the same transformation on the above still preserves the order, \\
hence: $ \mydownarrow \rho_{\vec{x}, \vec{y}}\overline{\exists}_{\vec{x}}(\Theta \cap S_G \eval{p(\vec{x})}) \supseteq \mydownarrow \rho_{\vec{x}, \vec{y}}\overline{\exists}_{\vec{x}}(\Theta) \cap S_G \eval{G_1}$ \\
and: $\mydownarrow \rho_{\vec{x}, \vec{y}}\overline{\exists}_{\vec{x}}( \Theta \cap S_G \eval{p(\vec{x})}) \supseteq \mydownarrow \rho_{\vec{x}, \vec{y}}\overline{\exists}_{\vec{x}}(\Theta) \cap S_G \eval{G_2,G_3}  \\
\indent \indent \indent = \mydownarrow \rho_{\vec{x}, \vec{y}}\overline{\exists}_{\vec{x}}(\Theta) \cap S_G \eval{G_2} \cap S_G \eval{G_3}$\\
and: $\mydownarrow \rho_{\vec{x}, \vec{y}}\overline{\exists}_{\vec{x}}( \Theta \cap S_G \eval{p(\vec{x})}) \supseteq  \mydownarrow \rho_{\vec{x}, \vec{y}}\overline{\exists}_{\vec{x}}(\Theta) \cap S_G \eval{G_4} $\\
by monotonicity of $\Tg$, therefore:\\
$\Tg \eval{G_1} \mu_k \ \mydownarrow \rho_{\vec{x}, \vec{y}}\overline{\exists}_{\vec{x}}([\Theta \cap S_G \eval{p(\vec{x})}]) \sqsupseteq \Tg \eval{G_1} \mu_k [\mydownarrow \rho_{\vec{x}, \vec{y}}\overline{\exists}_{\vec{x}}(\Theta) \cap S_G \eval{G_1} ]$\\
by assumption: $\Tg \eval{G_1} \mu_k [\mydownarrow \rho_{\vec{x}, \vec{y}}\overline{\exists}_{\vec{x}}(\Theta) \cap S_G \eval{G_1} ] \sqsupseteq \Tg \eval{G_1} \mu_k [\mydownarrow \rho_{\vec{x}, \vec{y}}\overline{\exists}_{\vec{x}}(\Theta) ]$\\
hence the following holds of the first part of the sequence:\\
$\Tg \eval{G_1} \mu_k \ \mydownarrow \rho_{\vec{x}, \vec{y}}\overline{\exists}_{\vec{x}}([\Theta \cap S_G \eval{p(\vec{x})}]) \sqsupseteq \Tg \eval{G_1} \mu_k \ \mydownarrow \rho_{\vec{x}, \vec{y}}\overline{\exists}_{\vec{x}}[\Theta ]$\\
and similarly: $\Tg \eval{G_4} \mu_k \ \mydownarrow \rho_{\vec{x}, \vec{y}}\overline{\exists}_{\vec{x}}([\Theta \cap S_G \eval{p(\vec{x})}]) \sqsupseteq \Tg \eval{G_4} \mu_k [\mydownarrow \rho_{\vec{x}, \vec{y}}\overline{\exists}_{\vec{x}}(\Theta) \cap S_G \eval{G_4} ]$\\
by assumption: $\Tg \eval{G_4} \mu_k [\mydownarrow \rho_{\vec{x}, \vec{y}}\overline{\exists}_{\vec{x}}(\Theta) \cap S_G \eval{G_4} ] \sqsupseteq \Tg \eval{G_4} \mu_k [\mydownarrow \rho_{\vec{x}, \vec{y}}\overline{\exists}_{\vec{x}}(\Theta) ]$\\
hence the parallel thing holds for the second possibility of the second part of the sequence:\\
$\Tg \eval{G_4} \mu_k \ \mydownarrow \rho_{\vec{x}, \vec{y}}\overline{\exists}_{\vec{x}}([\Theta \cap S_G \eval{p(\vec{x})}]) \sqsupseteq \Tg \eval{G_4} \mu_k \ \mydownarrow \rho_{\vec{x}, \vec{y}}\overline{\exists}_{\vec{x}}[\Theta ]$\\
As for the first possibility for the second part of the sequence, consider this:\\
by monotonicity of $\Tg$:\\
 $\Phi:\vec{\Phi} =  \Tg \eval{G_2} \mu_k \ \mydownarrow \rho_{\vec{x}, \vec{y}}\overline{\exists}_{\vec{x}}([\Theta \cap S_G \eval{p(\vec{x})}]) \\
\indent \indent \indent \sqsupseteq \Tg \eval{G_2} \mu_k [\mydownarrow \rho_{\vec{x}, \vec{y}}\overline{\exists}_{\vec{x}}(\Theta) \cap S_G \eval{G_2} \cap S_G \eval{G3} ]$\\
by assumption:\\
 $\Tg \eval{G_2} \mu_k [\mydownarrow \rho_{\vec{x}, \vec{y}}\overline{\exists}_{\vec{x}}(\Theta) \cap S_G \eval{G_2} \cap S_G \eval{G3} ] \sqsupseteq \Tg \eval{G_2} \mu_k [\mydownarrow \rho_{\vec{x}, \vec{y}}\overline{\exists}_{\vec{x}}(\Theta) \cap S_G \eval{G3} ]$\\
by Lemma 1: $\Tg \eval{G_2} \mu_k [\mydownarrow \rho_{\vec{x}, \vec{y}}\overline{\exists}_{\vec{x}}(\Theta) \cap S_G \eval{G3} ] = \Tg \eval{G_2} \mu_k [\mydownarrow \rho_{\vec{x}, \vec{y}}\overline{\exists}_{\vec{x}}[\Theta] \cap S_G \eval{G3} $\\
hence: $\Phi:\vec{\Phi} \sqsupseteq \Tg \eval{G_2} \mu_k [\mydownarrow \rho_{\vec{x}, \vec{y}}\overline{\exists}_{\vec{x}}[\Theta] \cap S_G \eval{G3}$ \\
now call the part of the sequence we are aiming for here $\Lambda : \vec{\Lambda} =\Tg \eval{ G_2} \mu_k\ \mydownarrow \rho_{\vec{x}, \vec{y}}\overline{\exists}_{\vec{x}}([\Theta ])$\\
then: $\Phi:\vec{\Phi} \sqsupseteq (\Lambda : \vec{\Lambda}) \cap S_G \eval{G_3}$\\
hence: $[\Phi] \sqsupseteq [\Lambda \cap S_G \eval{G_3}]$\\
hence: $\Tg \eval{G_3} \mu_k [\Phi] \sqsupseteq \Tg \eval{G_3} \mu_k [\Lambda \cap \eval{G_3}]$\\
by assumption: $\Tg \eval{G_3} \mu_k [\Lambda \cap \eval{G_3}] \sqsupseteq \Tg \eval{G_3} \mu_k [\Lambda]$\\
hence: $\Tg \eval{G_3} \mu_k [\Phi] \sqsupseteq \Tg \eval{G_3} \mu_k [\Lambda]$\\
\\
These last few lines show that each part of the sequence we are considering is greater than the sequence we are aiming for. Pulling these together, we arrive at:\\
$\mydownarrow \overline{\exists}_{\vec{y}} (\Tg \eval{G_1} \mu_k \ \mydownarrow \rho_{\vec{x}, \vec{y}}\overline{\exists}_{\vec{x}}([\Theta \cap S_G \eval{p(\vec{x})}]) : \vec{\Psi}) \sqsupseteq \ \mydownarrow \overline{\exists}_{\vec{y}} (\Tg \eval{G_1} \mu_k \ \mydownarrow \rho_{\vec{x}, \vec{y}}\overline{\exists}_{\vec{x}}([\Theta ]) : \vec{\Delta})$
\\
 where\\
$ \vec{\Psi} = \left\{ \begin{array}{l l}
\Tg \eval{G_3 } \mu_k [\Phi] 
& if \ \Tg \eval{ G_2} \mu_k\ \mydownarrow \rho_{\vec{x}, \vec{y}}\overline{\exists}_{\vec{x}}([\Theta \cap S_G \eval{p(\vec{x})}]) = \Phi : \vec{\Phi} \\
\Tg \eval{G_4} \mu_k\ \mydownarrow \rho_{\vec{x}, \vec{y}}\overline{\exists}_{\vec{x}}([\Theta \cap S_G \eval{p(\vec{x})}]) 
& if \  \Tg \eval{G_2} \mu_k\ \mydownarrow \rho_{\vec{x}, \vec{y}}\overline{\exists}_{\vec{x}}([\Theta \cap S_G \eval{p(\vec{x})}]) = [] \\
\end{array}	\right. \\$
and $\vec{\Delta} = \left\{ \begin{array}{l l}
\Tg \eval{G_3 } \mu_k [\Lambda] 
& if \ \Tg \eval{ G_2} \mu_k\ \mydownarrow \rho_{\vec{x}, \vec{y}}\overline{\exists}_{\vec{x}}([\Theta]) = \Lambda : \vec{\Lambda} \\
\Tg \eval{G_4} \mu_k\ \mydownarrow \rho_{\vec{x}, \vec{y}}\overline{\exists}_{\vec{x}}([\Theta]) 
& if \  \Tg \eval{G_2} \mu_k\ \mydownarrow \rho_{\vec{x}, \vec{y}}\overline{\exists}_{\vec{x}}([\Theta]) = [] \\
\end{array}	\right. $\\
\\
therefore: $\mu_{k+1} (p(\vec{y}))\ \mydownarrow \rho_{\vec{x}, \vec{y}}\overline{\exists}_{\vec{x}}([\Theta \cap S_G \eval{p(\vec{x})}]) \sqsupseteq \mu_{k+1} (p(\vec{y}))\ \mydownarrow \rho_{\vec{x}, \vec{y}}\overline{\exists}_{\vec{x}}([\Theta])$\\
applying the same renaming and projection to both sides preserves the order:\\ $\mydownarrow \rho_{\vec{y}, \vec{x}}\overline{\exists}_{\vec{y}} (\mu_{k+1} (p(\vec{y}))\ \mydownarrow \rho_{\vec{x}, \vec{y}}\overline{\exists}_{\vec{x}}([\Theta \cap S_G \eval{p(\vec{x})}])) \sqsupseteq \mydownarrow \rho_{\vec{y}, \vec{x}}\overline{\exists}_{\vec{y}}(\mu_{k+1} (p(\vec{y}))\ \mydownarrow \rho_{\vec{x}, \vec{y}}\overline{\exists}_{\vec{x}}([\Theta]))$\\
now name these two sequences: \\
$\mydownarrow \rho_{\vec{y}, \vec{x}}\overline{\exists}_{\vec{y}} (\mu_{k+1} (p(\vec{y}))\ \mydownarrow \rho_{\vec{x}, \vec{y}}\overline{\exists}_{\vec{x}}([\Theta \cap S_G \eval{p(\vec{x})}])) = \vec{\Psi'}$\\
and: $\mydownarrow \rho_{\vec{y}, \vec{x}}\overline{\exists}_{\vec{y}}(\mu_{k+1} (p(\vec{y}))\ \mydownarrow \rho_{\vec{x}, \vec{y}}\overline{\exists}_{\vec{x}}([\Theta])) = \vec{\Delta'}$\\
and notice the following two facts:\\
(1) $\vec{\Delta'} \cap \Theta = \Tg \eval{p(\vec{x})} \mu_{k+1} [\Theta]$\\
(2) $\vec{\Psi'} \cap S_G \eval{p(\vec{x})} \cap \Theta = \Tg \eval{p(\vec{x})} \mu_{k+1} [\Theta \cap S_G \eval{p(\vec{x})}]$ \\
then from above we have: $\vec{\Delta'} \sqsubseteq \vec{\Psi'}$\\
hence: $\vec{\Delta'} \cap \Theta \sqsubseteq \vec{\Psi'}$\\
by (1) and Theorem 1, therefore: $ \bigcup(\vec{\Delta'}) \cap \Theta = \bigcup(\vec{\Delta'} \cap \Theta) \subseteq \Theta \cap S_G \eval{p(\vec{x})}$\\
hence: $\bigcup (\vec{\Delta'}) \subseteq S_G \eval{p(\vec{x})} $\\
therefore for each $\Delta'$ in $\vec{\Delta'}$: $\Delta' \subseteq S_G \eval{p(\vec{x})}$ \\
hence for each $\Delta'$ in $\vec{\Delta'}$: $\Delta' \cap S_G \eval{p(\vec{x})} = \Delta'$ \\
hence: $\vec{\Delta'} \cap S_G \eval{p(\vec{x})} = \vec{\Delta'} $\\
hence: $\vec{\Delta'} \cap \Theta = \vec{\Delta'} \cap S_G \eval{p(\vec{x})} \cap \Theta  \sqsubseteq \vec{\Psi'} \cap S_G \eval{p(\vec{x})} \cap \Theta$\\
substituting using (2), we therefore arrive at:\\
$\Tg \eval{p(\vec{x})} \mu_{k+1} [\Theta] \sqsubseteq \Tg \eval{p(\vec{x})} \mu_{k+1} [\Theta \cap S_G \eval{p(\vec{x})}]$\\
\\
2.2 Induction Step: $G = G_1, G_2$\\
Assume: $\Tg \eval{G_1} \mu_{k+1} [\Theta_1] \sqsubseteq \Tg \eval{G_1} \mu_{k+1} [\Theta_1 \cap S_G \eval{G_1}]$\\
And: $\Tg \eval{G_2} \mu_{k+1} [\Theta_2] \sqsubseteq \Tg \eval{G_2} \mu_{k+1} [\Theta_2 \cap S_G \eval{G_2}]$\\
Show: $\Tg \eval{G_1, G_2} \mu_{k+1} [\Theta] \sqsubseteq \Tg \eval{G_1, G_2} \mu_{k+1} [\Theta \cap S_G \eval{G_1, G_2}]$\\
\\
$
\Tg \eval{G_1, G_2} \mu_{k+1} [\Theta] = \Tg \eval{G_2} \mu_{k+1} (\Tg \eval{G_1} \mu_{k+1} [\Theta])$\\
by assumption: $\Tg \eval{G_1} \mu_{k+1} [\Theta] \sqsubseteq \Tg \eval{G_1} \mu_{k+1} [\Theta \cap S_G \eval{G_1}]$ \\
hence: $\Tg \eval{G_2} \mu_{k+1} (\Tg \eval{G_1} \mu_{k+1} [\Theta]) \sqsubseteq \Tg \eval{G_2} \mu_{k+1} (\Tg \eval{G_1} \mu_{k+1} [\Theta \cap S_G \eval{G_1}])$\\
by assumption:\\
 $\Tg \eval{G_2} \mu_{k+1} (\Tg \eval{G_1} \mu_{k+1} [\Theta \cap S_G \eval{G_1}]) \sqsubseteq \Tg \eval{G_2} \mu_{k+1} ((\Tg \eval{G_1} \mu_{k+1} [\Theta \cap S_G \eval{G_1}]) \cap S_G \eval{G_2})$\\
by Lemma 1: $(\Tg \eval{G_1} \mu_{k+1} [\Theta \cap S_G \eval{G_1}]) \cap S_G \eval{G_2} = \Tg \eval{G_1} \mu_{k+1} [\Theta \cap S_G \eval{G_1} \cap S_G \eval{G_2}]$\\
hence: $\Tg \eval{G_2} \mu_{k+1} ((\Tg \eval{G_1} \mu_{k+1} [\Theta \cap S_G \eval{G_1}]) \cap S_G \eval{G_2})\\
\indent = \Tg \eval{G_2} \mu_{k+1} (\Tg \eval{G_1} \mu_{k+1} [\Theta \cap S_G \eval{G_1} \cap S_G \eval{G_2}])\\
\indent = \Tg \eval{G_1, G_2} \mu_{k+1} [\Theta \cap S_G \eval{G_1,G_2}]$\\
therefore: $\Tg \eval{G_1, G_2} \mu_{k+1} [\Theta ] \sqsubseteq \Tg \eval{G_1, G_2} \mu_{k+1} [\Theta \cap S_G \eval{G_1,G_2}]$\\
\\
Therefore since: 
(a) $\Tg \eval{G} \mu [\Theta \cap S_G \eval{G}] \sqsubseteq \Tg \eval{G} \mu [\Theta]$\\
and (b) $\Tg \eval{G} \mu [\Theta] \sqsubseteq \Tg \eval{G} \mu [\Theta \cap S_G \eval{G}]$,\\
it follows that:
$\indent \Tg \eval{G} \mu [\Theta] = \Tg \eval{G} \mu [\Theta \cap S_G \eval{G}]$\\
QED

\subsubsection{Proof of Theorem 2: For $\Theta \in \con$ and stratified $P = P_0 \cup \ldots \cup P_n$: $\Theta \subseteq \Dg \eval{G} \delta_P \Rightarrow |\Tg \eval{G} \mu_P [ \Theta ] | \leq 1$.}

First notice that the following things hold:  \\
(1) $\Theta \subseteq (\Phi \rightarrow \Psi) \Rightarrow \Theta \cap \Phi \subseteq \Psi$\\
(2) $\Theta \subseteq mux(\Phi, \Psi) \Rightarrow (\Theta \cap \Phi = \{false\}) \vee (\Theta \cap \Psi = \{false\})$\\
(3) $\Tg \eval{G} \mu \vec{\Theta} \subseteq \bigcup \vec{\Theta} \cap S_G \eval{G}$ \\
\indent for any $\mu$ constructed by application of $\Tprog \eval{P}$ to $\mu_{\bot}$\\
(4) $\overline{\forall}_{\vec{y}} (\Theta \cap \Phi) = \overline{\forall}_{\vec{y}} (\Theta) \cap \overline{\forall}_{\vec{y}} (\Phi)$\\
(5) $\Theta \subseteq \mydownarrow \rho_{\vec{y},\vec{x} } \overline{\forall}_{\vec{y}} (\Phi) \Rightarrow \mydownarrow \rho_{\vec{x}, \vec{y}}\overline{\exists}_{\vec{x}}(\Theta) \subseteq \Phi $\\ 
\indent This holds due to the following few lines of reasoning:\\
\indent $\Theta \subseteq \overline{\exists}_{\vec{x}} (\Theta)$ (since $\overline{\exists} $ is extensive)\\
\indent if $\Theta \subseteq \mydownarrow \rho_{\vec{y},\vec{x} } \overline{\forall}_{\vec{y}} (\Phi)$ \\
\indent then $\overline{\exists}_{\vec{x}} (\Theta) \subseteq \overline{\exists}_{\vec{x}} (\mydownarrow \rho_{\vec{y},\vec{x} } \overline{\forall}_{\vec{y}} (\Phi))$ \\
\indent \indent (by monotonicity of $\overline{\exists}$)\\
\indent then $\mydownarrow \rho_{\vec{x}, \vec{y}} \overline{\exists}_{\vec{x}} (\Theta) \subseteq \mydownarrow \rho_{\vec{x}, \vec{y}} \overline{\exists}_{\vec{x}} (\mydownarrow \rho_{\vec{y},\vec{x} } \overline{\forall}_{\vec{y}} (\Phi)) = \mydownarrow \rho_{\vec{x}, \vec{y}} (\mydownarrow \rho_{\vec{y},\vec{x} } \overline{\forall}_{\vec{y}} (\Phi))$ \\
\indent \indent (by monotonicity of $\mydownarrow$, $\rho$) \\
\indent $\mydownarrow \rho_{\vec{x},\vec{y} }\mydownarrow \rho_{\vec{y},\vec{x} }$ cancel out and $\overline{\forall}$ is reductive, hence:\\
\indent $\mydownarrow \rho_{\vec{x}, \vec{y}}\overline{\exists}_{\vec{x}}(\Theta) \subseteq \Phi$\\
(6) $\Tg \eval{G} \mu [\Theta] \sqsubseteq \Tg \eval{G} \mu (\Theta : \vec{\Theta})$\\
\indent again for any $\mu$ constructed by application of $\Tprog \eval{P}$ to $\mu_{\bot}$\\
(7) $\vec{\Theta}_1 \sqsubseteq \vec{\Theta}_2 \Rightarrow |\vec{\Theta}_1| <= |\vec{\Theta}_2|$\\
\\
Proof by nested induction on:\\
1. $\mu$, \\
2. structure of $G$:
\\
1 Base Case: $\mu = \mu_{\bot}$\\
show: $\Theta \subseteq \Dg \eval{G} \delta_P \Rightarrow |\Tg \eval{G} \mu_{\bot} [ \Theta ] | \leq 1$\\
\\
Induction on structure of $G$:\\
1.1 Two Base Cases: (1) $G = \post(\phi)$, (2) $G = p(\vec{x})$\\
\\
(1) $G = \post(\phi)$:\\
Show: $\Theta \subseteq \Dg \eval{\post(\phi)} \delta_P \Rightarrow |\Tg \eval{\post(\phi)} \mu_{\bot} [ \Theta ] | \leq 1$\\
\\
$\Tg \eval{\post(\phi)} \mu_{\bot} [\Theta] = trim([\closed{\phi} \cap \Theta])$\\
hence: $|\Tg \eval{\post(\phi)} \mu_{\bot} [ \Theta ] | = |trim([\closed{\phi} \cap \Theta])| \leq 1$\\
\\
(2) $G = p(\vec{x})$\\
Show: $\Theta \subseteq \Dg \eval{p(\vec{y})} \delta_P \Rightarrow |\Tg \eval{p(\vec{y})} \mu_{\bot} [ \Theta ] | \leq 1$\\
\\
$\Tg \eval{p(\vec{y})} \mu_{\bot} [ \Theta ] = \mydownarrow \rho_{\vec{y}, \vec{x}}\overline{\exists}_{\vec{y}}(\mu_{\bot}\ (p(\vec{y}))\ \mydownarrow \rho_{\vec{x}, \vec{y}}\overline{\exists}_{\vec{x}}([\Theta])) \cap \Theta : []\\
\mu_{\bot}\ (p(\vec{y}))\ \mydownarrow \rho_{\vec{x}, \vec{y}}\overline{\exists}_{\vec{x}}([\Theta]) = []$\\
hence: $\Tg \eval{p(\vec{y})} \mu_{\bot} [ \Theta ] = []$\\
hence: $|\Tg \eval{p(\vec{y})} \mu_{\bot} [ \Theta ]| = |[]| = 0$\\
\\
1.2 Induction Step:\\
$G = G_1,G_2:$\\
Assume: $\Theta_1 \subseteq \Dg \eval{G_1} \delta_P \Rightarrow |\Tg \eval{G_1} \mu_{\bot} [ \Theta_1 ] | \leq 1$\\
And: $\Theta_2 \subseteq \Dg \eval{G_2} \delta_P \Rightarrow |\Tg \eval{G_2} \mu_{\bot} [ \Theta_2] | \leq 1$\\
Show: $\Theta \subseteq \Dg \eval{G} \delta_P \Rightarrow |\Tg \eval{G} \mu_{\bot} [ \Theta ] | \leq 1$\\
\\
$\Dg \eval{G} \delta_P = (S_G \eval{ G_2} \rightarrow \Dg \eval{ G_1} \delta_P) \cap (S_G \eval{ G_1} \rightarrow \Dg \eval{ G_2} \delta_P)\\
\Theta \subseteq \Dg \eval{G} \delta_P \Rightarrow \Theta \subseteq (S_G \eval{ G_1} \rightarrow \Dg \eval{ G_2} \delta_P) \Rightarrow \Theta \cap S_G \eval{G_1} \subseteq \Dg \eval{G_2} \delta_P\\
\Theta \subseteq \Dg \eval{G} \delta_P \Rightarrow \Theta \subseteq (S_G \eval{ G_2} \rightarrow \Dg \eval{ G_1} \delta_P) \Rightarrow \Theta \cap S_G \eval{G_2} \subseteq \Dg \eval{G_1} \delta_P\\
\Tg \eval{G} \mu_{\bot} [\Theta] = \Tg \eval{G} \mu_{\bot} [\Theta \cap S_G \eval{G}]\ (by\ Lemma\ 2)\\
\Tg \eval{G} \mu_{\bot} [\Theta \cap S_G \eval{G}] = \Tg \eval{G_2} \mu_{\bot} (\Tg \eval{ G_1} \mu_{\bot} [\Theta \cap S_G \eval{G_1, G_2}])\\
\Theta \cap S_G \eval{G_1,G_2} = \Theta \cap S_G \eval{G_1} \cap S_G \eval{G_2} \subseteq \Theta \cap S_G \eval{G_2} \subseteq \Dg \eval{G_1} \delta_P$\\
hence by assumption: $|\Tg \eval{ G_1} \mu_{\bot} [\Theta \cap S_G \eval{G_1,G_2}]| <= 1$\\
distinguish two cases:\\
 (a) $|\Tg \eval{ G_1} \mu_{\bot} [\Theta \cap S_G \eval{G_1,G_2} ]| = 0$, \\
 (b) $|\Tg \eval{ G_1} \mu_{\bot} [\Theta \cap S_G \eval{G_1,G_2}]| = 1$\\
\\
(a) $|\Tg \eval{ G_1} \mu_{\bot} [\Theta \cap S_G \eval{G_1,G_2} ]| = 0$\\
$\Tg \eval{ G_1} \mu_{\bot} [\Theta \cap S_G \eval{G_1,G_2}] = []\\
\Tg \eval{G} \mu_{\bot} [\Theta \cap S_G \eval{G_1,G_2}] = \Tg \eval{G_2} \mu_{\bot} [] = []$\\
hence: $|\Tg \eval{G} \mu_{\bot} [\Theta \cap S_G \eval{G_1,G_2}]| <= 1$\\
by Lemma 2 (remembering $G = G_1,G_2$): $|\Tg \eval{G} \mu_{\bot} [\Theta]| <= 1$\\
\\
(b) $|\Tg \eval{ G_1} \mu_{\bot} [\Theta \cap S_G \eval{G_1,G_2}]| = 1$\\
$\Tg \eval{ G_1} \mu_{\bot} [\Theta \cap S_G \eval{G_1,G_2} ] = [\Psi]$\\
by Theorem 1: $\bigcup(\Tg \eval{ G_1} \mu_{\bot} [\Theta \cap S_G \eval{G_1,G_2}] )\\
\indent \indent \subseteq \Theta \cap S_G \eval{G_1,G_2} \cap S_G \eval{G_1} \\
\indent \indent \subseteq \Theta \cap S_G \eval{G_1}$\\
hence: $\Psi \subseteq \Theta \cap S_G \eval{G_1}$\\
hence: $\Psi \subseteq \Dg \eval{G_2} \delta_P$\\
hence by assumption: $|\Tg \eval{G_2} \mu_{\bot} [\Psi] | \leq 1$\\
hence (again by Lemma 2): $|\Tg \eval{G_2} \mu_{\bot} (\Tg \eval{ G_1} \mu_{\bot} [\Theta \cap S_G \eval{G_1,G_2}])|\\
\indent = |\Tg \eval{G} \mu_{\bot} [\Theta\cap S_G \eval{G_1,G_2}] | \\
\indent = |\Tg \eval{G} \mu_{\bot} [\Theta]| <= 1$\\
\\
2 Induction Step:\\
Assume: $X \subseteq \Dg \eval{H} \delta_P \Rightarrow |\Tg \eval{H} \mu_k [ X ] | \leq 1$\\
Show: $\Theta \subseteq \Dg \eval{G} \delta_P \Rightarrow |\Tg \eval{G} \mu_{k+1} [ \Theta ] | \leq 1$\\
where $\mu_{k+1} = \Tprog \eval{P} \mu_k$\\
\\
Induction on structure of $G$:\\
2.1 Two base cases: (1) $G = \post(\phi)$, (2) $G = p(\vec{x})$\\
\\
(1) $G = \post(\phi)$:\\
Show: $\Theta \subseteq \Dg \eval{\post(\phi)} \delta_P \Rightarrow |\Tg \eval{\post(\phi)} \mu_{k+1} [ \Theta ] | \leq 1$\\
\\
$\Tg \eval{\post(\phi)} \mu_{k+1} [\Theta] = trim([\closed{\phi} \cap \Theta])$\\
hence: $|\Tg \eval{\post(\phi)} \mu_{k+1} [ \Theta ] | = |trim([\closed{\phi} \cap \Theta])| \leq 1$\\
\\
(2) $G = p(\vec{x})$\\
Assume (without loss of generality): $p(\vec{y}) \neck G_1 ; G_2,!,G_3 ; G_4 \in P$\\
Show: $\Theta \subseteq \Dg \eval{p(\vec{x})} \delta_P \Rightarrow |\Tg \eval{p(\vec{x})} \mu_{k+1} [ \Theta ] | \leq 1$\\
\\
$\Tg \eval{p(\vec{x})} \mu_{k+1} [ \Theta ] = \mydownarrow \rho_{\vec{y}, \vec{x}}\overline{\exists}_{\vec{y}}(\mu_{k+1}\ (p(\vec{y}))\ \mydownarrow \rho_{\vec{x}, \vec{y}}\overline{\exists}_{\vec{x}}([\Theta])) \cap \Theta$ \\
hence: $|\Tg \eval{p(\vec{x})} \mu_{k+1} [ \Theta ] | = |\mydownarrow \rho_{\vec{y}, \vec{x}}\overline{\exists}_{\vec{y}}(\mu_{k+1}\ (p(\vec{y}))\ \mydownarrow \rho_{\vec{x}, \vec{y}}\overline{\exists}_{\vec{x}}([\Theta])) \cap \Theta | = |\mu_{k+1}\ (p(\vec{y}))\ \mydownarrow \rho_{\vec{x}, \vec{y}}\overline{\exists}_{\vec{x}}([\Theta])|$\\
and: $\mu_{k+1}\ (p(\vec{y}))\ \mydownarrow \rho_{\vec{x}, \vec{y}}\overline{\exists}_{\vec{x}}([\Theta])  = \mydownarrow \overline{\exists}_{\vec{y}} (\Tg \eval{G_1} \mu_k \mydownarrow \rho_{\vec{x}, \vec{y}}\overline{\exists}_{\vec{x}}([\Theta]) : \vec{\Psi})$
\\ where  $\vec{\Psi} = \left\{ \begin{array}{l l}
\Tg \eval{G_3 } \mu_k [\Phi] 
& if \ \Tg \eval{ G_2} \mu_k \mydownarrow \rho_{\vec{x}, \vec{y}}\overline{\exists}_{\vec{x}}([\Theta]) = \Phi : \vec{\Phi} \\
\Tg \eval{G_4} \mu_k \mydownarrow \rho_{\vec{x}, \vec{y}}\overline{\exists}_{\vec{x}}([\Theta]) 
& if \  \Tg \eval{G_2} \mu_k \mydownarrow \rho_{\vec{x}, \vec{y}}\overline{\exists}_{\vec{x}}([\Theta]) = [] \\
\end{array}	\right.\\
\\ 
|\mydownarrow \overline{\exists}_{\vec{y}} (\Tg \eval{G_1} \mu_k \mydownarrow \rho_{\vec{x}, \vec{y}}\overline{\exists}_{\vec{x}}([\Theta]) : \vec{\Psi})| = |\Tg \eval{G_1} \mu_k \mydownarrow \rho_{\vec{x}, \vec{y}}\overline{\exists}_{\vec{x}}([\Theta]) : \vec{\Psi}|$\\
\\
Show $\Theta \subseteq \Dg \eval{p(\vec{x})} \delta_P \Rightarrow |\Tg \eval{G_1} \mu_k \mydownarrow \rho_{\vec{x}, \vec{y}}\overline{\exists}_{\vec{x}}([\Theta]) : \vec{\Psi}| \leq 1$ in two steps:\\
\\
1 Show that each component cannot be longer than 1:\\
1a Show: $\Theta \subseteq \Dg \eval{p(\vec{x})} \delta_P \Rightarrow |\Tg \eval{G_1} \mu_k \mydownarrow \rho_{\vec{x}, \vec{y}}\overline{\exists}_{\vec{x}}([\Theta])| \leq 1$\\
1b Show: $\Theta \subseteq \Dg \eval{p(\vec{x})} \delta_P \Rightarrow |\Tg \eval{G_4} \mu_k \mydownarrow \rho_{\vec{x}, \vec{y}}\overline{\exists}_{\vec{x}}([\Theta])| \leq 1$\\
1c Show: $\Theta \subseteq \Dg \eval{p(\vec{x})} \delta_P \Rightarrow |\Tg \eval{G_3} \mu_k [\Phi]| \leq 1$\\
where $\Tg \eval{ G_2} \mu_k\ \mydownarrow \rho_{\vec{x}, \vec{y}}\overline{\exists}_{\vec{x}}([\Theta]) = \Phi : \vec{\Phi}$ \\
\\
2 Show that only one component can be longer than 0:\\
 $\Theta \subseteq \Dg \eval{p(\vec{x})}  \delta_P \Rightarrow \neg(|\Tg \eval{G_1} \mu_k \mydownarrow \rho_{\vec{x}, \vec{y}}\overline{\exists}_{\vec{x}}([\Theta])| \neq 0 \wedge |\vec{\Psi}| \neq 0)$\\
This is done thus:\\
2a Show:\\
 $\Theta \subseteq \Dg \eval{p(\vec{x})}  \delta_P \Rightarrow \neg(|\Tg \eval{G_1} \mu_k \mydownarrow \rho_{\vec{x}, \vec{y}}\overline{\exists}_{\vec{x}}([\Theta])| \neq 0 \wedge |\Tg \eval{G_4} \mu_k \mydownarrow \rho_{\vec{x}, \vec{y}}\overline{\exists}_{\vec{x}}([\Theta])| \neq 0)$\\
2b Show:\\
 $\Theta \subseteq \Dg \eval{p(\vec{x})} \delta_P \Rightarrow \neg(|\Tg \eval{G_1} \mu_k \mydownarrow \rho_{\vec{x}, \vec{y}}\overline{\exists}_{\vec{x}}([\Theta])| \neq 0 \wedge |\Tg \eval{G_3} \mu_k [\Phi]| \neq 0)\\
\indent where\ \Tg \eval{ G_2} \mu_k \mydownarrow \rho_{\vec{x}, \vec{y}}\overline{\exists}_{\vec{x}}([\Theta]) = \Phi : \vec{\Phi}$ \\
\\
$\Dg \eval{p(\vec{x})} \delta_P = \mydownarrow \rho_{\vec{y},\vec{x} } ( \overline{\forall}_{\vec{y}}(\delta_P (p(\vec{y})))) \\
 = \mydownarrow \rho_{\vec{y},\vec{x} } ( \overline{\forall}_{\vec{y}}(\mydownarrow \overline{\forall}_{\vec{y}}( \Dg \eval{ G_1} \delta_P \cap (S_G \eval{ G_2} \rightarrow \Dg \eval{G_3} \delta_P) \cap \Dg \eval{G_4} \delta_P \cap \Theta_1 \cap \Theta_2 )))\\
 = \mydownarrow \rho_{\vec{y},\vec{x} } \overline{\forall}_{\vec{y}}( \Dg \eval{ G_1} \delta_P \cap (S_G \eval{ G_2} \rightarrow \Dg \eval{G_3} \delta_P) \cap \Dg \eval{G_4} \delta_P \cap \Theta_1 \cap \Theta_2)\\
 \indent where\ \Theta_1 = mux( S_G \eval{G_1} , S_G \eval{G_4})\\  
 \indent and\ \Theta_2 = mux( S_G \eval{G_1}, S_G \eval{ G_2, G_3})\\
 = \mydownarrow \rho_{\vec{y},\vec{x} } \overline{\forall}_{\vec{y}}(\Dg \eval{ G_1} \delta_P)\\
 \indent \cap \mydownarrow \rho_{\vec{y},\vec{x} } \overline{\forall}_{\vec{y}} (S_G \eval{G_2} \rightarrow \Dg \eval{G_3} \delta_P)\\
 \indent \cap \mydownarrow \rho_{\vec{y},\vec{x} } \overline{\forall}_{\vec{y}} (\Dg \eval{G_4} \delta_P)\\
 \indent \cap \mydownarrow \rho_{\vec{y},\vec{x} } \overline{\forall}_{\vec{y}} (mux( S_G \eval{G_1} , S_G \eval{G_4}))\\
 \indent \cap \mydownarrow \rho_{\vec{y},\vec{x} } \overline{\forall}_{\vec{y}} (mux( S_G \eval{G_1}, S_G \eval{ G_2,G_3}))$\\
\\
1a Show: $\Theta \subseteq \Dg \eval{p(\vec{x})} \delta_P \Rightarrow |\Tg \eval{G_1} \mu_k \mydownarrow \rho_{\vec{x}, \vec{y}}\overline{\exists}_{\vec{x}}([\Theta])| \leq 1$\\
$\Theta \subseteq \Dg \eval{p(\vec{x})} \delta_P \Rightarrow \Theta \subseteq \mydownarrow \rho_{\vec{y},\vec{x} } \overline{\forall}_{\vec{y}}(\Dg \eval{ G_1}) \delta_P$\\
hence (by (5) stated above): $\mydownarrow \rho_{\vec{x}, \vec{y}}\overline{\exists}_{\vec{x}}(\Theta) \subseteq \Dg \eval{G_1} \delta_P$\\
hence by assumption: $|\Tg \eval{G_1} \mu_k \mydownarrow \rho_{\vec{x}, \vec{y}}\overline{\exists}_{\vec{x}}([\Theta])| \leq 1$\\
\\
1b Show: $\Theta \subseteq \Dg \eval{p(\vec{x})} \delta_P \Rightarrow |\Tg \eval{G_4} \mu_k \mydownarrow \rho_{\vec{x}, \vec{y}}\overline{\exists}_{\vec{x}}([\Theta])| \leq 1$\\
$\Theta \subseteq \Dg \eval{p(\vec{x})} \delta_P \Rightarrow \Theta \subseteq	\mydownarrow \rho_{\vec{y},\vec{x} } \overline{\forall}_{\vec{y}} (\Dg \eval{G_4} \delta_P)$\\
hence (again by (5) above): $\mydownarrow \rho_{\vec{x}, \vec{y}}\overline{\exists}_{\vec{x}}(\Theta) \subseteq \Dg \eval{G_4} \delta_P$\\
hence by assumption: $|\Tg \eval{G_4} \mu_k \mydownarrow \rho_{\vec{x}, \vec{y}}\overline{\exists}_{\vec{x}}([\Theta])| \leq 1$\\
\\
1c Show: $ \Theta \subseteq \Dg \eval{p(\vec{x})} \delta_P \Rightarrow |\Tg \eval{G_3} \mu_k [\Phi])| \leq 1$\\ % spurious bracket!?
where $\Tg \eval{ G_2} \mu_k \mydownarrow \rho_{\vec{x}, \vec{y}}\overline{\exists}_{\vec{x}}([\Theta]) = \Phi : \vec{\Phi} \\
\Theta \subseteq \Dg \eval{p(\vec{x})}  \delta_P\Rightarrow \Theta \subseteq \mydownarrow \rho_{\vec{y},\vec{x} } \overline{\forall}_{\vec{y}} (S_G \eval{G_2} \rightarrow \Dg \eval{G_3} \delta_P)$\\
hence (again by (5) above): $\mydownarrow \rho_{\vec{x}, \vec{y}}\overline{\exists}_{\vec{x}}(\Theta) \subseteq (S_G \eval{G_2} \rightarrow \Dg \eval{G_3} \delta_P)$\\
hence (by (1) stated above): $\mydownarrow \rho_{\vec{x}, \vec{y}}\overline{\exists}_{\vec{x}}(\Theta) \cap S_G \eval{G_2} \subseteq \Dg \eval{G_3} \delta_P$\\
by Theorem 1: $\bigcup(\Phi:\vec{\Phi})= \bigcup (\Tg \eval{ G_2} \mu_k \mydownarrow \rho_{\vec{x}, \vec{y}}\overline{\exists}_{\vec{x}}([\Theta])) \subseteq \mydownarrow \rho_{\vec{x}, \vec{y}}\overline{\exists}_{\vec{x}}([\Theta]) \cap S_G \eval{G_2}$\\
therefore (since $\Phi \subseteq \bigcup(\Phi : \vec{\Phi})$):
$\Phi \subseteq \mydownarrow \rho_{\vec{x}, \vec{y}}\overline{\exists}_{\vec{x}}([\Theta]) \cap S_G \eval{G_2} \subseteq \Dg \eval{G_3} \delta_P$\\
by assumption: $ |\Tg \eval{G_3} \mu_k [\Phi]| <= 1$\\
\\
2a Show:\\
 $\Theta \subseteq \Dg \eval{p(\vec{x})} \delta_P \Rightarrow \neg(|\Tg \eval{G_1} \mu_k \mydownarrow \rho_{\vec{x}, \vec{y}}\overline{\exists}_{\vec{x}}([\Theta])| \neq 0 \wedge |\Tg \eval{G_4} \mu_k \mydownarrow \rho_{\vec{x}, \vec{y}}\overline{\exists}_{\vec{x}}([\Theta])| \neq 0)\\
\Theta \subseteq \Dg \eval{p(\vec{x})} \delta_P \Rightarrow \Theta \subseteq \mydownarrow \rho_{\vec{y},\vec{x} } \overline{\forall}_{\vec{y}} (mux( S_G \eval{G_1} , S_G \eval{G_4}))$\\
hence (by (5) stated above): $\mydownarrow \rho_{\vec{x}, \vec{y}}\overline{\exists}_{\vec{x}}(\Theta) \subseteq mux( S_G \eval{G_1} , S_G \eval{G_4})$\\
hence (by (2) stated above):\\
 $(\mydownarrow \rho_{\vec{x}, \vec{y}}\overline{\exists}_{\vec{x}}(\Theta) \cap S_G \eval{G_1} = \{false\}) \vee (\mydownarrow \rho_{\vec{x}, \vec{y}}\overline{\exists}_{\vec{x}}(\Theta) \cap S_G \eval{G_4} = \{false\})$\\
by Theorem 1: $\mydownarrow \rho_{\vec{x}, \vec{y}}\overline{\exists}_{\vec{x}}(\Theta) \cap S_G \eval{G_1} = \{false\} \Rightarrow \Tg \eval{G_1} \mu_k\ \mydownarrow \rho_{\vec{x}, \vec{y}}\overline{\exists}_{\vec{x}}([\Theta]) = []$\\
hence: $\mydownarrow \rho_{\vec{x}, \vec{y}}\overline{\exists}_{\vec{x}}(\Theta) \cap S_G \eval{G_1} = \{false\} \Rightarrow |\Tg \eval{G_1} \mu_k\ \mydownarrow \rho_{\vec{x}, \vec{y}}\overline{\exists}_{\vec{x}}([\Theta])| = 0$\\
similarly: $\mydownarrow \rho_{\vec{x}, \vec{y}}\overline{\exists}_{\vec{x}}(\Theta) \cap S_G \eval{G_4} = \{false\} \Rightarrow \Tg \eval{G_4} \mu_k\ \mydownarrow \rho_{\vec{x}, \vec{y}}\overline{\exists}_{\vec{x}}([\Theta]) = []$\\
hence: $\mydownarrow \rho_{\vec{x}, \vec{y}}\overline{\exists}_{\vec{x}}(\Theta) \cap S_G \eval{G_4} = \{false\} \Rightarrow |\Tg \eval{G_4} \mu_k\ \mydownarrow \rho_{\vec{x}, \vec{y}}\overline{\exists}_{\vec{x}}([\Theta])| = 0$\\
therefore: $(|\Tg \eval{G_1} \mu_k\ \mydownarrow \rho_{\vec{x}, \vec{y}}\overline{\exists}_{\vec{x}}([\Theta])| = 0) \vee (|\Tg \eval{G_4} \mu_k \mydownarrow \rho_{\vec{x}, \vec{y}}\overline{\exists}_{\vec{x}}([\Theta])| = 0)$\\
hence: $\neg((|\Tg \eval{G_1} \mu_k\ \mydownarrow \rho_{\vec{x}, \vec{y}}\overline{\exists}_{\vec{x}}([\Theta])| \neq 0) \wedge (|\Tg \eval{G_4} \mu_k\ \mydownarrow \rho_{\vec{x}, \vec{y}}\overline{\exists}_{\vec{x}}([\Theta])| \neq 0))$\\
\\
2b Show: $\Theta \subseteq \Dg \eval{p(\vec{x})} \delta_P \Rightarrow \neg(|\Tg \eval{G_1} \mu_k \mydownarrow \rho_{\vec{x}, \vec{y}}\overline{\exists}_{\vec{x}}([\Theta])| \neq 0 \wedge |\Tg \eval{G_3} \mu_k [\Phi]| \neq 0)$\\
where $\Tg \eval{ G_2} \mu_k \mydownarrow \rho_{\vec{x}, \vec{y}}\overline{\exists}_{\vec{x}}([\Theta]) = \Phi : \vec{\Phi}$ \\
$\Theta \subseteq \Dg \eval{p(\vec{x})} \delta_P \Rightarrow \Theta \subseteq \mydownarrow \rho_{\vec{y},\vec{x} } \overline{\forall}_{\vec{y}} (mux( S_G \eval{G_1}, S_G \eval{ G_2, G_3}))$\\
hence (again by (5) above): $\mydownarrow \rho_{\vec{x}, \vec{y}}\overline{\exists}_{\vec{x}}(\Theta) \subseteq mux( S_G \eval{G_1}, S_G \eval{ G_2,G_3})$\\
hence (again by (2) above):\\
 $(\mydownarrow \rho_{\vec{x}, \vec{y}}\overline{\exists}_{\vec{x}}(\Theta) \cap S_G \eval{G_1} = \{false\}) \vee (\mydownarrow \rho_{\vec{x}, \vec{y}}\overline{\exists}_{\vec{x}}(\Theta) \cap S_G \eval{G_2,G_3} = \{false\})$\\
by Theorem 1: $\Phi \subseteq \bigcup (\Tg \eval{ G_2} \mu_k\ \mydownarrow \rho_{\vec{x}, \vec{y}}\overline{\exists}_{\vec{x}}([\Theta])) \subseteq \mydownarrow \rho_{\vec{x}, \vec{y}}\overline{\exists}_{\vec{x}}(\Theta) \cap S_G \eval{G_2}$ \\
by Theorem 1: $\bigcup (\Tg \eval{G_3} \mu_k\ [\Phi]) \subseteq \Phi \cap S_G \eval{G_3} \subseteq \mydownarrow \rho_{\vec{x}, \vec{y}}\overline{\exists}_{\vec{x}}(\Theta) \cap S_G \eval{G_2} \cap S_G \eval{G_3}$\\
hence: $\bigcup (\Tg \eval{G_3} \mu_k [\Phi]) \subseteq \mydownarrow \rho_{\vec{x}, \vec{y}}\overline{\exists}_{\vec{x}}(\Theta) \cap S_G \eval{G_2,G_3}$\\
hence: $\mydownarrow \rho_{\vec{x}, \vec{y}}\overline{\exists}_{\vec{x}}(\Theta) \cap S_G \eval{G_2,G_3} = \{false\} \Rightarrow \Tg \eval{G_3} \mu_k [\Phi] = []$\\
hence: $ \mydownarrow \rho_{\vec{x}, \vec{y}}\overline{\exists}_{\vec{x}}(\Theta) \cap S_G \eval{G_2,G_3} = \{false\} \Rightarrow |\Tg \eval{G_3} \mu_k [\Phi]| = 0$\\
also (by Theorem 1): $\mydownarrow \rho_{\vec{x}, \vec{y}}\overline{\exists}_{\vec{x}}(\Theta) \cap S_G \eval{G_1} = \{false\} \Rightarrow \Tg \eval{G_1} \mu_k\ \mydownarrow \rho_{\vec{x}, \vec{y}}\overline{\exists}_{\vec{x}}([\Theta]) = []$\\
hence: $\mydownarrow \rho_{\vec{x}, \vec{y}}\overline{\exists}_{\vec{x}}(\Theta) \cap S_G \eval{G_1} = \{false\} \Rightarrow |\Tg \eval{G_1} \mu_k\ \mydownarrow \rho_{\vec{x}, \vec{y}}\overline{\exists}_{\vec{x}}([\Theta])| = 0$\\
hence:$ (|\Tg \eval{G_1} \mu_k \mydownarrow \rho_{\vec{x}, \vec{y}}\overline{\exists}_{\vec{x}}([\Theta])| = 0) \vee (|\Tg \eval{G_3} \mu_k\ [\Phi]| = 0)$\\
hence: $\neg((|\Tg \eval{G_1} \mu_k\ \mydownarrow \rho_{\vec{x}, \vec{y}}\overline{\exists}_{\vec{x}}([\Theta])| \neq 0) \vee (|\Tg \eval{G_3} \mu_k [\Phi]| \neq 0))$\\
\\
2.2 Induction Step:\\
$G = G1, G2:$\\
Assume: $\Theta_1 \subseteq \Dg \eval{G_1} \delta_P \Rightarrow |\Tg \eval{G_1} \mu_{k+1} [ \Theta_1 ] | \leq 1$\\
And: $\Theta_2 \subseteq \Dg \eval{G_2} \delta_P \Rightarrow |\Tg \eval{G_2} \mu_{k+1} [ \Theta_2 ] | \leq 1$\\
Show: $\Theta \subseteq \Dg \eval{G} \Rightarrow |\Tg \eval{G} \mu_{k+1} [ \Theta ] | \leq 1$\\
where $\mu_{k+1} = \Tprog \eval{P} \mu_k$\\
\\
$\Dg \eval{G} \delta_P = (S_G \eval{ G_2} \rightarrow \Dg \eval{ G_1} \delta_P) \cap (S_G \eval{ G_1} \rightarrow \Dg \eval{ G_2} \delta_P)$\\
therefore if $\Theta \subseteq \Dg \eval{G} \delta_P$\\
\indent then $\Theta \subseteq (S_G \eval{ G_1} \rightarrow \Dg \eval{ G_2} \delta_P)$\\
\indent and hence $\Theta \cap S_G \eval{G_1} \subseteq \Dg \eval{G_2} \delta_P$\\
similarly if $\Theta \subseteq \Dg \eval{G} \delta_P$\\
\indent then $\Theta \subseteq (S_G \eval{ G_2} \rightarrow \Dg \eval{ G_1} \delta_P)$\\
\indent and hence $\Theta \cap S_G \eval{G_2} \subseteq \Dg \eval{G_1} \delta_P$\\
by Lemma 2: $\Tg \eval{G} \mu_{k+1} [\Theta] = \Tg \eval{G} \mu_{k+1} [\Theta \cap S_G \eval{G}]$\\
applying the definition of $\Tg$:\\
$\Tg \eval{G} \mu_{k+1} [\Theta \cap S_G \eval{G}] = \Tg \eval{G_2} \mu_{k+1} (\Tg \eval{ G_1} \mu_{k+1} [\Theta \cap S_G \eval{G_1, G_2}])$\\
now notice that: $\Theta \cap S_G \eval{G_1,G_2}\\
\indent \indent = \Theta \cap S_G \eval{G_1} \cap S_G \eval{G_2}\\
\indent \indent \subseteq \Theta \cap S_G \eval{G_2} \\
\indent \indent \subseteq \Dg \eval{G_1} \delta_P$\\
hence by assumption: $|\Tg \eval{ G_1} \mu_{k+1} [\Theta \cap S_G \eval{G_1,G_2}]| <= 1$\\
\\
distinguish two cases:\\
 (a) $|\Tg \eval{ G_1} \mu_{k+1} [\Theta \cap S_G \eval{G_1,G_2} ]| = 0$,\\
 (b) $|\Tg \eval{ G_1} \mu_{k+1} [\Theta \cap S_G \eval{G_1,G_2}]| = 1$\\
\\
(a) $|\Tg \eval{ G_1} \mu_{k+1} [\Theta \cap S_G \eval{G_1,G_2} ]| = 0\\
\Tg \eval{ G_1} \mu_{k+1} [\Theta \cap S_G \eval{G_1,G_2}] = []\\
\Tg \eval{G} \mu_{k+1} [\Theta \cap S_G \eval{G_1,G_2}] = \Tg \eval{G_2} \mu_{k+1} [] = []$\\
hence: $|\Tg \eval{G} \mu_{k+1} [\Theta \cap S_G \eval{G_1,G_2}]| <= 1$\\
hence by Lemma 2 (remembering $G = G_1, G_2$): $|\Tg \eval{G} \mu_{k+1} [\Theta]| <= 1$\\
\\
(b) $|\Tg \eval{ G_1} \mu_{k+1} [\Theta \cap S_G \eval{G_1,G_2}]| = 1\\
\Tg \eval{ G_1} \mu_{k+1} [\Theta \cap S_G \eval{G_1,G_2} ] = [\Psi]$\\
therefore: $\bigcup(\Tg \eval{ G_1} \mu_{k+1} [\Theta \cap S_G \eval{G_1,G_2}] ) = \Psi$\\
by Theorem 1: $\Psi \subseteq \Theta \cap S_G \eval{G_1,G_2} \cap S_G \eval{G_1} \subseteq \Theta \cap S_G \eval{G_1}$\\
hence since $\Theta \cap S_G \eval{G_1} \subseteq \Dg \eval{G_2} \delta_P$ (see above): $\Psi \subseteq \Dg \eval{G_2} \delta_P$\\
hence by assumption: $|\Tg \eval{G_2} \mu_{k+1} [\Psi] | \leq 1$\\
hence (again using Lemma 2): $|\Tg \eval{G_2} \mu_{k+1} (\Tg \eval{ G_1} \mu_{k+1} [\Theta \cap S_G \eval{G_1,G_2}])| \\
\indent = |\Tg \eval{G} \mu_{k+1} [\Theta\cap S_G \eval{G_1,G_2}] | \\
\indent = |\Tg \eval{G} \mu_{k+1} [\Theta]| <= 1$\\
QED   \\   
\\
\subsection{Abstraction Proofs}
\subsubsection{Proposition 1: 
\text{If
$\Theta_1 \subseteq \concr{\vec{x}}{f_1}$
and
$\concr{\vec{x}}{f_1} \subseteq \Theta_2$
then
$\concr{\vec{x}}{f_1 \Rightarrow f_2} \subseteq \Theta_1 \rightarrow \Theta_2$}}

$
\concr{\vec{x}}{f_1 \Rightarrow f_2} \\
= \bigcup \{ \concr{\vec{x}}{f} \mid f \models f_1 \Rightarrow f_2 \} \\
= \bigcup \{ \Theta \mid \abstr{\vec{x}}{\Theta} \models f_1 \Rightarrow f_2 \} \\
= \bigcup \{ \Theta \mid (\abstr{\vec{x}}{\Theta} \models f_1) \Rightarrow (\abstr{\vec{x}}{\Theta} \models f_2) \} \\
= \bigcup \{ \Theta \mid (\Theta \subseteq \concr{\vec{x}}{f_1}) \Rightarrow (\Theta \subseteq \concr{\vec{x}}{f_2}) \} \\
\subseteq \bigcup \{ \Theta \mid (\Theta \subseteq \Theta_1) \Rightarrow (\Theta \subseteq \Theta_2) \} \\
= \bigcup \{ \Theta \mid (\Theta \subseteq \Theta_1 \cap \Theta_2) \vee (\Theta \not\subseteq \Theta_1) \} \\
= \bigcup \{ \Theta \mid \Theta \subseteq (\Theta_1 \cap \Theta_2) \cup (Con \setminus \Theta_1) \} \\
= \bigcup \{ \Theta \mid \Theta \cap \Theta_1 \subseteq \Theta_2 \} \\
= \Theta_1 \rightarrow \Theta_2
$ 

\subsubsection{Proposition 2: $\concr{\vec{x}}{\abs{mux_{\vec{x}}} (\dk{\Theta_1}, \dk{\Theta_2})} \subseteq mux(\Theta_1, \Theta_2) $}

Proof:\\
First notice that by the definition of the Galois connection (i.e. of $\concr{}{}$ and $\abstr{}{}$ the following:
$\concr{\vec{x}}{\abs{mux_{\vec{x}}} (\Theta_1^{DK}, \Theta_2^{DK})} \subseteq mux(\Theta_1, \Theta_2)$ \\
is equivalent to: $\abstr{\vec{x}}{\Psi} \models \abs{mux_{\vec{x}}} (\Theta_1^{DK}, \Theta_2^{DK}) \rightarrow \Psi \subseteq mux(\Theta_1, \Theta_2)$
\\[2ex]
Now: $\abstr{\vec{x}}{\Psi} \models \abs{mux_{\vec{x}}} (\Theta_1^{DK}, \Theta_2^{DK})$ iff for each clause\ in $\abstr{\vec{x}}{\Psi}$ there is a clause in $\abs{mux_{\vec{x}}} (\Theta_1^{DK}, \Theta_2^{DK})$ that is entailed by it, ie:\\ 
$\forall \psi \in \Psi _{.} \exists Y \subseteq vars(\vec{x}) _{.} ( \forall \theta_1 \in {\dk{\Theta_1}}_{.} \forall \theta_2 \in {\dk{\Theta_2}}_{.}\\
\indent \indent ( \overline{\exists}_Y (\theta_1) \wedge \overline{\exists}_Y (\theta_2) = false ) \wedge \abstr{\vec{x}}{\psi} \models \bigwedge Y )$ \\
Since $\abs{mux_{\vec{x}}} (\Theta_1^{DK}, \Theta_2^{DK})$ contains only positive (ie non-negated) literals, only the positive literals entailed by $\abstr{\vec{x}}{\psi}$ are relevant.\\
Now, the positive literals entailed by $\abstr{\vec{x}}{\psi}$ are exactly $vars(\vec{x}) \cap fix(\psi)$.\\
Therefore: $\psi \in \concr{\vec{x}}{\abs{mux_{\vec{x}}} (\Theta_1^{DK}, \Theta_2^{DK})}$\\ 
\indent iff $\exists Y \subseteq (vars(\vec{x}) \cap fix(\psi)) _{.} (\forall \theta_1 \in {\dk{\Theta_1}}_{.} \forall \theta_2 \in {\dk{\Theta_2}}_{.} ( \overline{\exists}_Y (\theta_1) \wedge \overline{\exists}_Y (\theta_2) = false ) )$\\
\\
Now observe that the following three things hold:\\ 
(1) $\forall \phi \in \Phi _{.} \exists \phi' \in \dk{\Phi} _{.} (\phi \models \phi')$\\
(2) $((f_1 \models f_1') \wedge (f_2 \models f_2') \wedge (f_1' \wedge f_2' = false ) ) \rightarrow f_1 \wedge f_2 = false\\$
(3) $\phi \models \phi' \rightarrow \overline{\exists}_Y (\phi) \models \overline{\exists}_Y (\phi')$\\
Therefore from $\forall \theta_1' \in {\dk{\Theta_1}}_{.} \forall \theta_2' \in {\dk{\Theta_2}}_{.} ( \overline{\exists}_Y (\theta_1') \wedge \overline{\exists}_Y (\theta_2') = false )$ \\
it follows: $\forall \theta_1 \in {\Theta_1} _{.} \forall \theta_2 \in {\Theta_2} _{.} (\overline{\exists}_Y (\theta_1) \wedge \overline{\exists}_Y (\theta_2) = false)$\\
And thus: $ \overline{\exists}_Y (\Theta_1) \cap \overline{\exists}_Y (\Theta_2)  = \{ false \}$\\
Hence the following entailment holds:\\
 $\forall \phi _{.}( \exists Y \subseteq (vars(\vec{x}) \cap fix(\psi)) _{.} (\forall \theta_1 \in {\dk{\Theta_1}}_{.} \forall \theta_2 \in {\dk{\Theta_2}}_{.} ( \overline{\exists}_Y (\theta_1) \wedge \overline{\exists}_Y (\theta_2) = false ) )$ \\
 $\models$ \\
\indent $\exists Y \subseteq fix(\phi) _{.} (\overline{\exists}_Y (\Theta_1) \cap \overline{\exists}_Y (\Theta_2)  = \{ false \}) )$\\
Therefore: $\forall \phi _{.} (\phi \in \concr{\vec{x}}{\abs{mux_{\vec{x}}} (\Theta_1^{DK}, \Theta_2^{DK})} \rightarrow \phi \in mux(\Theta_1, \Theta_2))$\\
From which it follows: $\concr{\vec{x}}{\abs{mux_{\vec{x}}} (\dk{\Theta_1}, \dk{\Theta_2})} \subseteq mux(\Theta_1, \Theta_2)
$\\

\subsection{Theorem 3:  $\forall i \in \mathbb{N}: \concr{vars(G)}{\abs{\Dg} \eval{G} \adelta_i} \subseteq \Dg \eval{G} \delta_i$ where $\adelta_i / \delta_i$ are the results of $i$ applications of $\abs{\Dp} \eval{P} / \Dp \eval{P}$ to $ \adelta_{\top}/ \delta_{\top}$ respectively.} 

Proof by nested induction on:\\
1. $i$,\\
2. the structure of $G$:\\
\\
notice first that:
$\concr{vars(\vec{x})}{ \abs{\rho_{\vec{y},\vec{x} } } \abs{\overline{\forall}_{\vec{y}}}(f)} \subseteq  \rho_{\vec{y},\vec{x}} \overline{\forall}_{\vec{y}} ( \concr{vars(\vec{y})} {f})  $
\\
1 Base Case: $i = 0$\\
$
\adelta_0 = \adelta_{\top}\\
\delta_0 = \delta_{\top}\\
$
Show:  $\concr{vars(G)}{\abs{\Dg} \eval{G} \adelta_{\top}} \subseteq \Dg \eval{G} \delta_{\top}$
\\
Induction on structure of $G$:\\
1.1 Two base cases: (1) $G = post(\phi)$, (2) $G = p(\vec{x})$
\\
(1) $G = post(\phi)$\\
$
\concr{vars(\phi)}{\abs{\Dg} \eval{post(\phi)} \adelta_{\top}} \\
= \concr{vars(\phi)}{true} \\
= \closed{true}\\
= \Dg \eval{post(\phi)} \delta_{\top} \\
$
hence: $\concr{vars(\phi)}{\abs{\Dg} \eval{post(\phi)} \adelta_{\top}} \subseteq \Dg \eval{post(\phi)} \delta_{\top}$
\\
\\
(2) $G = p(\vec{x})$\\
$
\concr{vars(\vec{x})}{ \abs{\Dg} \eval{p(\vec{x})} \adelta_{\top}}\\
= \concr{vars(\vec{x})}{\abs{\rho_{\vec{y},\vec{x} }}  \abs{\overline{\forall}_{\vec{y}}} (true)}\\
\subseteq \rho_{\vec{y},\vec{x}} \overline{\forall}_{\vec{y}} ( \concr{vars(\vec{y})} {true})\\
= \rho_{\vec{y},\vec{x}} \overline{\forall}_{\vec{y}} (\closed{true})\\
= \Dg \eval{p(\vec{x})} \delta_{\top} 
$
\\
\\
1.2 Induction step: $G = G_1, G_2$\\
Assume: $\concr{vars(G_{1/2})}{\abs{\Dg} \eval{G_{1/2}} \adelta_{\top}} \subseteq \Dg \eval{G_{1/2}} \delta_{\top}$\\

$\concr{vars(G_1,G_2)}{\abs{\Dg} \eval{G_1,G_2} \adelta_{\top}} \\
= \concr{vars(G_1,G_2)}{(\abs{S_G} \eval{ G_2} \Rightarrow \abs{\Dg} \eval{ G_1} \adelta_{\top} ) \wedge (\abs{S_G} \eval{ G_1} \Rightarrow \abs{\Dg} \eval{ G_2} \adelta_{\top})}\\
\subseteq \concr{vars(G_1,G_2)}{\abs{S_G} \eval{ G_2} \Rightarrow \abs{ \Dg} \eval{ G_1} \adelta_{\top}} \cap \concr{vars(G_1,G_2)}{\abs{S_G} \eval{ G_1} \Rightarrow \abs{\Dg} \eval{ G_2} \adelta_{\top}} \\ 
\indent \text{(by monotonicity i.e. $\concr{vars(G_1, G_2)}{f_1 \wedge f_2} \subseteq \concr{vars(G_1, G_2)}{f_i}$)} \\
\subseteq (S_G \eval{G_2} \rightarrow \Dg \eval{G_1} \delta_{\top}) \cap (S_G \eval{G_1} \rightarrow \Dg \eval{G_2} \delta_{\top}) \\
\indent \text{(by Proposition 1 and Proposition 3
and the induction assumption)}\\ 
= \Dg \eval{G_1,G_2} \delta_{\top}
$
\\
\\
2 Induction step: $ i = k+1$\\
Assume: $ \concr{vars(G)}{\abs{\Dg} \eval{G} \adelta_k} \subseteq \Dg \eval{G} \delta_k$\\
Show: $ \concr{vars(G)}{\abs{\Dg} \eval{G} \adelta_{k+1}} \subseteq \Dg \eval{G} \delta_{k+1}$\\
where $\delta_{k+1} = \Dp \eval{P} \delta_k$ and $\adelta_{k+1} = \abs{\Dp} \eval{P} \adelta_k$\\
\\
Induction on structure of G:\\
2.1 Two base cases: (1) $G = post(\phi)$, (2) $G = p(\vec{x})$
\\
(1) $G = post(\phi)$\\
$
\concr{vars(\phi)}{\abs{\Dg} \eval{post(\phi)} \adelta_{k+1}} \\
= \concr{vars(\phi)}{true} \\
= \closed{true}\\
= \Dg \eval{post(\phi)} \delta_{k+1} \\
$
hence: $\concr{vars(\phi)}{\abs{\Dg} \eval{post(\phi)} \adelta_{\top}} \subseteq \Dg \eval{post(\phi)} \delta_{\top}$
\\
\\
(2) $G = p(\vec{x})$\\
Assume (without loss of generality): $p(\vec{y}) \neck G_1 ; G_2,!,G_3 ; G_4 \in P$\\
\\
$\concr{vars(\vec{x})}{\abs{\Dg} \eval{p(\vec{x})} \adelta_{k+1}} \\
= \concr{vars(\vec{x})}{ \abs{\rho_{\vec{y},\vec{x} }} ( \abs{\overline{\forall}_{\vec{y}}}  ( \abs{\Dg} \eval{p(\vec{y})} \adelta_{k+1}))}\\
= \concr{vars(\vec{x})}{ \abs{\rho_{\vec{y},\vec{x} }} ( \abs{\overline{\forall}_{\vec{y}}}  (  \abs{\overline{\forall}_{\vec{y}}}( \abs{\Dg} \eval{ G_1} \adelta_{k} \wedge (\abs{S_G} \eval{ G_2} \Rightarrow \abs{\Dg} \eval{G_3} \adelta_{k}) \wedge \abs{\Dg} \eval{G_4} \adelta_{k} \\
\indent \wedge \abs{mux_{vars(\vec{y})}} ( \dk{S_G} \eval{G_1} , \dk{S_G} \eval{G_4}) \\
\indent \wedge \abs{mux_{vars(\vec{y})} }( \dk{S_G} \eval{G_1}, \dk{S_G} \eval{ G_2, G_3})) )  )}\\
= \concr{vars(\vec{x})}{ \abs{\rho_{\vec{y},\vec{x} }} ( \abs{\overline{\forall}_{\vec{y}}}  (   \abs{\Dg} \eval{ G_1} \adelta_{k} \wedge (\abs{S_G} \eval{ G_2} \Rightarrow \abs{\Dg} \eval{G_3} \adelta_{k}) \wedge \abs{\Dg} \eval{G_4} \adelta_{k} \\
\indent \wedge \abs{mux_{vars(\vec{y})}} ( \dk{S_G} \eval{G_1} , \dk{S_G} \eval{G_4}) \\
\indent \wedge \abs{mux_{vars(\vec{y})} }  ( \dk{S_G} \eval{G_1}, \dk{S_G} \eval{ G_2, G_3}) )  )}\\
\subseteq \rho_{\vec{y},\vec{x} } \overline{\forall}_{\vec{y}} (\concr{vars(\vec{y})}{  \abs{\Dg} \eval{ G_1} \adelta_{k} \wedge (\abs{S_G} \eval{ G_2} \Rightarrow \abs{\Dg} \eval{G_3} \adelta_{k}) \wedge \abs{\Dg} \eval{G_4} \adelta_{k} \\
\indent \wedge \abs{mux_{vars(\vec{y})}} ( \dk{S_G} \eval{G_1} , \dk{S_G} \eval{G_4}) \\
\indent \wedge \abs{mux_{vars(\vec{y})} }  ( \dk{S_G} \eval{G_1}, \dk{S_G} \eval{ G_2, G_3}) } )\\
\subseteq \rho_{\vec{y},\vec{x} } \overline{\forall}_{\vec{y}} (\concr{vars(\vec{y})}{  \abs{\Dg} \eval{ G_1} \adelta_{k}} \cap (\concr{vars(\vec{y})}{ \abs{S_G} \eval{ G_2}} \rightarrow \concr{vars(\vec{y})}{ \abs{\Dg} \eval{G_3} \adelta_{k}}) \cap \concr{vars(\vec{y})}{ \abs{\Dg} \eval{G_4} \adelta_{k}} \\
\indent \cap \concr{vars(\vec{y})}{  \abs{mux_{vars(\vec{y})}} ( \dk{S_G} \eval{G_1} , \dk{S_G} \eval{G_4})}  ) \\
\indent \cap \concr{vars(\vec{y})}{  \abs{mux_{vars(\vec{y})}} ( \dk{S_G} \eval{G_1} , \dk{S_G} \eval{G_2,G_3})}  ) \\
\subseteq \rho_{\vec{y},\vec{x} } \overline{\forall}_{\vec{y}} ( \Dg \eval{ G_1} \delta_k \cap (S_G \eval{ G_2} \rightarrow \Dg \eval{G_3} \delta_k) \cap \Dg \eval{G_4} \delta_k \\
\indent \cap mux( {S_G} \eval{G_1} , {S_G} \eval{G_4}) \\
\indent \cap mux( {S_G} \eval{G_1},{S_G} \eval{ G_2, G_3})) \\
= \Dg \eval{p(\vec{x})} \delta_{k+1}
$
\\
\\
2.2 Induction step: $G = G_1, G_2$\\
Assume: $\concr{vars(G_{1/2})}{\abs{\Dg} \eval{G_{1/2}}} \subseteq \Dg \eval{G_{1/2}}$\\
again, notice that: (1) $A \subseteq B \Rightarrow \concr{B}{f} \subseteq \concr{A}{f} $\\
and: $ vars(G_1,G_2) = vars(G_1) \cup vars(G_2)$\\
and hence: ($2^1$) $vars(G_1) \subseteq vars(G_1,G_2)$\\
and similarly: ($2^2$) $vars(G_2) \subseteq vars(G_1,G_2)$

$\concr{vars(G_1,G_2)}{\abs{\Dg} \eval{G_1,G_2} \adelta_{k+1}} \\
= \concr{vars(G_1,G_2)}{(\abs{S_G} \eval{ G_2} \Rightarrow \abs{\Dg} \eval{ G_1} \adelta_{k+1} ) \wedge (\abs{S_G} \eval{ G_1} \Rightarrow \abs{\Dg} \eval{ G_2} \adelta_{k+1})}\\
\subseteq \concr{vars(G_1,G_2)}{(\abs{S_G} \eval{ G_2} \Rightarrow \abs{\Dg} \eval{ G_1} \adelta_{k+1} )} \cap \concr{vars(G_1,G_2)}{(\abs{S_G} \eval{ G_1} \Rightarrow \abs{\Dg} \eval{ G_2} \adelta_{k+1})}\\
\indent \text{(by monotonicity i.e. $\concr{vars(G_1, G_2)}{f_1 \wedge f_2} \subseteq \concr{vars(G_1, G_2)}{f_i}$)} \\
\subseteq \concr{vars(G_1,G_2)}{\abs{S_G} \eval{ G_2} } \rightarrow \concr{vars(G_1,G_2)}{\abs{ \Dg} \eval{ G_1} \adelta_{k+1}} 
\cap \concr{vars(G_1,G_2)}{ \abs{S_G} \eval{ G_1} } \rightarrow \concr{vars(G_1,G_2)}{ \abs{\Dg} \eval{ G_2} \adelta_{k+1}}\\
\indent \text{(by Proposition 1 and Proposition 3
and the induction assumption)}\\ 
\subseteq \concr{vars(G_2)}{\abs{S_G} \eval{ G_2} } \rightarrow \concr{vars(G_1)}{\abs{ \Dg} \eval{ G_1} \adelta_{k+1}} \cap \concr{vars(G_1)}{ \abs{S_G} \eval{ G_1} } \rightarrow \concr{vars(G_2)}{ \abs{\Dg} \eval{ G_2} \adelta_{k+1}}\\
\indent \text{(by (1), ($2^1$) and ($2^2$) above)}\\
\subseteq S_G \eval{G_2} \rightarrow \Dg \eval{G_1} \delta_{k+1} \cap S_G \eval{G_1} \rightarrow \Dg \eval{G_2} \delta_{k+1}\\
= \Dg \eval{G_1,G_2} \delta_{k+1}
$\\
QED

%\newpage
%\section{Appendix B - Changes}
%\input{appB}

\end{document}